\documentclass[traditabstract]{aa} 
\usepackage{graphicx}
\usepackage{natbib}

\usepackage{savesym}
\usepackage{txfonts}
\savesymbol{iint}
\savesymbol{iiint}
\savesymbol{iiiint}
\savesymbol{idotsint}
\usepackage{amsmath}
\restoresymbol{AMS}{iint}
\restoresymbol{AMS}{iiint}
\restoresymbol{AMS}{iiiint}
\restoresymbol{AMS}{idotsint}

\usepackage{aas_macros}
\usepackage{dsfont}
\usepackage{fixltx2e}
\begin{document}
\title{Unbiased flux calibration methods for spectral-line radio observations}
\author{B. Winkel
          \and
         A. Kraus
          \and
         U. Bach
       }

\institute{Max-Planck-Institut f\"{u}r Radioastronomie (MPIfR), 
              Auf dem H\"{u}gel\,69, 53121 Bonn, Germany;
              \email{bwinkel@mpifr.de}
}

\date{??; ??}

\abstract{Position and frequency switching techniques used for the removal of the bandpass dependence of radio astronomical spectra are presented and discussed in detail. Both methods are widely used, although the frequency dependence of the system temperature and/or noise diode is often neglected. This leads to systematic errors in the calibration that potentially have a significant impact on scientific results, especially when using large-bandwidth receivers or performing statistical analyses. We present methods to derive an unbiased calibration using a noise diode, which is part of many heterodyne receivers. We compare the proposed methods and describe the advantages and bottlenecks of the various approaches. Monte Carlo simulations are used to qualitatively investigate both systematics and the error distribution of the reconstructed flux estimates about the correct flux values for the new methods but also the `classical' case. Finally, the determination of the frequency-dependent noise temperature of the calibration diode using hot--cold measurements or observations of well-known continuum sources is also briefly discussed.}


\maketitle
%

\section{Introduction}
Spectroscopic data obtained from radio astronomical observations typically need to be post-processed in several steps before one can proceed with the scientific analysis. The uncalibrated flux values not only have to be converted from device units (counts) to physical quantities (e.g. Kelvin or Jy/beam), but one must also deal with the frequency-dependent gain (bandpass curve), radio frequency interference, removal of any residual baselines, and so on. While several methods are discussed in the literature to solve the bandpass issue \citep[][and references therein]{heiles07}, the flux calibration is often handled using only a very simplified approach. In particular, the widely used position and frequency switching techniques \citep[see][for a review]{heiles07} suffer from an improper treatment of calibration. In this paper we describe the underlying problem, present possible solutions, and assess the quality of the various approaches in terms of flux calibration accuracy. We compare the proposed methods with the `standard' approach, and show that the latter is subject to bias effects that can have a large impact on scientific results.

The paper is organised as follows.  Section\,\ref{sec:basicequations} introduces the basic equations and definitions necessary to describe the problem. Sections\,\ref{sec:pswitch} and \ref{sec:fswitch} present the position and frequency switching procedures, along with several methods to correctly calibrate the measured spectra. An alternative approach is shown in Section\,\ref{sec:directmethod} which may be of interest in cases where neither of the switching schemes are applicable. To illustrate the calibration methods, artificial spectra were generated with well-defined input quantities. Section\,\ref{sec:errordistribution} discusses the various approaches and their ability to reconstruct the simulated flux values. The flux calibration presented here relies on the signal of a noise diode fed into the receiver. Two possibilities for measuring the spectrum of this calibration signal are discussed in Section\,\ref{sec:tcalmeasurement}. We conclude with a summary in Section\,\ref{sec:summary}.

\section{Basic equations and definitions}\label{sec:basicequations}
The output signal, $P$, of a spectroscopic (heterodyne) receiving system can be described with the simple formula
\begin{equation}
P^\mathrm{[cal]}(\nu)=G_\mathrm{RF}(\nu)G_\mathrm{IF}(\nu)\left( T_\mathrm{sou}(\nu)+T_\mathrm{sys}^\mathrm{[cal]}(\nu)\right)\label{eqbasiceq},
\end{equation}
where the telescope itself and the receiving system exhibit a frequency-dependent gain, $G\equiv G_\mathrm{RF}G_\mathrm{IF}$, on the incident radiation. Often referred to as $T_A$, $T_\mathrm{sou}$, is the antenna temperature caused by the flux density of the astronomical source of interest, which potentially incorporates line emission and a continuum contribution, i.e.
$T_\mathrm{sou}=T^\mathrm{line}_\mathrm{sou}+T^\mathrm{cont}_\mathrm{sou}$. Apart from the flux of the astronomical source, $T_\mathrm{sou}$, that we are interested in, there are several noise contributions, which we subsume in the so-called system temperature,
$T_\mathrm{sys}^\mathrm{[cal]}$.\footnote{Note that sometimes the system temperature is defined differently to include $T_\mathrm{sou}$.} This consists of
\begin{equation}
T_\mathrm{sys}^\mathrm{[cal]}=T_\mathrm{bg}+T_\mathrm{atm}+T_\mathrm{spill}+T_\mathrm{sw}+T_\mathrm{loss}+T_\mathrm{rx}[+T_\mathrm{cal}]\label{eqtsys}
\end{equation}
\begin{align*}
T_\mathrm{bg} &\ldots \mathrm{microwave~and~galactic~backgrounds}\\
T_\mathrm{atm} &\ldots \mathrm{atmospheric~emission}\\
T_\mathrm{spill} &\ldots \mathrm{ground~radiation~(spillover~and~scattering)}\\
T_\mathrm{sw} &\ldots \mathrm{standing~wave~pattern~(often~in~secondary~focus)}\\
T_\mathrm{loss} &\ldots \mathrm{losses~in~feed,~ohmic~losses}\\
T_\mathrm{rx} &\ldots \mathrm{receiver~noise~temperature}\\
T_\mathrm{cal} &\ldots \mathrm{injected~noise~using~a~noise~tube/diode}.
\end{align*}
Standing waves (SW) are produced when the incident radio signal is reflected at the aperture of the antenna leading to interference patterns that often exhibit a sinusoidal shape. For a more detailed discussion, we refer the reader to \citet{rohlfs04}.

In this paper we make extensive use of the calibration signal $T_\mathrm{cal}$, which is a part of many heterodyne radio receivers. Its purpose is to provide a well-defined intensity standard that can be used to calibrate the measured spectra in terms of the antenna temperature. In practice, there are two common techniques to make use of $T_\mathrm{cal}$. One is to rapidly switch the $T_\mathrm{cal}$ signal on and off on time-scales of the order of tens of milliseconds up to a few seconds throughout the observation, leading to two so-called switching phases \textit{cal} and \textit{non-cal}. This method is used at the 100-m Effelsberg telescope, for example. In the following, we adopt the notation $P$ and $P^\mathrm{cal}$ when referring to \textit{non-cal} and \textit{cal} phases respectively, whilst $P^\mathrm{[cal]}$ indicates either of the two. The alternative method in use slowly cycles between the two phases 
during a dedicated measurement of a calibration source (or blank sky) every few hundred seconds, e.g. at the Arecibo telescope. In this paper, we assume a rapid $T_\mathrm{cal}$ switching for the sake of a simpler presentation of the calibration methods. However, our work may easily be extended to the slow cycling scheme. We would also like to point out that many (sub-)mm-wave telescopes make use of chopper wheels or hot-cold loads to improve the calibration. It should again be straightforward to adapt our methods to these techniques, though this is beyond the scope of this paper.

A spectroscopic back-end usually records the spectral density function, $P$, in arbitrary units (e.g. \textit{counts}) such that the raw spectra need to be calibrated in terms of both flux density values and also with respect to the frequency-dependent bandpass shape $G$. We note that for simplification we incorporate the coefficient that converts the antenna temperature to \textit{counts} into the gain factor $G$. Using a heterodyne receiver, one has to distinguish between the gain in the receiver/radio frequency (RF) and in the intermediate frequency (IF) part of the receiving system.

One major problem in the reduction of spectroscopic data is that the bandpass curve(s), $G$, must be disentangled from the input signal, $T$. From Eq.\,(\ref{eqbasiceq}), it is clear that one has to divide the measured signal, $P$, by the bandpass, $G$, to obtain the input temperature spectra. Unfortunately, the bandpass curve is unknown\footnote{Even if it had been measured, it is still subject to instabilities, e.g. caused by temperature drifts, such that frequent re-evaluation would be needed.}. To cope with this issue, one often applies a switching technique to obtain a reference spectrum, e.g. position or frequency switching as explained below. Dividing by this reference removes the frequency-dependent gain. Residual baselines caused by SW, for example,
may then be subtracted. We note that one can neither bandpass-calibrate a measurement by simply subtracting a reference spectrum nor remove residual baselines by division of a baseline model.  

Even after the (successful) removal of the gain curve, the extracted $T_\mathrm{sou}$ spectrum is not equal to the true brightness temperature, $T_\mathrm{sou}^\ast$, of the astronomical object. Both the atmospheric dampening and aperture efficiency, $\eta_\mathrm{ap}$, which act on $T_\mathrm{sou}^\ast$ prior to amplification, must also be taken into account
\begin{equation}
T_\mathrm{sou}=\eta_\mathrm{ap} T_\mathrm{sou}^\ast\exp\left(-\tau_0 \mathrm{AM}\right)\label{eqapeff},
\end{equation}
where $\tau_0$ is the zenith opacity and $\mathrm{AM}=\sin^{-1}(\mathrm{El})$ the elevation-dependent airmass. \footnote{Note that $\sin^{-1}(\mathrm{El})$ does not accurately describe the airmass at low elevation angles and must be replaced by a site-dependent model, including vertical values of density and refraction index \citep[e.g.,][]{maddalena05}.}
The aperture efficiency, $\eta_\mathrm{ap}$, is the product of several other efficiencies. The antenna pattern and blockage of the aperture play a role, as well as the surface accuracy and to some extent ohmic losses. This cannot be calculated theoretically, especially for the large single-dish telescopes.
It is possible to model opacity \citep[see][]{kraus09techreport}. In the best case, an atmospheric model makes use of simultaneous measurements of water vapour to achieve greater accuracy \citep[e.g. using a water vapour radiometer,][]{wvrtechreport}. For the remainder of this paper, we neglect these corrections for simplification as they are completely independent of the calibration methods presented here.

\section{Position switching}\label{sec:pswitch}
In position switching, one points the telescope first to a reference position (\textsc{Off}) before measuring toward the true target (\textsc{On}). Applying Eq.\,(\ref{eqbasiceq}), we obtain
\begin{align}
P_\mathrm{on}^\mathrm{[cal]}&=G_\mathrm{RF}G_\mathrm{IF}T_\mathrm{on}^\mathrm{[cal]}=G_\mathrm{RF}G_\mathrm{IF}\left( T_\mathrm{sou}+T_\mathrm{sys,on}^\mathrm{[cal]}\right),\\
P_\mathrm{off}^\mathrm{[cal]}&=G_\mathrm{RF}G_\mathrm{IF}T_\mathrm{off}^\mathrm{[cal]}=G_\mathrm{RF}G_\mathrm{IF}T_\mathrm{sys,off}^\mathrm{[cal]}\label{pswitchbasicpoff}.
\end{align}
The aim of position switching is to remove the bandpass curve. In the ideal case, $G_\mathrm{RF}G_\mathrm{IF}$ is equal for both positions 
and dividing both spectra leads to
\begin{equation}
\frac{P_\mathrm{on}^\mathrm{[cal]}}{P_\mathrm{off}^\mathrm{[cal]}}=\frac{T_\mathrm{sou}+T_\mathrm{sys,on}^\mathrm{[cal]}}{T_\mathrm{sys,off}^\mathrm{[cal]}}.
\end{equation}
This can also be written as
\begin{equation}
\frac{P_\mathrm{on}^\mathrm{[cal]}-P_\mathrm{off}^\mathrm{[cal]}}{P_\mathrm{off}^\mathrm{[cal]}}=\frac{T_\mathrm{sou}+\Delta T_\mathrm{sys}}{T_\mathrm{sys,off}^\mathrm{[cal]}}\label{eqonoff},
\end{equation}
where we introduce as
\begin{equation}
\Delta T_\mathrm{sys}\equiv T_\mathrm{sys,on}^\mathrm{[cal]}-T_\mathrm{sys,off}^\mathrm{[cal]}\approx \Delta T_\mathrm{bg}+\Delta T_\mathrm{sky}+\Delta T_\mathrm{spill}+\Delta T_\mathrm{sw}\label{eqdeltatsys},
\end{equation}
the difference between the system temperatures in the \textsc{On} and \textsc{Off} positions. The 
receiving system is usually stable for the duration of a single position switch and only a few contributors must be taken into account. While $T_\mathrm{sky}$ and $T_\mathrm{spill}$ are elevation dependent, $T_\mathrm{sw}$ may be a function of the incident continuum flux. In some cases, SW would only be noticeable if one pointed toward strong continuum sources, e.g. calibrators, and in other cases during daytime because of the sun, regardless of the observed source. The Galactic background continuum radiation depends slightly on the pointing position of the telescope. Using an appropriate \textsc{Off} position, for instance at constant elevation angle, one should usually be able to minimise $\Delta T_\mathrm{sys}$. The \textsc{Off} position should obviously not contain a continuum source itself. The equations could otherwise still be interpreted in a way where $T_\mathrm{cont}^\mathrm{sou}$ is not the true continuum flux of the \textsc{On} position.
Re-arranging Eq.\,(\ref{eqonoff}), we obtain
\begin{equation}
T_\mathrm{sou}+\Delta T_\mathrm{sys}=T_\mathrm{sys,off}^\mathrm{[cal]}\frac{P_\mathrm{on}^\mathrm{[cal]}-P_\mathrm{off}^\mathrm{[cal]}}{P_\mathrm{off}^\mathrm{[cal]}}\label{eqpswitchbase}
\end{equation}
or alternatively
\begin{equation}
T_\mathrm{sou}+T_\mathrm{sys,off}^\mathrm{[cal]}+\Delta T_\mathrm{sys}=T_\mathrm{sys,off}^\mathrm{[cal]}\frac{P_\mathrm{on}^\mathrm{[cal]}}{P_\mathrm{off}^\mathrm{[cal]}}.\label{eqpswitchbasealt}
\end{equation}

\textit{It is important to realise that $T_\mathrm{sys,off}^\mathrm{[cal]}$ is a frequency-dependent function. One can ensure a proper flux calibration over the complete spectrum only if this is determined correctly.} In the remaining part of this section, we develop two methods to determine $T_\mathrm{sys,off}^\mathrm{[cal]}(\nu)$. The first is easier to apply and more robust. The second may be of interest under particular circumstances when using frequency switching and is also discussed in the case of position switching for completeness. Both approaches make use of the signal of the calibration diode (or noise tube). Furthermore, we discuss the `classical' method, which assumes that $T_\mathrm{sys}\equiv\mathrm{constant}$ and $T_\mathrm{cal}\equiv\mathrm{constant}$, as a simplification of the first method.

\subsection{Setting up simulations}\label{subsec:simpleexample}
In the following subsections, we present the three alternative calibration methods. 
Artificial spectra were generated to enable a direct comparison of input and output spectral line intensities. In Section\,\ref{sec:errordistribution}, we statistically assess the quality of these methods to evaluate their general robustness and error distribution.

\begin{figure}[!t]
\centering
\includegraphics[width=0.45\textwidth,bb=26 42 521 392,clip=]{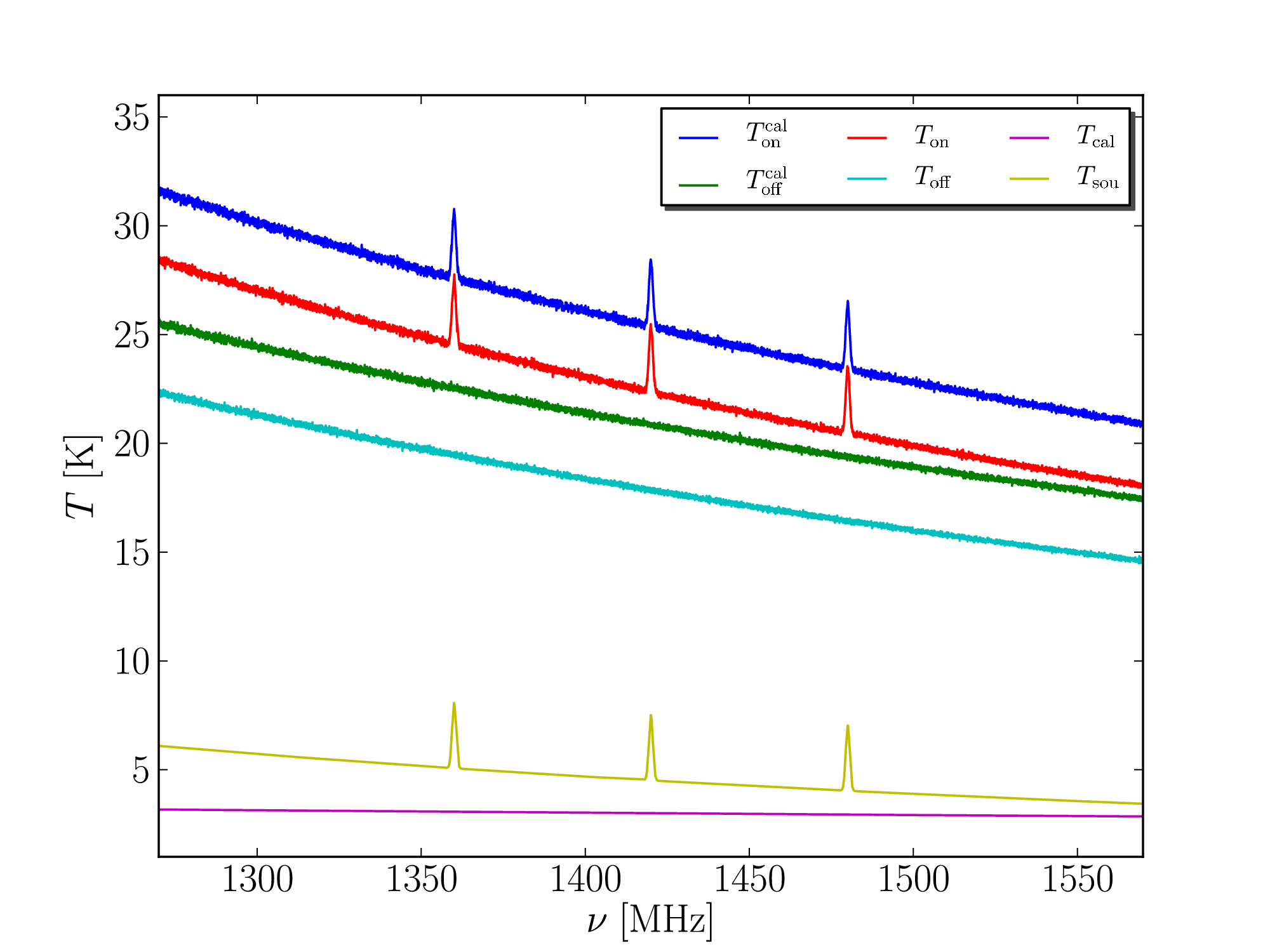}\\[0ex]
\includegraphics[width=0.45\textwidth,bb=26 1 521 392,clip=]{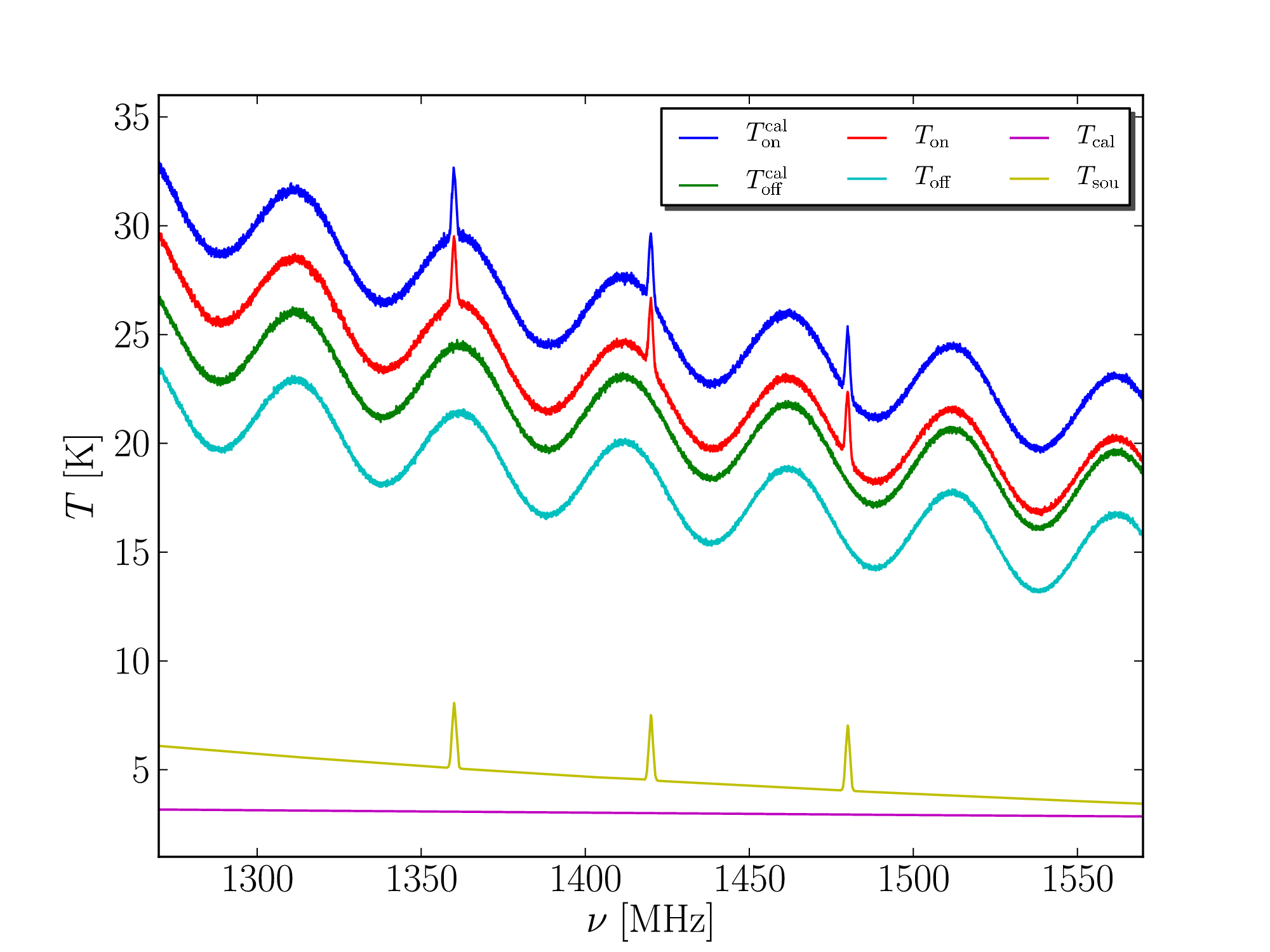}
\caption{To test the influence of the different calibration schemes on the reconstructed flux values, synthetic spectra were produced. The \textbf{upper panel} shows the input power spectral densities. The antenna temperature on-source ($T_\mathrm{sou}$) is a superposition of three Gaussians with an amplitude of 3\,K on top of a continuum spectrum with a spectral index of $-0.7$, i.e. $T_\mathrm{sou}\sim\nu^{-2.7}$. For the system temperature, $T_\mathrm{sys,off}\sim\nu^{-2.1}$ is used for convenience. We also simulate a slight frequency-dependence of the calibration diode using the power law $T_\mathrm{cal}\sim\nu^{-0.5}$. Gaussian noise was added to the four measurement phases according to their total temperature values (see text). Note that each synthetic spectrum contains a different noise realisation. In addition, a more complicated case is included with a standing wave contribution, which is parametrised as a monochromatic sine wave component \textbf{(lower panel)}.}%
\label{fig:syntheticspectraT}%
\end{figure}

The test spectrum uses a relatively simple set-up. The continuum flux of the \textsc{Off} position is described as a single-slope power law $T_\mathrm{sys,off}=400(\nu/\nu_0)^{-2.1}\,\mathrm{[K]}$, where $\nu_0=300\,\mathrm{MHz}$. This is a strong simplification of reality where several different power laws (e.g. from the galactic background, sky, and ground radiation) add to a more or less `arbitrary' temperature contribution from the receiver.  For the astronomical source, $T_\mathrm{sou}$, a set of three Gaussians superposed on the power law $T^\mathrm{cont}_\mathrm{sou}=200(\nu/\nu_0)^{-2.7}\,\mathrm{[K]}$ (where $\nu_0=300\,\mathrm{MHz}$) was used. Each Gaussian has equivalent line parameters (3\,K amplitude, 1.4\,MHz line width, FWHM) but are added at different frequencies. These are used as `probes' to measure the flux calibration quality at different frequencies after data reduction. We chose to model a possible frequency dependence of the noise diode using a shallow power law $T_\mathrm{cal}=3(\nu/\nu_0)^{-0.5}\,\mathrm{[K]}$ (where $\nu_0=1420\,\mathrm{MHz}$). In reality $T_\mathrm{cal}$ is usually a much more complicated function especially at higher frequencies, but this setup is suitable for the main purpose of our simulations, which is to illustrate the various steps in the data processing as simply as possible. Section\,\ref{subsec:pswitchrealisticexample} illustrates a more realistic test case with respect to the functional form of the system temperature and calibration diode. The resulting temperature spectra are plotted in Fig.\,\ref{fig:syntheticspectraT} (upper panel). 

Gaussian noise was added to each component according to the root mean square (RMS) noise calculated using the radiometer equation \citep[see][]{rohlfs04}
\begin{equation}
T_\mathrm{rms}\equiv \Delta T=\frac{k T_\mathrm{input}}{\sqrt{\Delta f \tau}}\label{eqradiometer},
\end{equation}
where $k$ is a constant factor depending on, e.g. the input quantisation of the signals, $\Delta f$ is the bandwidth of a spectral channel, and $\tau$ is the integration time. In the simulations, we assume the total number of spectral channels to be 16k over a bandwidth of $300\,\mathrm{MHz}$ and an integration time (per spectral dump or measurement phase) of $5s$. For example, a system temperature of 25\,K  would result in $T_\mathrm{rms}\approx83\,\mathrm{mK}$ (per spectral channel). Eq.\,(\ref{eqradiometer}) implies that noise is a function of frequency because the temperature values, $T_\mathrm{input}$, for each measurement phase vary over the observed band. Likewise, the additional temperature components of the continuum emission of the observed source and the calibration diode increase the noise. It is also possible to calculate the theoretical RMS level expected in the final (reduced) spectrum using the noise values of the individual input spectra (see Appendix\,\ref{sec:appendixnoise}).

Finally, for a more complicated test case, we simulated (mono-modal) SW by adding a sine wave component $T_\mathrm{sw}=2\sin(2\pi (\nu/\nu_\mathrm{sw}))\,\mathrm{[K]}$ to $T_\mathrm{sys,off}$ (see Fig.\,\ref{fig:syntheticspectraT}, lower panel).\footnote{Note that this set-up does not account for all possible situations. For example, if strong continuum sources alone produce SW, then they contribute to $T_\mathrm{cont}^\mathrm{sou}$ only.  However, this is only a minor problem as the computation of $T_\mathrm{sys}$ is not affected. Nevertheless, one has to account for such a case if one uses continuum sources for calibration.} For the position switching test, $\nu_\mathrm{sw}=50\,\mathrm{MHz}$.

\begin{figure}[!t]
   \centering%
  \includegraphics[width=0.45\textwidth,bb=18 1 521 392,clip=]{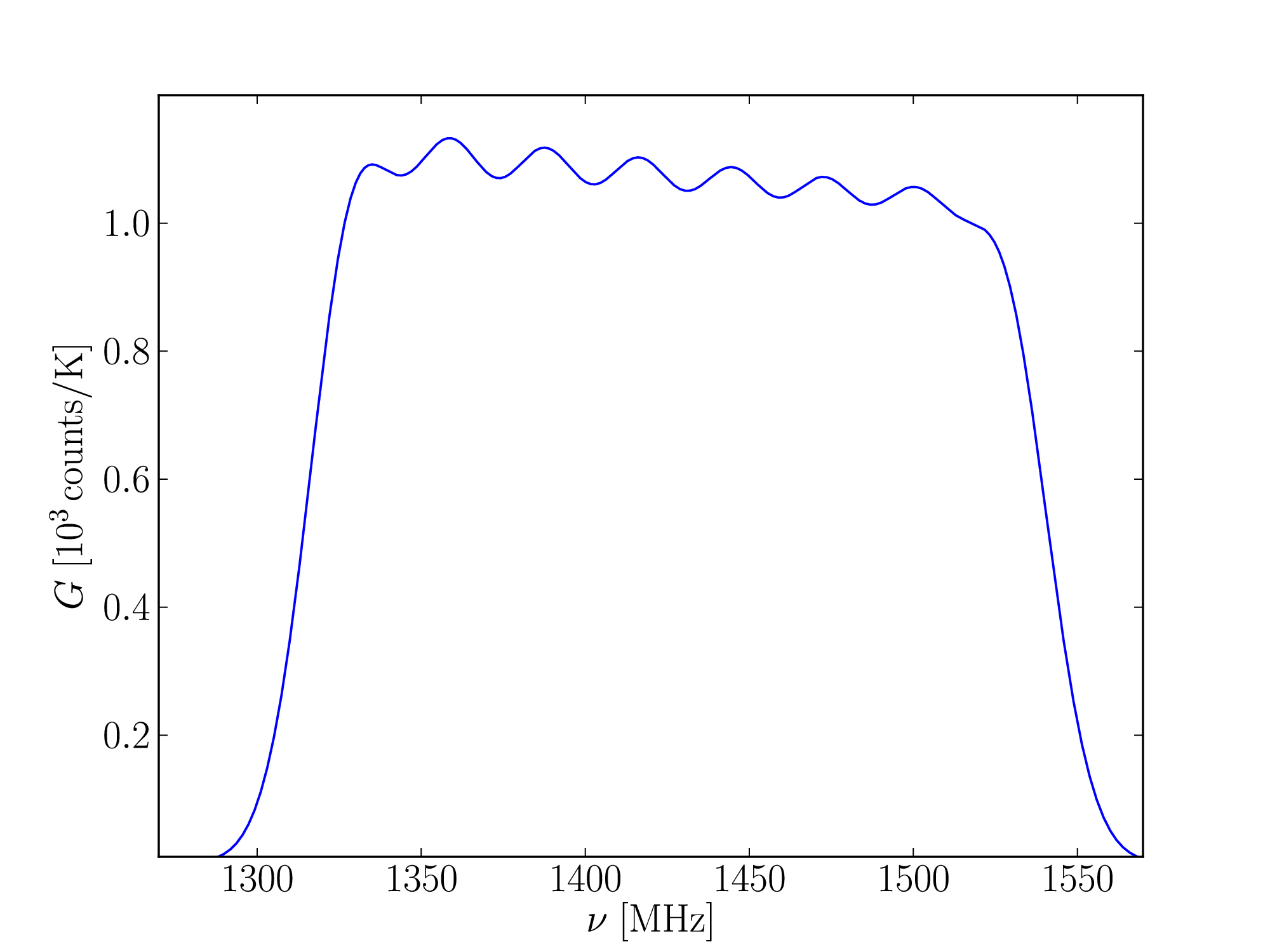}
  \caption{The bandpass curve, $G=G_\mathrm{RF}G_\mathrm{IF}$, used for the simulations.}%
   \label{fig:syntheticbandpass}%
\end{figure}

\begin{figure}[!t]
\centering
\includegraphics[width=0.45\textwidth,bb=26 42 521 392,clip=]{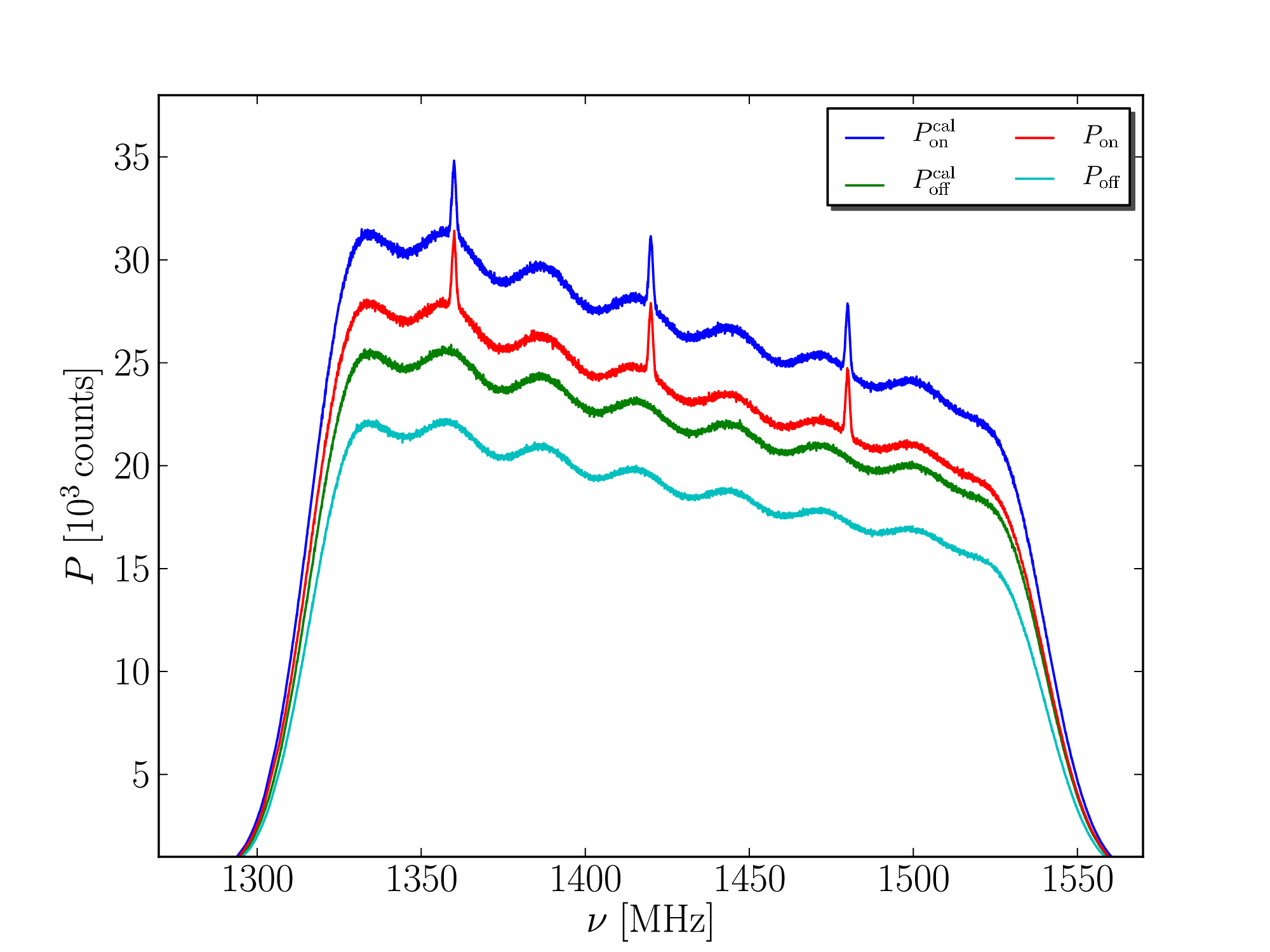}\\[0ex]
\includegraphics[width=0.45\textwidth,bb=26 1 521 392,clip=]{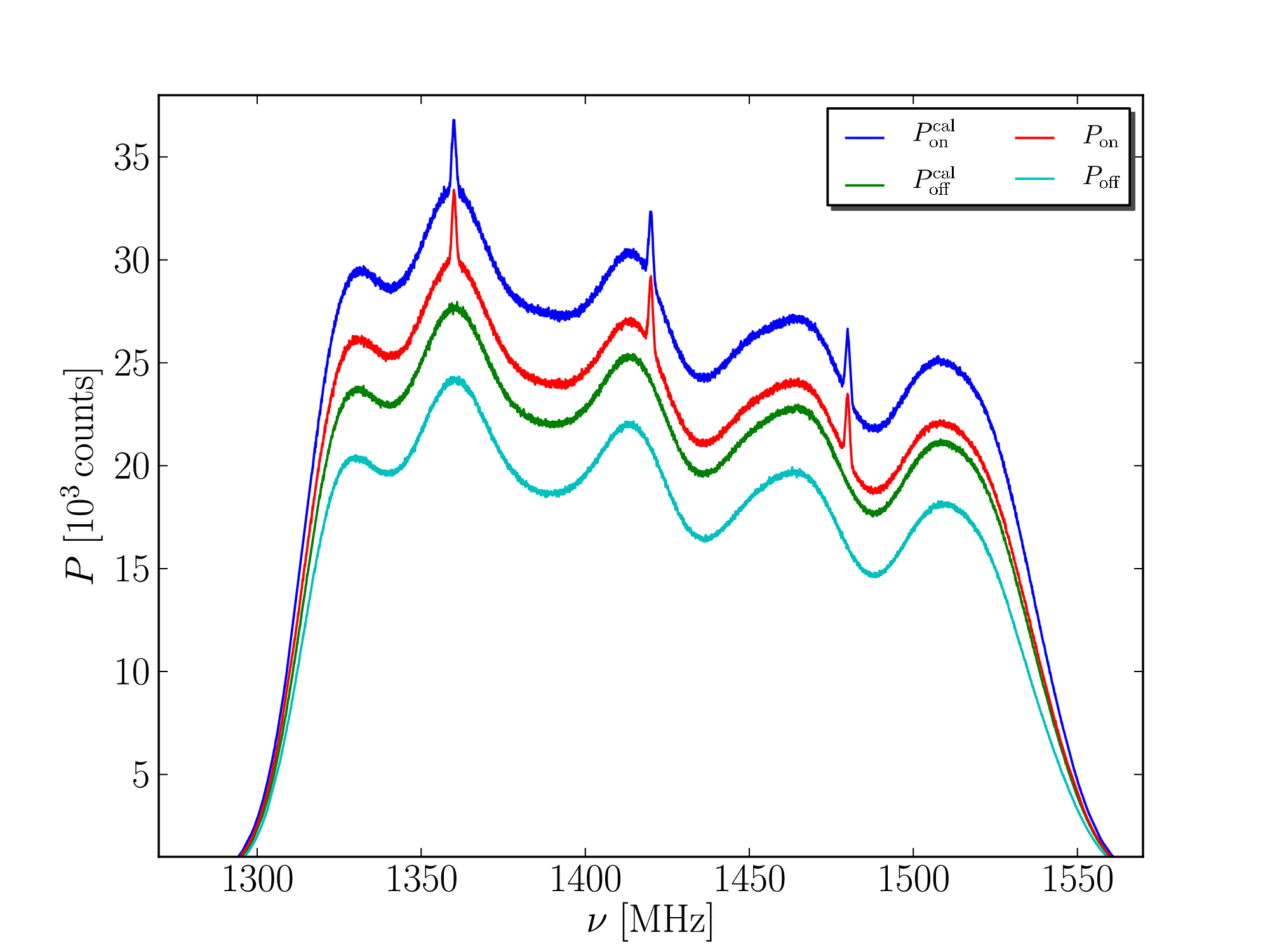}
\caption{Multiplying the input spectral densities with the bandpass shape shown in Fig.\,\ref{fig:syntheticspectraT} returns the 
`measured' spectra $P$. The \textbf{upper panel} shows the results for the simpler case, while the \textbf{lower panel} includes 
the standing wave contribution.}%
\label{fig:pswitchsyntheticspectraP}%
\end{figure}

The input temperature spectra were multiplied with a simple bandpass curve shown in Fig.\,\ref{fig:syntheticbandpass}. The resulting power spectral densities (in arbitrary units) are plotted in Fig.\,\ref{fig:pswitchsyntheticspectraP}. The superposition of the SW and the ripples in the bandpass shown in the lower panel of Fig.\,\ref{fig:pswitchsyntheticspectraP} creates a rather complex pattern.

\subsection{Using the \textsc{Off} position to obtain $T_\mathrm{sys}$}\label{subsec:pswitchmethod1}
Computing
\begin{equation}
\frac{P_\mathrm{off}^\mathrm{cal}}{P_\mathrm{off}}=\frac{T_\mathrm{sys,off}+T_\mathrm{cal}}{T_\mathrm{sys,off}}=\frac{T_\mathrm{cal}}{T_\mathrm{sys,off}}+1\label{eqpswitchtsystcal}
\end{equation}
leads to an equation to infer the system temperature of the \textsc{Off} position, which depends only on the temperature of the noise diode, $T_\mathrm{cal}$. If the latter is known, then
\begin{equation}
T_\mathrm{sys,off}=T_\mathrm{cal}\left[\frac{P_\mathrm{off}^\mathrm{cal}}{P_\mathrm{off}}-1\right]^{-1}.
\end{equation}
However, the noise diode temperature is a frequency-dependent quantity. It can be measured with good precision using the hot--cold method (compare also Section\,\ref{subsec:hotcoldtheory} and Appendix\,\ref{subsec:tcalhotcold}), which, unfortunately, is a time-consuming procedure. One solution is to establish a catalogue of astronomical calibrators, i.e. bright continuum sources, which serve as reference to (re-)calibrate the $T_\mathrm{cal}$ spectrum on appropriate timescales (see Section\,\ref{subsec:contcal} and Appendix\,\ref{subsec:tcalcontinuum}). This approach was also proposed by \citet{maddalena05}. Since $T_\mathrm{cal}$ is a time-dependent quantity (e.g. owing to a change in the environmental conditions), it is always a potential source of (systematic) error. Nevertheless, a temporal stability on the order of 1\% can typically be expected on the timescale of one hour.

Unfortunately, even if one has a good model of $T_\mathrm{cal}$, the measured quantity $P_\mathrm{off}^\mathrm{cal}/P_\mathrm{off}$ is still subject to noise and will substantially increase the noise in the final reduced spectra. One solution is to suppress the noise in the obtained $T_\mathrm{sys,off}$ spectrum before substituting it into Eq.\,(\ref{eqpswitchbase}).

In simpler cases, namely in the absence of a standing wave contribution and a relatively flat $T_\mathrm{cal}(\nu)$, the quantity
\begin{equation}
\left[\frac{P_\mathrm{off}^\mathrm{cal}}{P_\mathrm{off}}-1\right]^{-1}=\frac{T_\mathrm{sys,off}}{T_\mathrm{cal}}\equiv \kappa_\mathrm{off}(\nu).\label{eqpswitchkappa}
\end{equation}
is approximately proportional to $T_\mathrm{sys,off}$, which can be described by a power law. Hence, a low-order polynomial might already suffice to describe the quantity $\kappa_\mathrm{off}(\nu)$.

Unfortunately, at higher frequencies in particular, it is almost impossible to engineer a sufficiently flat $T_\mathrm{cal}(\nu)$. In such cases, a filtering approach or the use of high-order polynomials is required to suppress noise (see Section\,\ref{subsec:pswitchrealisticexample} for a more realistic example). In most cases, it is easier to model $\kappa_\mathrm{off}^{-1}$ for numerical stability, as usually $T_\mathrm{sys,off}\gg T_\mathrm{cal}$. The noise-free (or low-noise) model can then safely be inverted to obtain $\kappa_\mathrm{off}$.

\begin{figure}[!t]
\centering
\includegraphics[width=0.45\textwidth,bb=9 42 521 392,clip=]{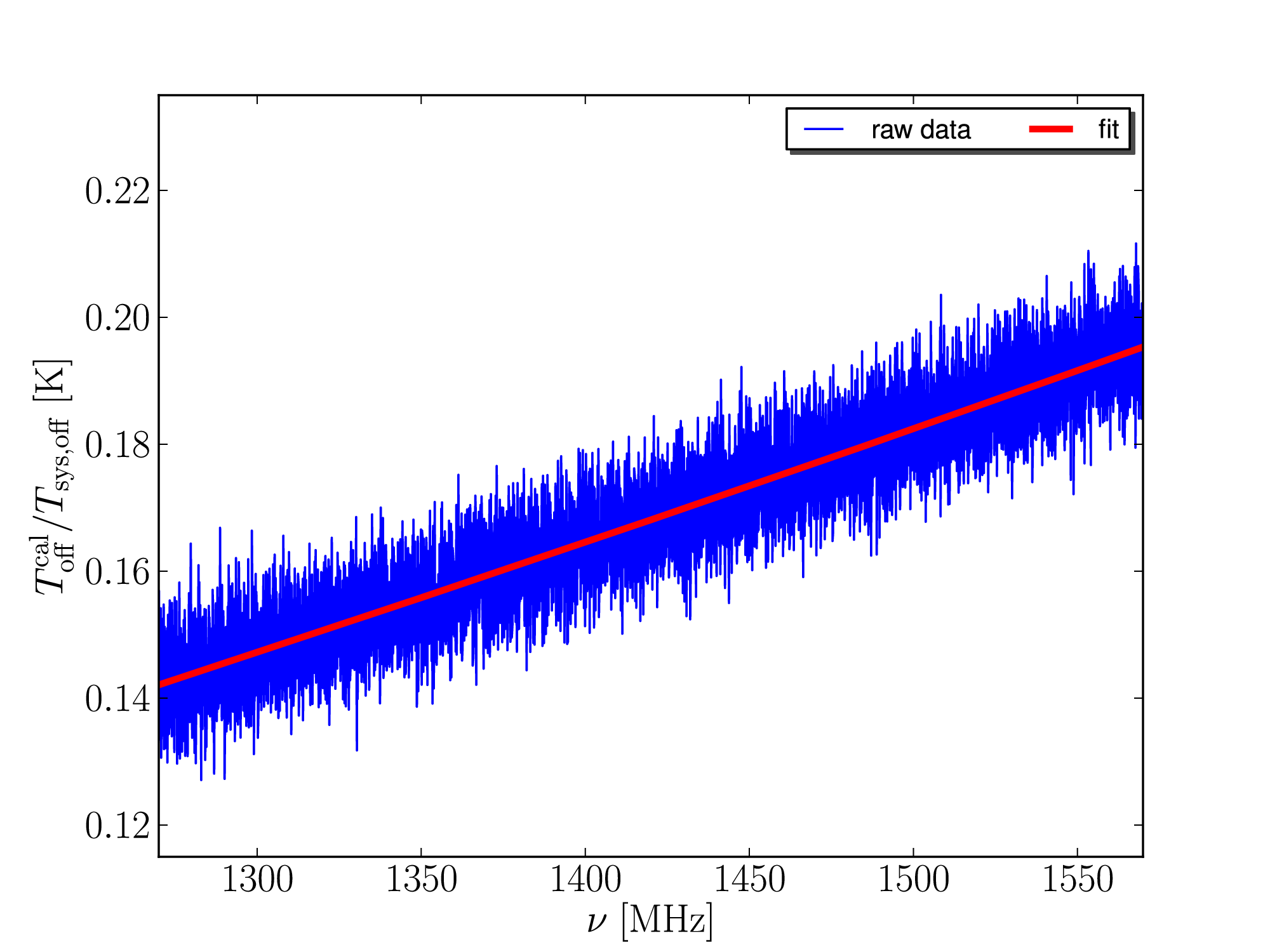}\\[0ex]
\includegraphics[width=0.45\textwidth,bb=9 1 521 392,clip=]{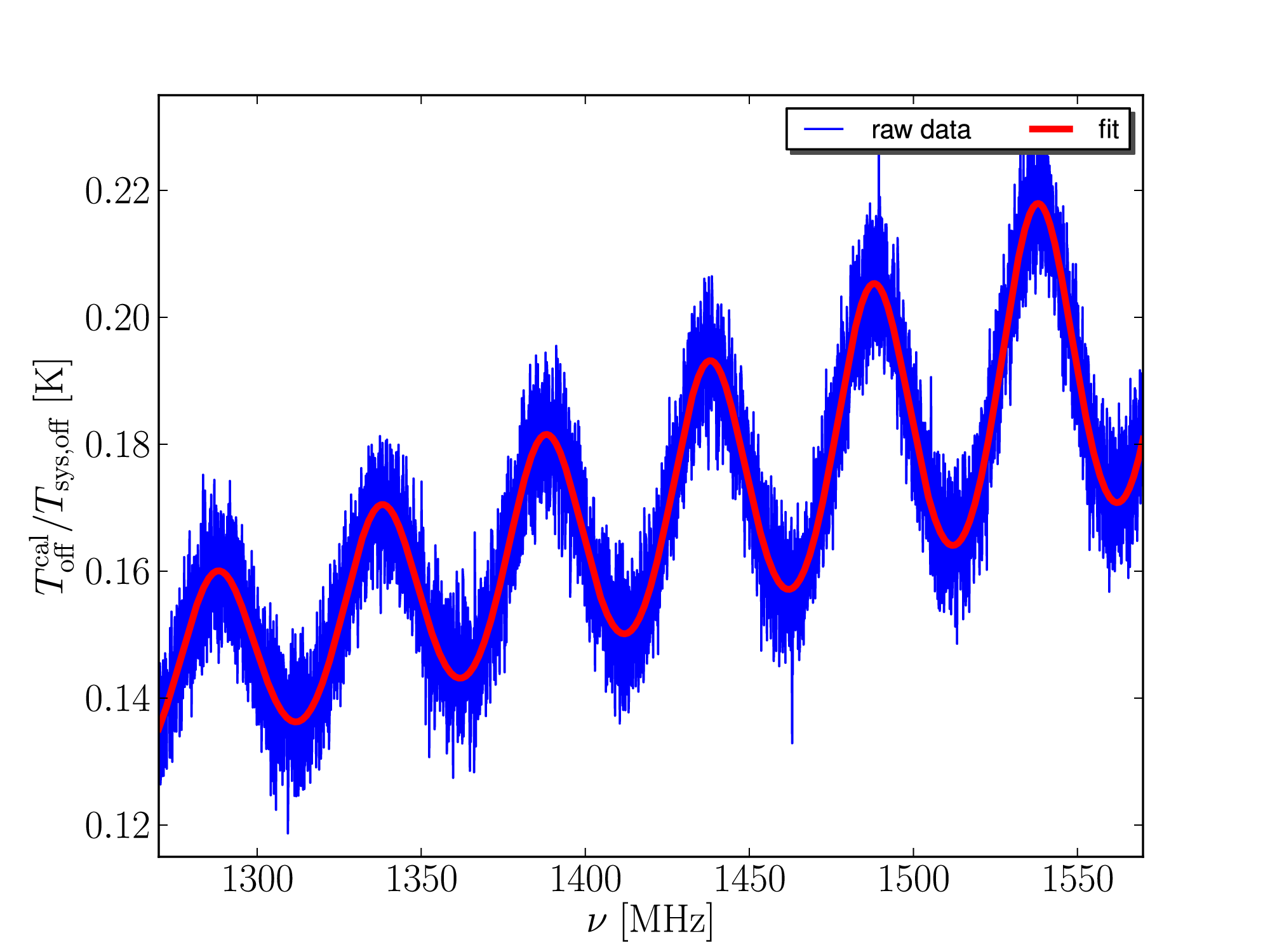}
\caption{Spectrum of $\kappa_\mathrm{off}^{-1}$ for the simpler (\textbf{upper panel}) and standing wave (\textbf{lower panel}) cases
and the fitting models inferred. }%
\label{fig:pswitch_method1_modelfits}%
\end{figure}

Fig.\,\ref{fig:pswitch_method1_modelfits} illustrates $\kappa_\mathrm{off}^{-1}(\nu)=T_\mathrm{cal}/T_\mathrm{sys,off}$ for the two test cases with and without SW. In the former case (upper panel), a third-order polynomial was used to describe $\kappa_\mathrm{off}^{-1}$. For the latter, a more complicated model is necessary
\begin{equation}
\kappa_\mathrm{off}^{-1}=\frac{ L_1(n_1,\nu)}{L_2(n_2,\nu)+A\sin(a\nu+b)},
\end{equation}
where $L_i(n_i,\nu)$ are polynomial functions of degree $n_i=3$. This result was obtained only after providing suitable initial fit parameter values and is shown in the lower panel. An entirely automated procedure would most likely have difficulties in handling SW --- even for the simplest scenario of monochromatic SW.

Substituting Eq.\,(\ref{eqpswitchkappa}) into Eq.\,(\ref{eqpswitchbase}) leads to
\begin{equation}
T_\mathrm{sou}+\Delta T_\mathrm{sys}=\kappa_\mathrm{off}T_\mathrm{cal}\frac{P_\mathrm{on}-P_\mathrm{off}}{P_\mathrm{off}}=(\kappa_\mathrm{off}+1)T_\mathrm{cal}\frac{P_\mathrm{on}^\mathrm{cal}-P_\mathrm{off}^\mathrm{cal}}{P_\mathrm{off}^\mathrm{cal}}.\label{eqpswitchbaseusingkappa}
\end{equation}
We note that the continuum flux of the source is also reconstructed. To obtain an optimal signal-to-noise ratio (S/N), the reduced \textit{cal} and \textit{non-cal} spectra should be averaged. 
\begin{figure}[!t]
\centering
\includegraphics[width=0.45\textwidth,bb=7 42 521 392,clip=]{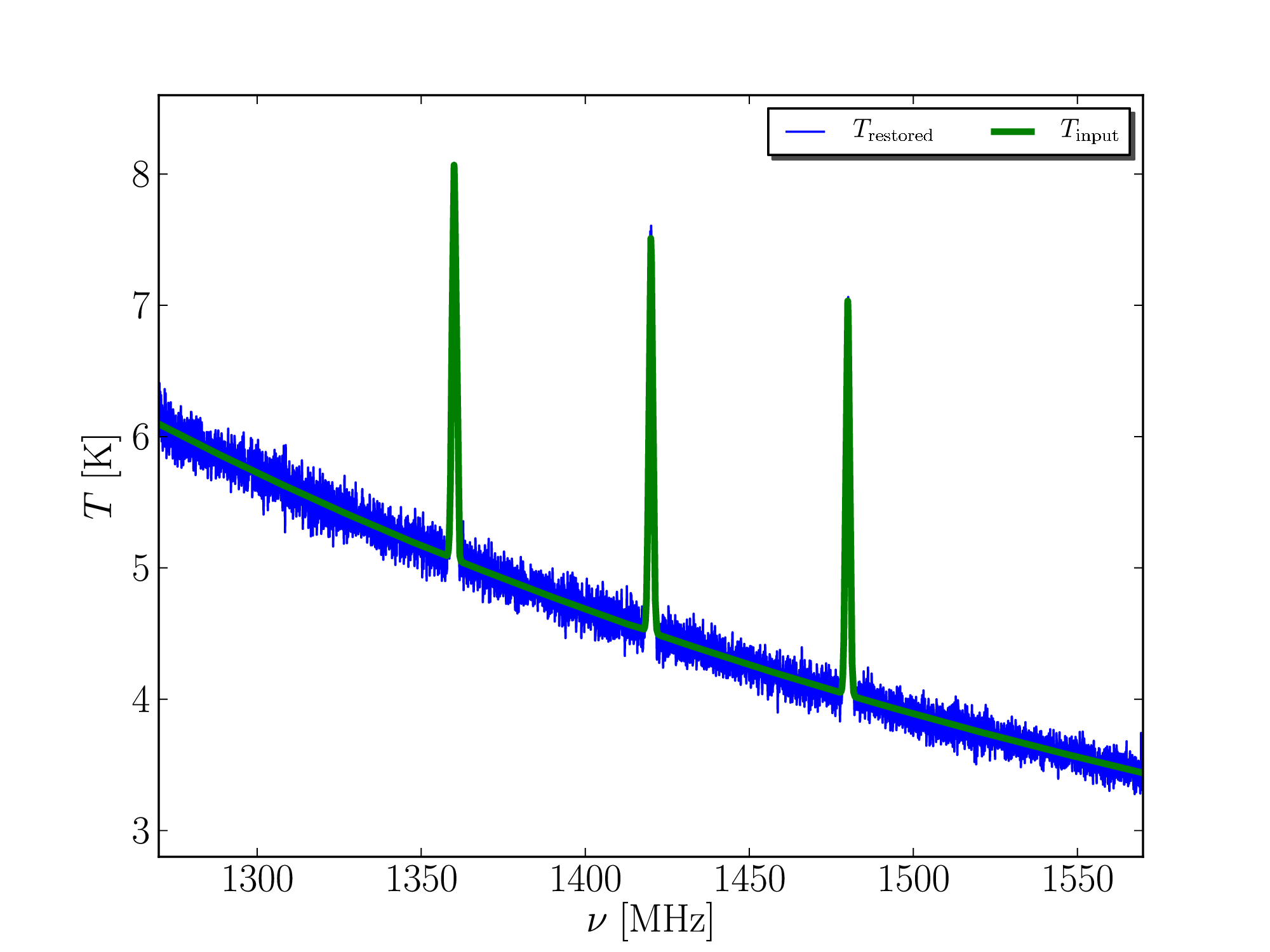}\\[0ex]
\includegraphics[width=0.45\textwidth,bb=7 1 521 392,clip=]{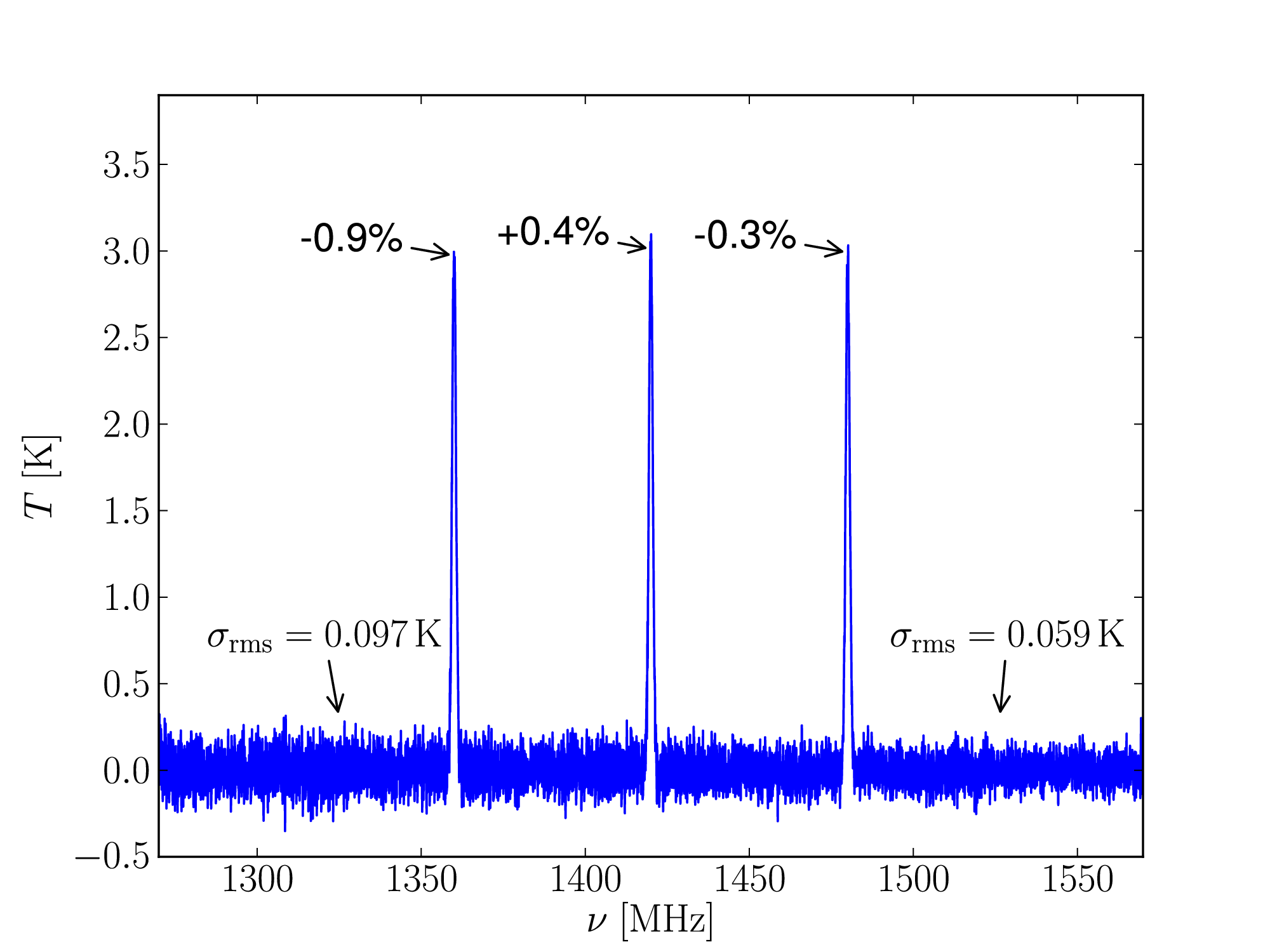}
\caption{\textbf{Upper panel:} Resulting spectrum for the SW case after applying Eq.\,(\ref{eqpswitchbaseusingkappa}), using the model fit shown in Fig.\,\ref{fig:pswitch_method1_modelfits}. The continuum contribution of the source is recovered. A baseline is subtracted in the \textbf{lower panel} and spectral line intensities were fitted locally. The recovered intensities match the input very well. The noise values in different spectral regions are also consistent.}%
\label{fig:pswitch_method1_resultswithsw}%
\end{figure}

After applying Eq.\,(\ref{eqpswitchbaseusingkappa}) to the simulated data (using a polynomial model for $\kappa_\mathrm{off}$; see Fig.\,\ref{fig:pswitch_method1_modelfits}), the spectra shown in Fig.\,\ref{fig:pswitch_method1_resultswithsw} are obtained. To measure the spectral line intensities, we fit a Gaussian superposed on a third-order polynomial to each spectral line. We note that Fig.\,\ref{fig:pswitch_method1_resultswithsw} shows the reduced spectra for the SW case. The non-SW scenario produces very similar results. The upper panel shows the direct result and the lower panel contains the spectrum after baseline removal. The noise values and the recovered line intensities match the expected results. The residual noise is higher than one might naively expect (the RMS in the final spectrum is around the same level as for each of the four contributing phases).  This is due to the division by the reference spectrum that does not contain signal but adds noise (see also Appendix\,\ref{subsec:appendixpswitch}).

It is possible  to smooth the reference spectrum prior to division greatly reducing the RMS in the final spectrum by almost a factor of $\sqrt{2}$.  However, as \citet{gbtidl} points out, this procedure can have unwanted side-effects, for example, degradation of the baseline or emphasis of narrow features present in the reference, e.g. radio frequency interference (RFI), and results in (locally) correlated noise.

One drawback of the proposed method is that the \textsc{Off} position alone is utilised to infer $T_\mathrm{sys}$. As a consequence, a factor of $\sqrt{2}$ in sensitivity is lost in calculating $\kappa_\mathrm{off}$. However, the final result is unaffected so long as $\kappa_\mathrm{off}$ is described by a (smooth) model. If sensitivity is not a major concern, even the original (noisy) $\kappa_\mathrm{off}$ spectrum, i.e., without modelling, may be used.  This is advantageous since the solution is then unaffected by the choice of a given model.

\subsection{Using  \textsc{On} and \textsc{Off} positions to obtain $T_\mathrm{sys}$}\label{subsec:pswitchmethod2}

The second method utilises Eq.\,(\ref{eqonoff}) and uses continuum contributions only, to define the quantities
\begin{align}
f(\nu)&\equiv  \left.\frac{P_\mathrm{on}^\mathrm{\phantom{cal}}-P_\mathrm{off}^\mathrm{\phantom{cal}}}{P_\mathrm{off}^\mathrm{\phantom{cal}}}\right|_\mathrm{cont}= \frac{T_\mathrm{cont}^\mathrm{sou}+\Delta T_\mathrm{sys}}{T_\mathrm{sys,off}} \label{eqpswitchf},\\
f^\mathrm{cal}(\nu)&\equiv\left.\frac{P_\mathrm{on}^\mathrm{cal}-P_\mathrm{off}^\mathrm{cal}}{P_\mathrm{off}^\mathrm{cal}}\right|_\mathrm{cont} =\frac{T_\mathrm{cont}^\mathrm{sou}+\Delta T_\mathrm{sys}}{T_\mathrm{sys,off}+T_\mathrm{cal}}.    \label{eqpswitchfcal}
\end{align}
Since the \textsc{On} spectra are involved, it is important that the fitting algorithm excludes spectral lines. In simple cases, the use of relatively low-order polynomial functions should suffice to provide a good model. If SW occur and only few modes (i.e. SW frequencies) are present, one should be able to model these with superposed sine-waves. In principle, one could even try to use a filtering approach to obtain a noise-free model $f$.

\begin{figure}[!t]
   \centering%
   \includegraphics[width=0.45\textwidth,bb=34 42 521 392,clip=]{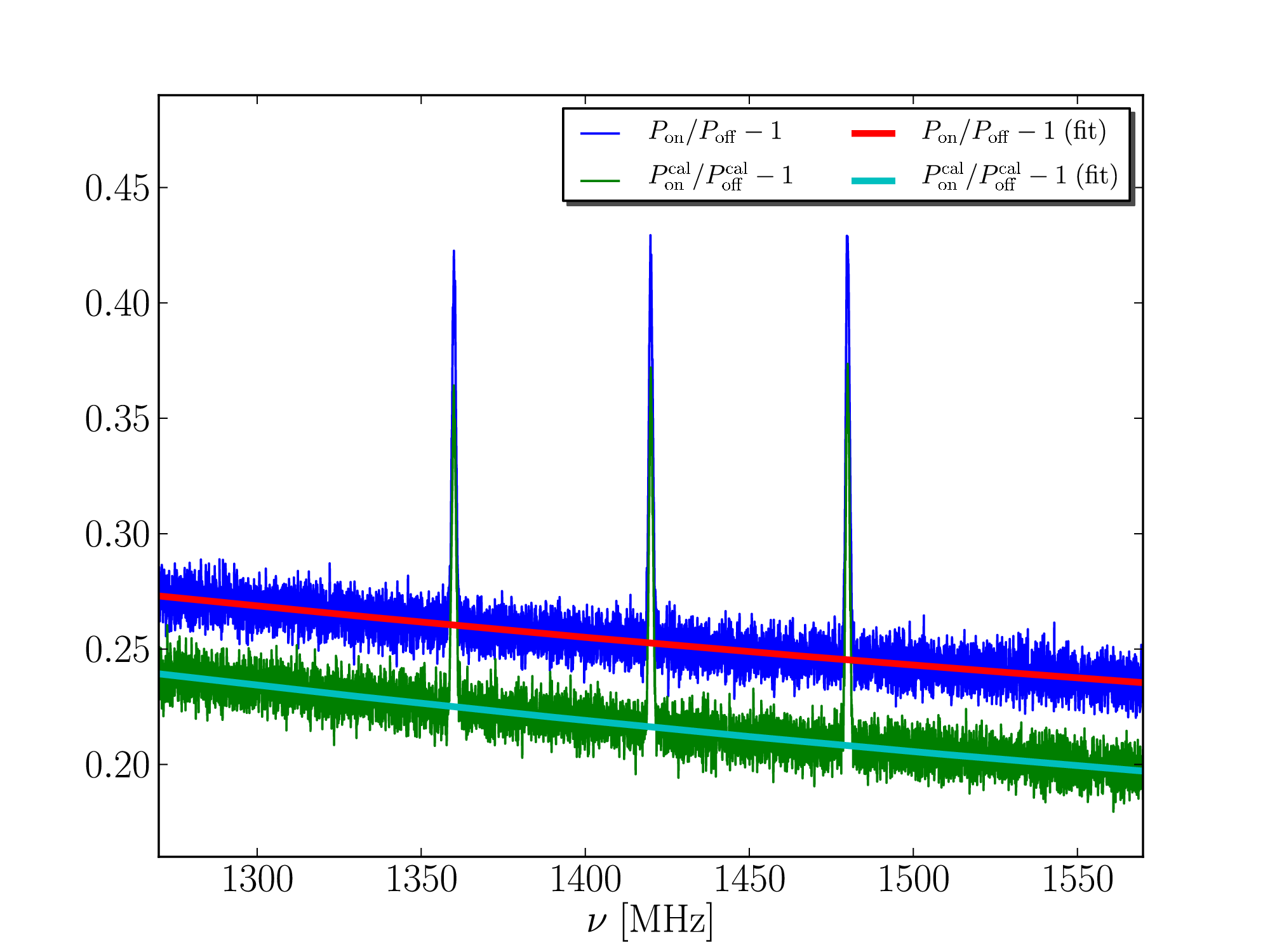}\\[0ex]
   \includegraphics[width=0.45\textwidth,bb=34 1 521 392,clip=]{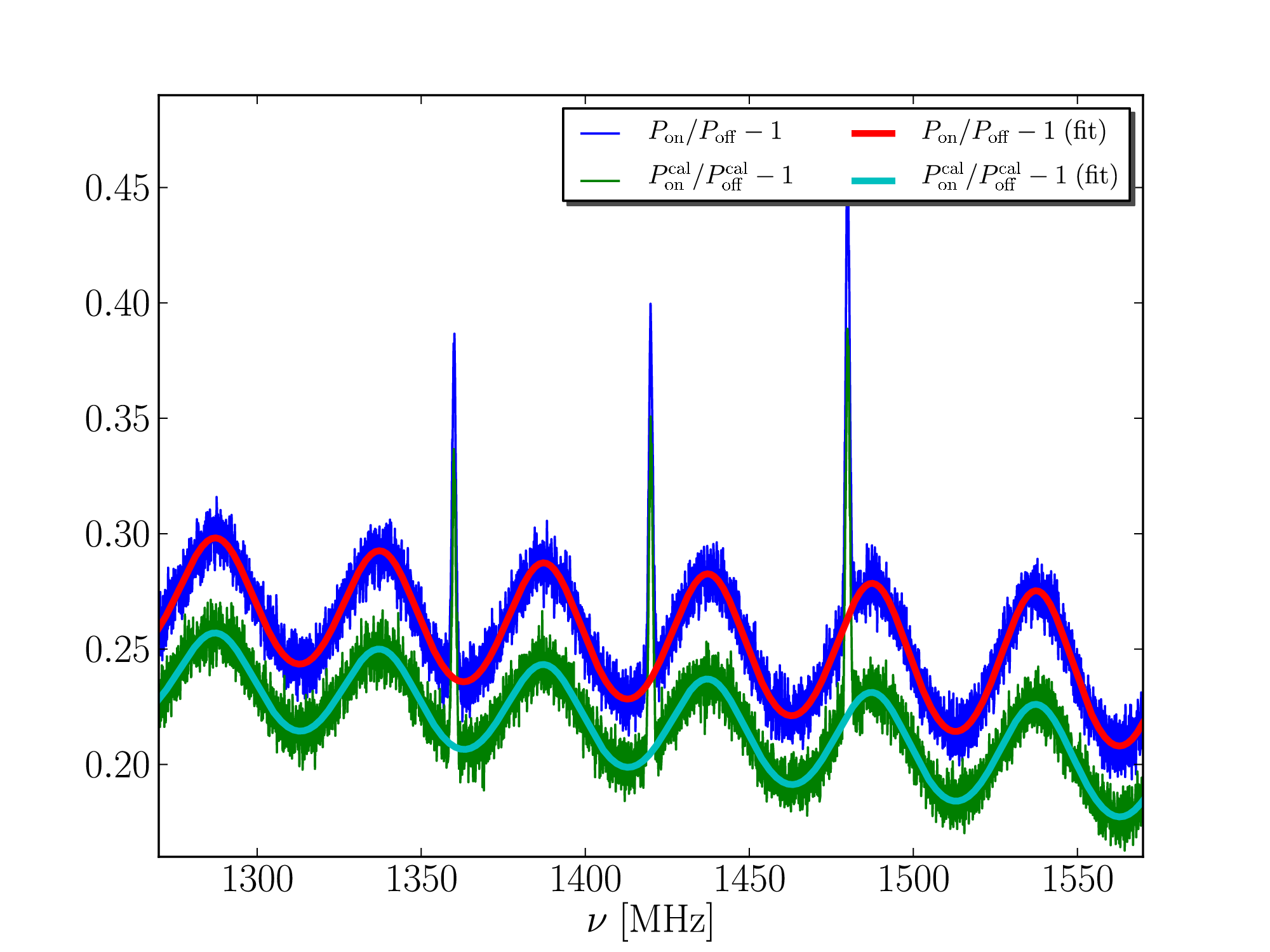}
  \caption{To correctly reconstruct the original fluxes, the second proposed method uses models of the intermediate spectra $(P_\mathrm{on}^\mathrm{[cal]}-P_\mathrm{off}^\mathrm{[cal]})/P_\mathrm{off}^\mathrm{[cal]}$ (see Eq.\,(\ref{eqpswitchf}) and (\ref{eqpswitchfcal})). Appropriate windows around the spectral lines should be set  because one is only interested in the continuum contribution. The \textbf{upper panel} shows the results of the simpler case, utilising second-order polynomials sufficient to describe the baseline. For the standing wave case (\textbf{lower panel}), a more complicated fitting model, $L_1(n_1,\nu)/(L_2(n_2,\nu)+A\sin(a\nu+b))$, was applied, where $L_i(n_i,\nu)$ are polynomial functions of degree $n_i$. In this example, we use $n_1=n_2=3$.}%
   \label{fig:pswitch_method2_fit_through_div}%
\end{figure}

The fits calculated for the example spectrum are shown in Fig.\,\ref{fig:pswitch_method2_fit_through_div}. For the non-SW case, a third-order polynomial model is used.  The SW case uses
\begin{equation}
f(\nu)\equiv\frac{ L_1(n_1,\nu)}{L_2(n_2,\nu)+A\sin(a\nu+b)},
\end{equation}
where $L_i(n_i,\nu)$ are polynomial functions of degree $n_i$.

Substituting Eq.\,(\ref{eqpswitchf}) into Eq.\,(\ref{eqpswitchfcal}) gives
\begin{align}
T_\mathrm{sys,off}&=T_\mathrm{cal}\frac{f^\mathrm{cal}(\nu)}{f(\nu)-f^\mathrm{cal}(\nu)} \label{eqpswitchtsysfromf},\\
T_\mathrm{sys,off}+T_\mathrm{cal}&=T_\mathrm{cal}\frac{f(\nu)}{f(\nu)-f^\mathrm{cal}(\nu)}\label{eqpswitchtsysfromfcal},
\end{align}
such that $T_\mathrm{sys,off}(\nu)$ is a function of $T_\mathrm{cal}$. Eq.\,(\ref{eqpswitchbase}) now becomes
\begin{equation}
\begin{split}
T^\mathrm{line}_\mathrm{sou}+T^\mathrm{cont}_\mathrm{sou}+\Delta T_\mathrm{sys}&= T_\mathrm{cal}\frac{f^\mathrm{cal}(\nu)}{f(\nu)-f^\mathrm{cal}(\nu)}  \frac{P_\mathrm{on}-P_\mathrm{off}}{P_\mathrm{off}}\\
& = T_\mathrm{cal} \frac{f(\nu)}{f(\nu)-f^\mathrm{cal}(\nu)}  \frac{P_\mathrm{on}^\mathrm{cal}-P_\mathrm{off}^\mathrm{cal}}{P_\mathrm{off}^\mathrm{cal}}.\label{eqpswitchfinal}
\end{split}
\end{equation}

\begin{figure}[!t]
   \centering%
   \includegraphics[width=0.45\textwidth,bb=34 42 521 392,clip=]{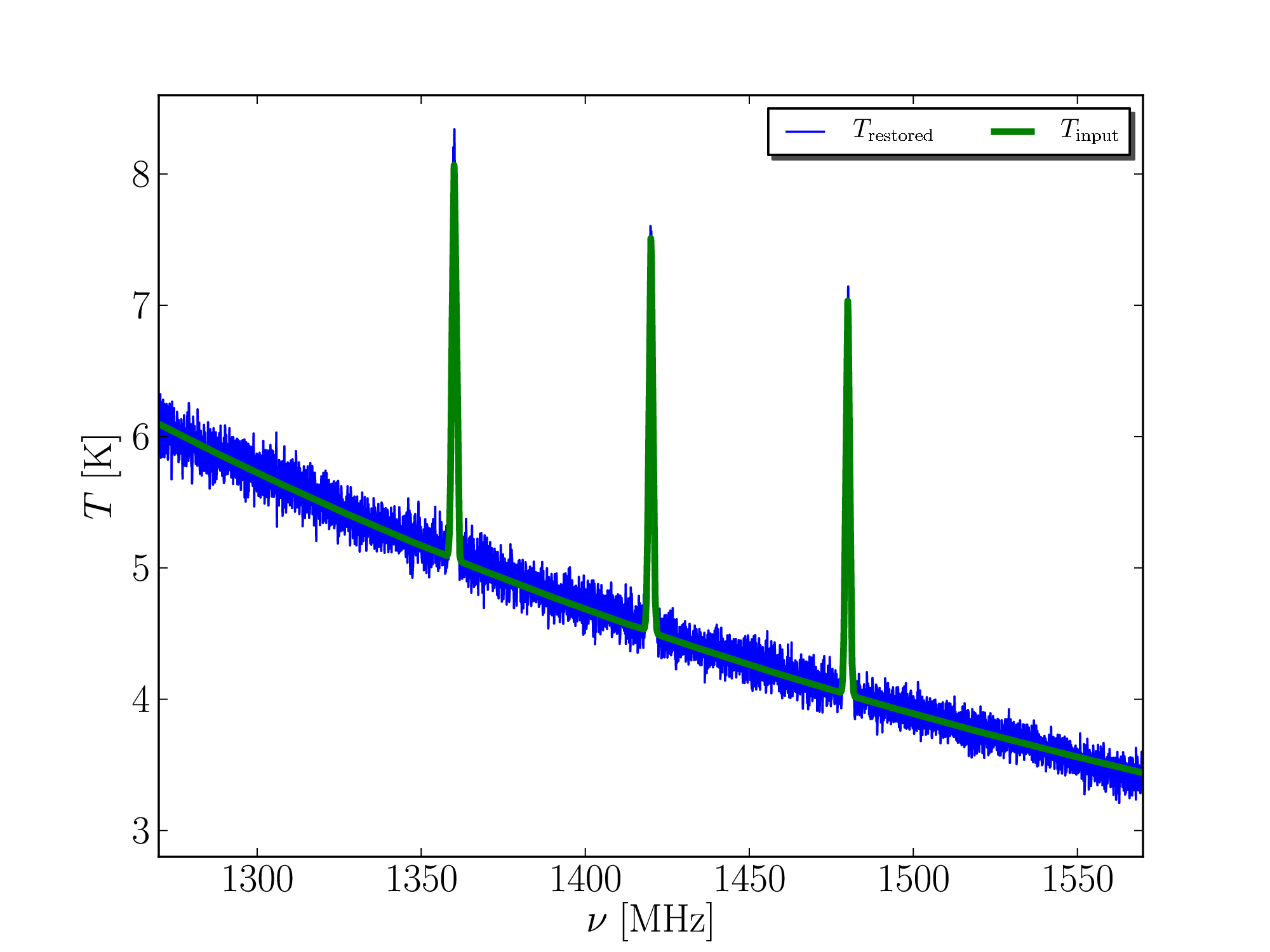}\\[0ex]
   \includegraphics[width=0.45\textwidth,bb=34 1 521 392,clip=]{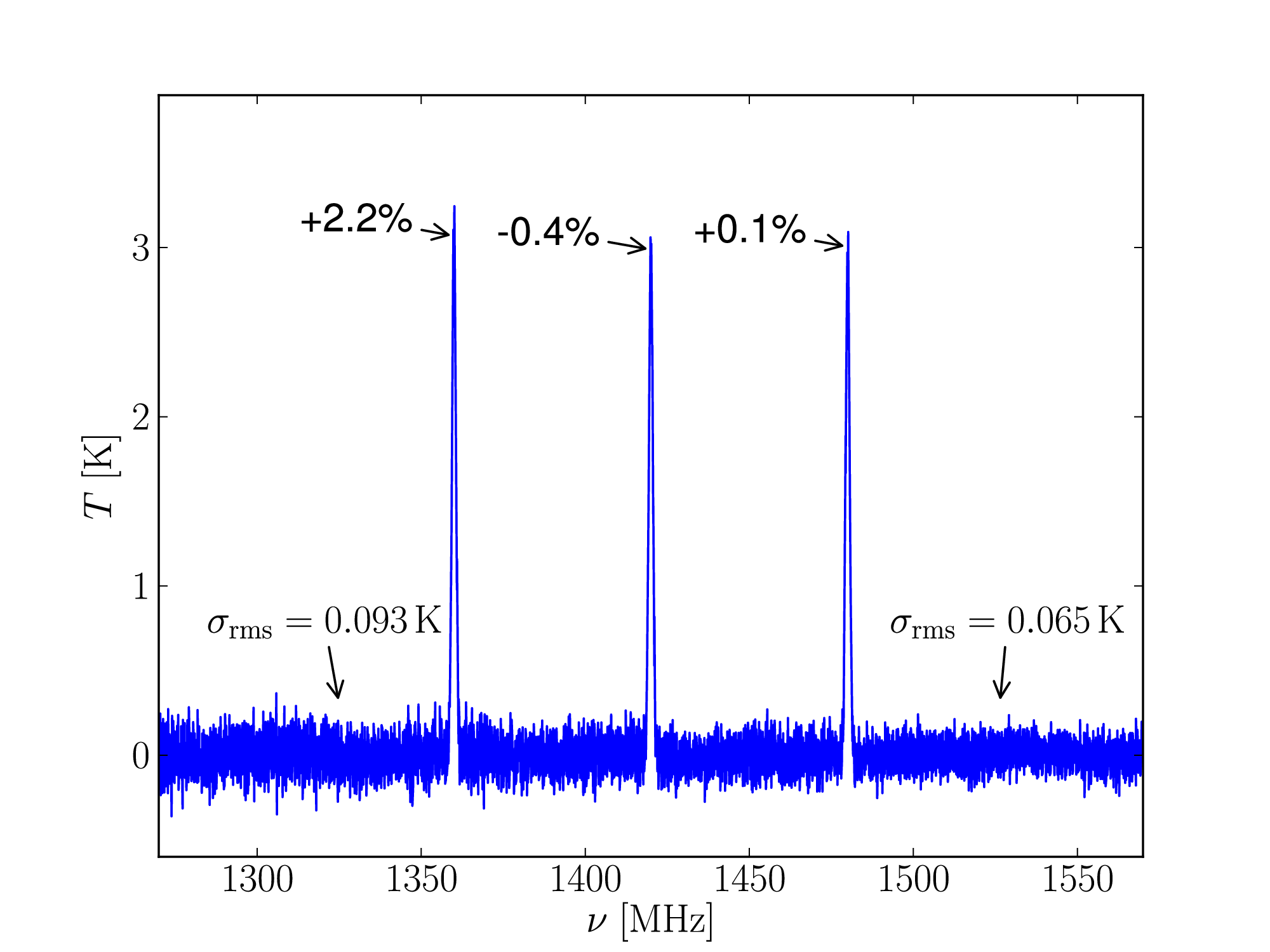}
  \caption{Inserting the model $f^\mathrm{[cal]}(\nu)$ into Eq.\,(\ref{eqpswitchfinal}) results in the correct flux calibration (\textbf{upper panel}). The \textbf{lower panel} shows the spectrum after baseline subtraction. Note that the correct continuum contribution of the source was also reconstructed (upper panel). The example was computed for the SW case.}%
   \label{fig:pswitch_method2_result}%
\end{figure}

Using the calculated model fits and applying Eq.\,(\ref{eqpswitchfinal}), one obtains the reduced spectrum. Fig.\,\ref{fig:pswitch_method2_result} shows the standing wave example where the upper panel shows the reconstructed spectrum, $T_\mathrm{sou}$, and the lower panel contains the result after baseline subtraction. The continuum contribution, $T^\mathrm{cont}_\mathrm{sou}$, of the source is correctly reproduced. In some cases however, the baseline may contain a residual imprint caused by standing waves, which affects the quality of the model fitting to some extent, especially in the presence of noise. In the worst case, this might also affect the flux calibration.  The more complex the required model, e.g. in the presence of SW, the larger the potential error in flux calibration.

We note that this method \textit{may not be applicable in all cases}. Inspecting Eq.\,(\ref{eqpswitchfinal}) in detail reveals that the denominator of the term 
\begin{equation}
\frac{f^\mathrm{cal}(\nu)}{f(\nu)-f^\mathrm{cal}(\nu)} 
\end{equation}
can become zero. This occurs if $T_\mathrm{cont}^\mathrm{sou}+T_\mathrm{sys,on}\approx T_\mathrm{sys,off}$, i.e. for sources with low continuum flux. In these cases, a nearby calibrator may be used for an independent reference position in order to obtain a well-behaved correction term.

\subsection{Determination of $T_\mathrm{sys}$ neglecting any dependence on frequency --- the `classical' approach}\label{subsec:pswitchclassicalmethod}
The `classical' case where $T_\mathrm{sys}$ and $T_\mathrm{cal}$ are both treated as constants is clearly a simplified case of the first method (presented in Section\,\ref{subsec:pswitchmethod1}) and of course causes errors in the flux calibration. This simplification was justifiable to some extent until about two decades ago when broadband receiving systems started to become more commonly used. Today, this approach is still widely used, e.g. within the online reduction pipeline at the 100-m telescope in Effelsberg \citep[see][]{kraus09techreport}, in order to provide a fast and robust online display of measured spectra.

Eq.\,(\ref{eqpswitchtsystcal}) may be rewritten as
\begin{equation}
\frac{T_\mathrm{sys,off}}{T_\mathrm{cal}}=\frac{P_\mathrm{off}}{(P_\mathrm{off}^\mathrm{cal}-P_\mathrm{off})}=\frac{(P_\mathrm{off}^\mathrm{cal}+P_\mathrm{off})-(P_\mathrm{off}^\mathrm{cal}-P_\mathrm{off})}{2(P_\mathrm{off}^\mathrm{cal}-P_\mathrm{off})}.
\end{equation}
In principle, this provides a direct means of calculating $T_\mathrm{sys,off}$ from $T_\mathrm{cal}$. Unfortunately, a simple evaluation of the above equation is numerically unstable because the denominator can have values close to zero or even become negative owing to noise. If we treat $T_\mathrm{sys,off}$ and $T_\mathrm{cal}$ as a constant with respect to time and \textit{frequency}, we find that
\begin{equation}
\frac{T_\mathrm{sys,off}}{T_\mathrm{cal}} \approx \frac{(P_\mathrm{off}^\mathrm{cal}+P_\mathrm{off})-\langle P_\mathrm{off}^\mathrm{cal}-P_\mathrm{off}\rangle_\nu}{2\langle P_\mathrm{off}^\mathrm{cal}-P_\mathrm{off}\rangle_\nu}\label{eqsimpletsys},
\end{equation}
where $\langle P^\mathrm{cal}-P\rangle_\nu$ is obtained by calculating the average of the inner 50\% of the difference spectrum in the online reduction pipeline at the 100-m telescope at Effelsberg. The current data reduction pipeline at the GBT \citep[GBTIDL;][]{gbtidl}\footnote{\texttt{http://gbtidl.nrao.edu/}} uses the equivalent method except using the inner 80\% of the spectra. For future versions of GBTIDL, a vectorised approach is planned whereby averaging is performed on certain bins.

The parameter $\overline T_\mathrm{sys}$ is calculated by the same averaging procedure, such that
\begin{equation}
\begin{split}
\overline T_\mathrm{sys,off}&=\langle T_\mathrm{sys,off}\rangle_\nu \\
&= T_\mathrm{cal} \left\langle\frac{(P_\mathrm{off}^\mathrm{cal}+P_\mathrm{off})-\langle P_\mathrm{off}^\mathrm{cal}-P_\mathrm{off}\rangle_\nu}{2\langle P_\mathrm{off}^\mathrm{cal}-P_\mathrm{off}\rangle_\nu} \right\rangle_\nu.\label{eqeffelsbergpipelinetsys}
\end{split}
\end{equation}
Using a scalar value for $T_\mathrm{cal}$, roughly known for each receiver from Eq.\,(\ref{eqsimpletsys}), we obtain the `calibrated' spectrum

\begin{equation}
T_\mathrm{sou}=\overline T_\mathrm{sys,off}\frac{ P_\mathrm{on}- P_\mathrm{off}}{ P_\mathrm{off}}=\left(\overline T_\mathrm{sys,off}+T_\mathrm{cal}\right)\frac{ P_\mathrm{on}^\mathrm{cal}- P_\mathrm{off}^\mathrm{cal}}{ P_\mathrm{off}^\mathrm{cal}}\label{eqeffelsbergpipeline}.
\end{equation}
Both, \textit{cal} and \textit{non-cal} spectra, should again be added to reduce noise in the final spectrum.

Although this approach is much easier to implement in software, the reader should be warned that neglecting the frequency dependence of $T_\mathrm{sys}$ introduces systematic errors, which can be a serious problem. This is especially the case when computing line ratios or dealing with statistical analyses of large samples of sources at various radial velocities/frequencies (see Section\,\ref{sec:errordistribution}).

\begin{figure*}[!t]
   \centering%
   \includegraphics[width=0.48\textwidth,bb=26 42 521 392,clip=]{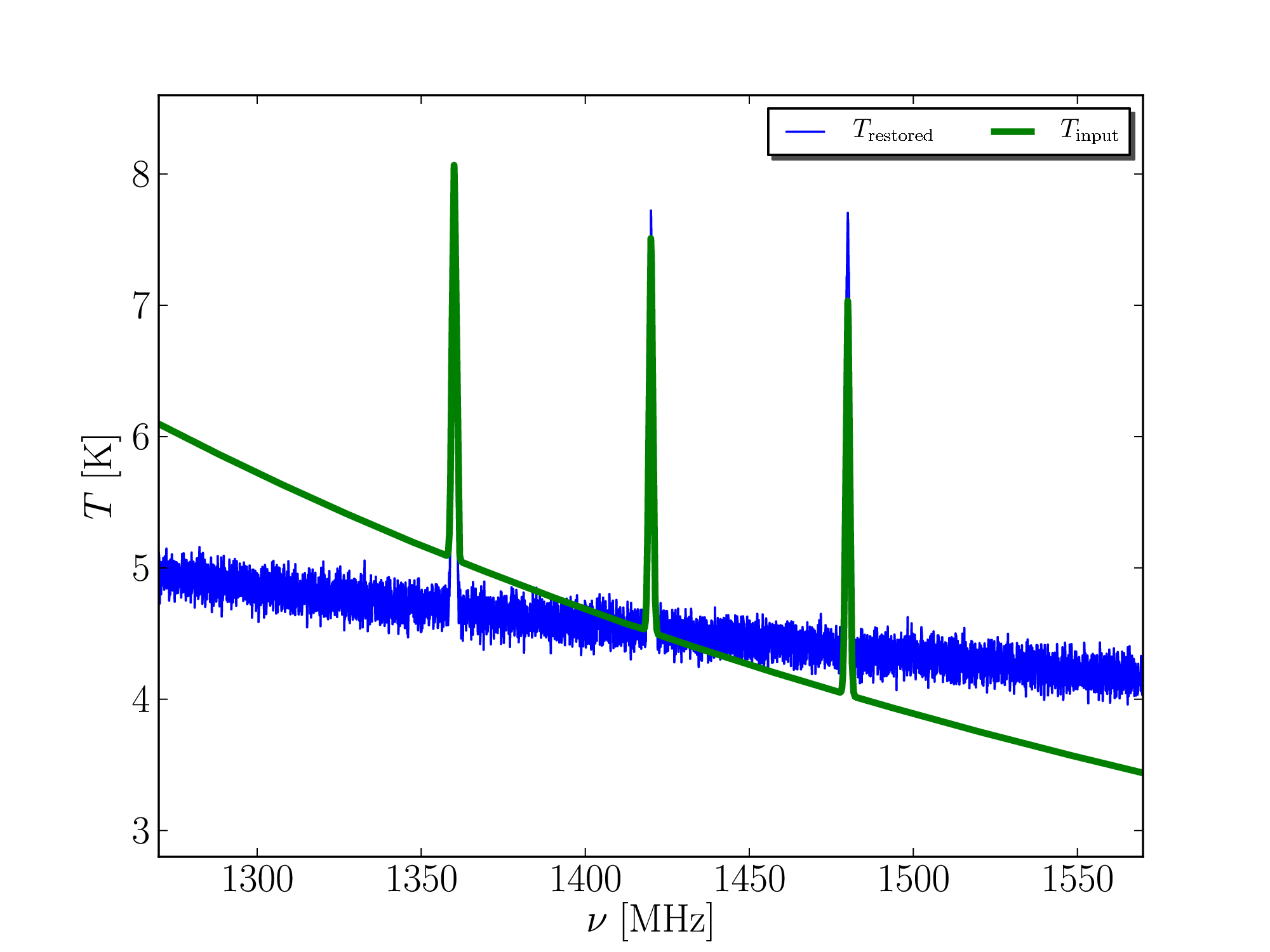}\quad
   \includegraphics[width=0.48\textwidth,bb=26 42 521 392,clip=]{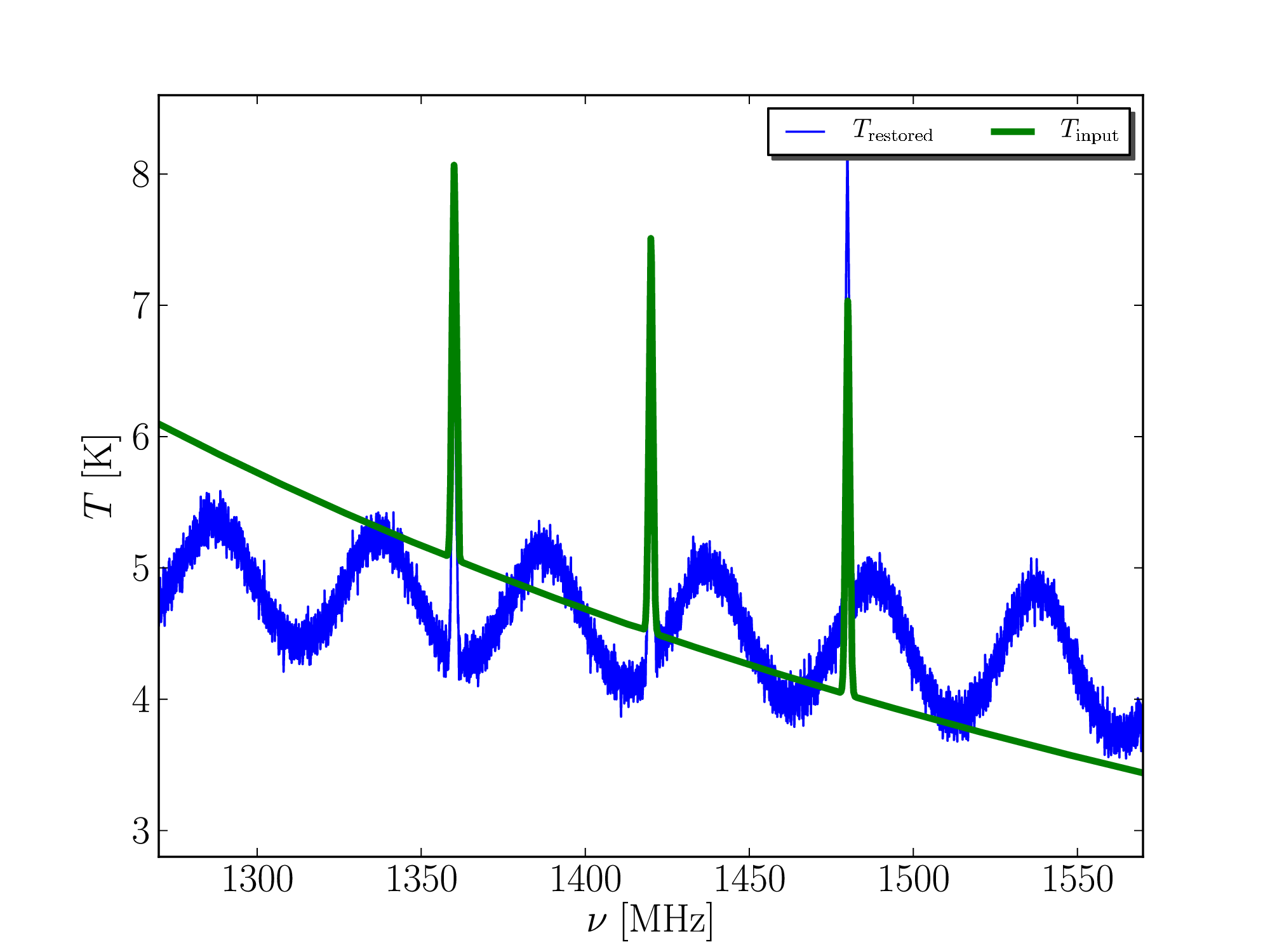}\\[0ex]
   \includegraphics[width=0.48\textwidth,bb=26 1 521 392,clip=]{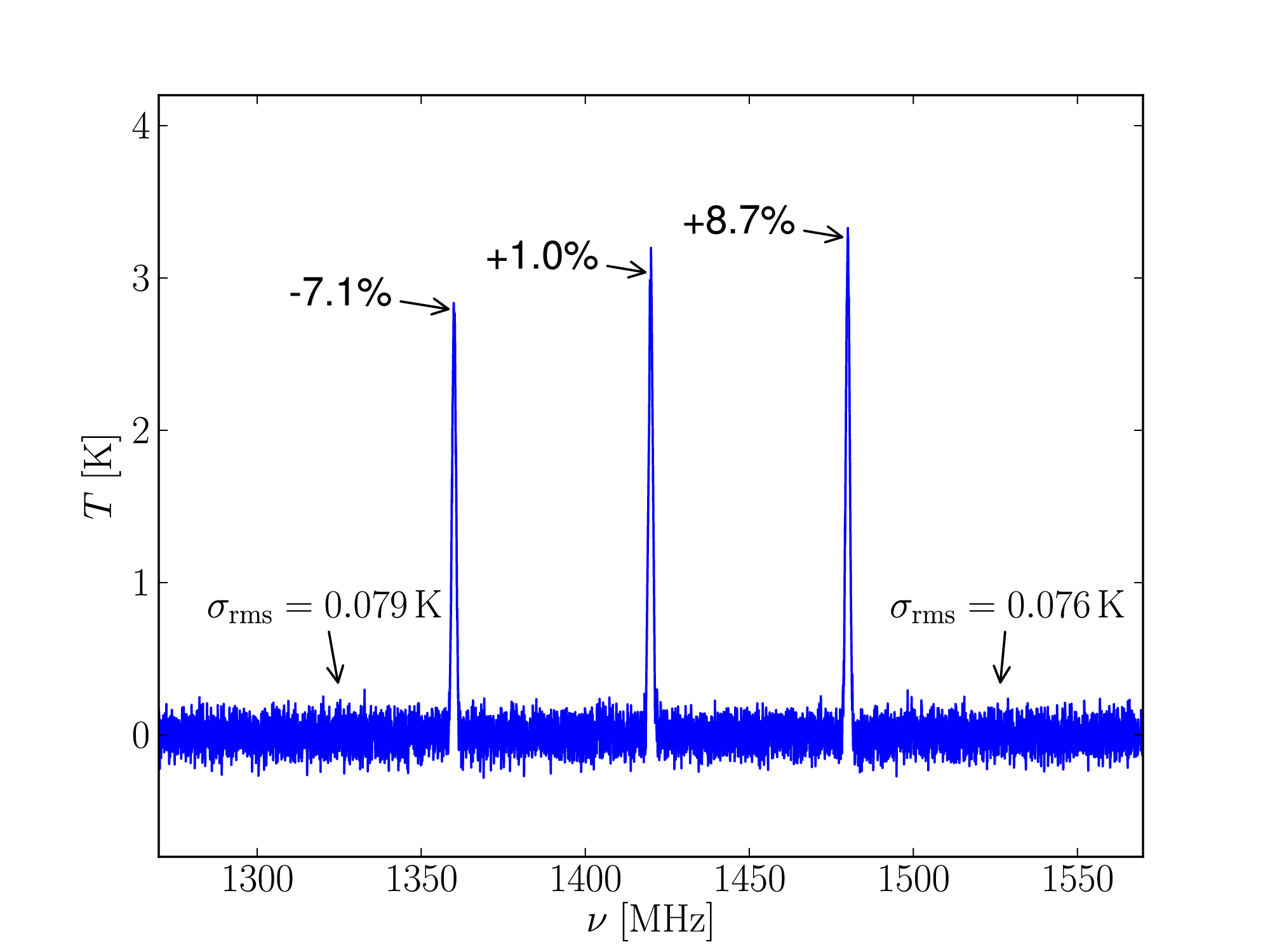}\quad
   \includegraphics[width=0.48\textwidth,bb=26 1 521 392,clip=]{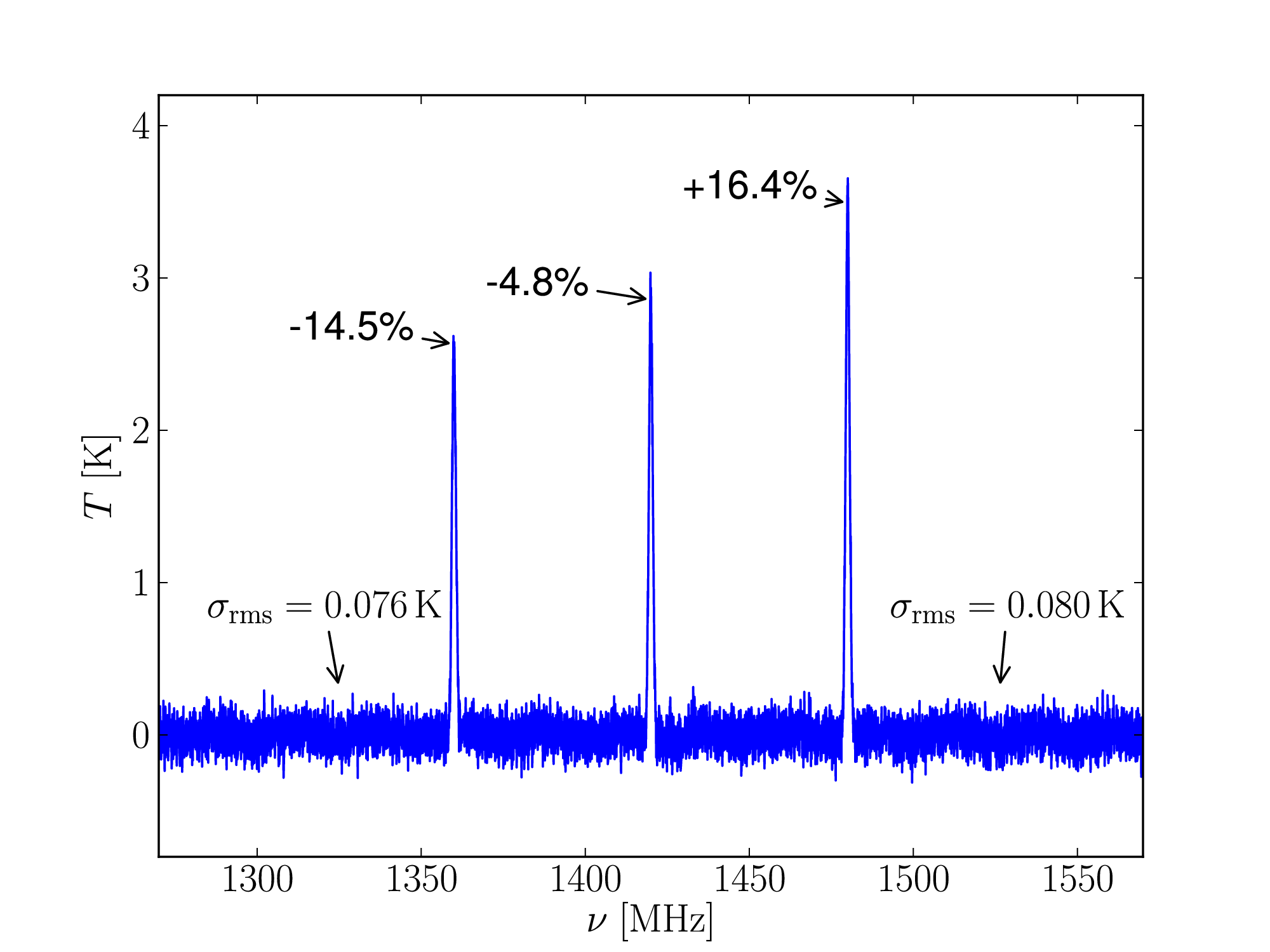}
  \caption{Applying the `classical' calibration scheme that is currently in use at the 100-m telescope (for online display purposes), using Eq.\,(\ref{eqeffelsbergpipeline}), we obtain the reduced spectra (\textbf{upper row}). The input spectra $T_\mathrm{input}$ (green) are also shown for reference. Not only are the flux values of the Gaussians improperly reconstructed, which becomes visible after baseline subtraction  (\textbf{lower row}), the inferred continuum fluxes of the source are also incorrect. The \textbf{left panels} show the results for the simpler case, while the \textbf{right panels} includes the standing wave contribution.  }%
   \label{fig:pswitch_classic_result}%
\end{figure*}

Figure~\ref{fig:pswitch_classic_result} (upper panels) shows the resulting spectra after applying Eq.\,(\ref{eqeffelsbergpipeline}). We note that the baseline does not match the input baseline at all. For reference, we subtracted a baseline (which proved to be rather complicated in the SW case) from the resulting spectra. The result clearly shows a systematic flux calibration error and is displayed in Fig.\,\ref{fig:pswitch_classic_result} (lower panels).

\subsection{Computing the bandpass curve $G=G_\mathrm{IF}G_\mathrm{RF}$}

\begin{figure}[!t]
\centering
\includegraphics[width=0.48\textwidth,bb=7 1 523 394,clip=]{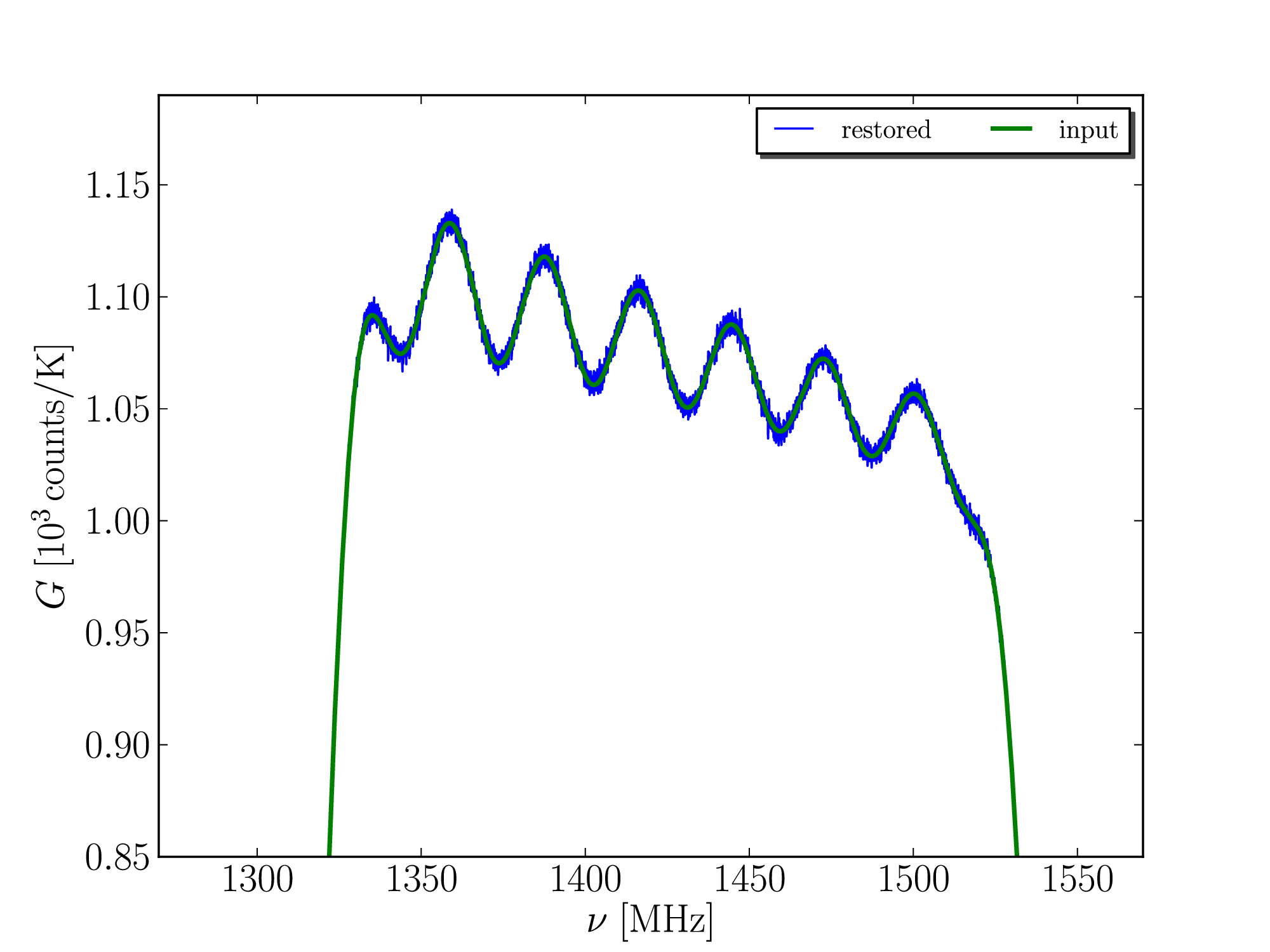}
\caption{For position switching, one can determine the gain curve. The plot shows the input and reconstructed bandpass.}%
\label{fig:pswitch_method1_reconstructedbandpass}%
\end{figure}

From Eq.\,(\ref{pswitchbasicpoff}), it directly follows that
\begin{equation}
G=G_\mathrm{IF}G_\mathrm{RF}=\frac{P_\mathrm{off}}{T_\mathrm{sys,off}}= \frac{P_\mathrm{off}^\mathrm{cal}}{(\kappa_\mathrm{off}+1) T_\mathrm{cal}}\,\label{eqpswitchreconstructedbandpass}.
\end{equation}
where $T_\mathrm{sys,off}$ can be calculated, for example, via Eq.\,(\ref{eqpswitchkappa}), $T_\mathrm{sys,off}=\kappa_\mathrm{off} T_\mathrm{cal}$, or using Eq.\,(\ref{eqpswitchtsysfromf}) and (\ref{eqpswitchtsysfromfcal}). This can be of great practical use. While the inferred $G$ from a single measurement (see Fig.\,\ref{fig:pswitch_method1_reconstructedbandpass}) is rather noisy, one may use several observations, even of different sources, to average the gain curve and suppress noise. The calculated $G$ could then in turn be applied to all involved data sets. In practice, the gain is usually imperfectly stable in time. Eq.\,(\ref{eqpswitchreconstructedbandpass}) enables us to monitor the time evolution of the receiving system. It is even likely that temporal drifts can be approximated to some degree with a low-order polynomial (spectral channel-wise) and can then be used to improve the accuracy of the data reduction.

\subsection{A realistic case}\label{subsec:pswitchrealisticexample}
\begin{figure}[!t]
\centering%
\includegraphics[width=0.48\textwidth,bb=15 1 522 392,clip=]{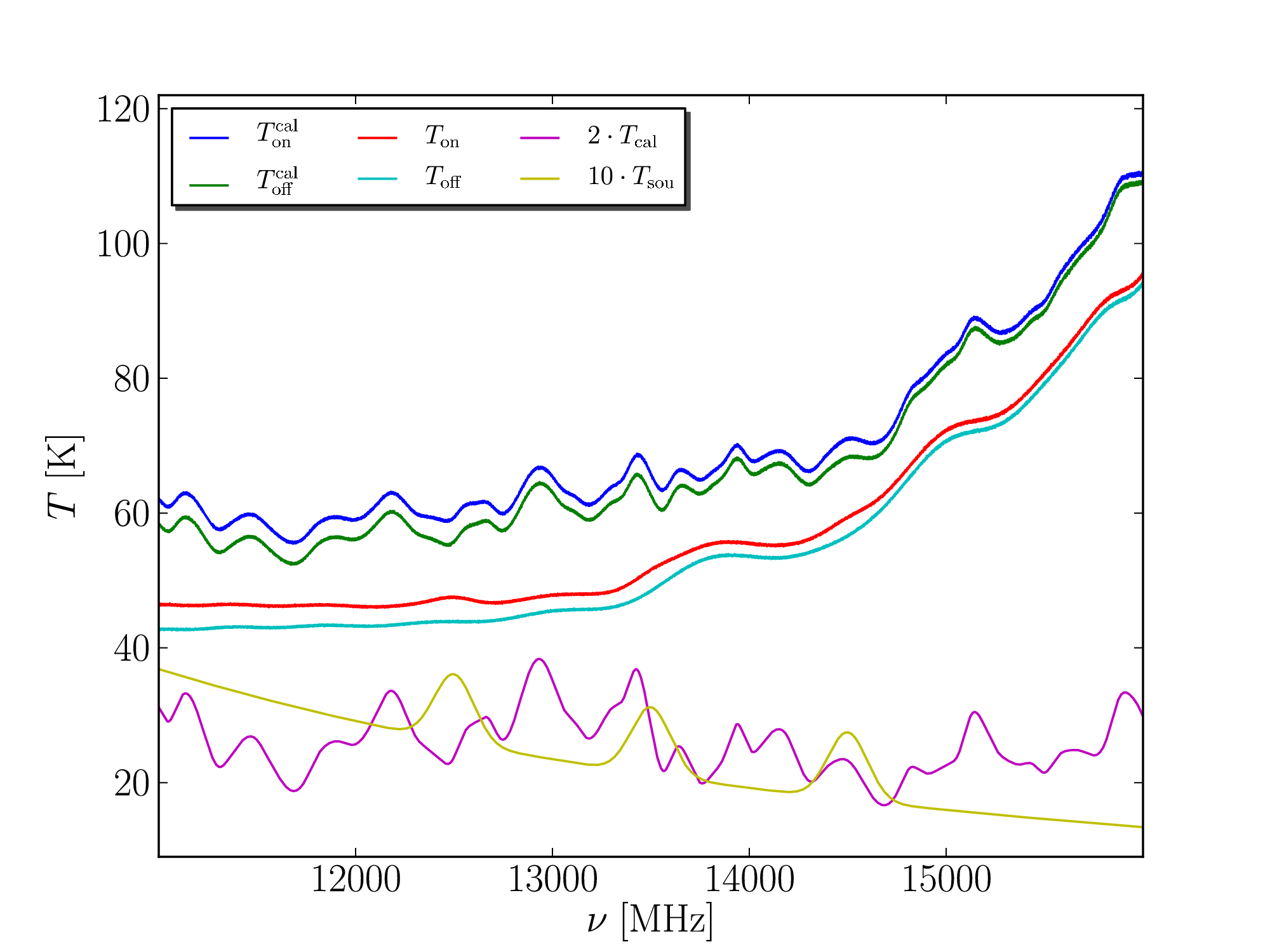}
\caption{Temperature inputs for a more realistic simulation. The three Gaussian emission lines have a 1\,K amplitude and a line width of 240\,MHz (FWHM). The number of spectral channels is 128k.}%
\label{fig:pswitchcomplicatedtcalinput}%
\end{figure}

We present the results of a more complex simulation that resembles realistic observations. In comparison to the previous example, both $T_\mathrm{sys}$ and $T_\mathrm{cal}$ now have more structure. Both quantities were modelled to imitate the outcome of a measurement with the 1.9-cm primary-focus receiver at the 100-m telescope. The resulting spectrum is illustrated in Fig.\,\ref{fig:pswitchcomplicatedtcalinput}. $T_\mathrm{sys}$ shows a steep increase towards higher frequency, attributed to the water vapour line at $22\,\mathrm{GHz}$ (during the observations, weather conditions were rather bad).

\begin{figure}[!t]
\centering%
\includegraphics[width=0.48\textwidth,bb=10 1 522 392,clip=]{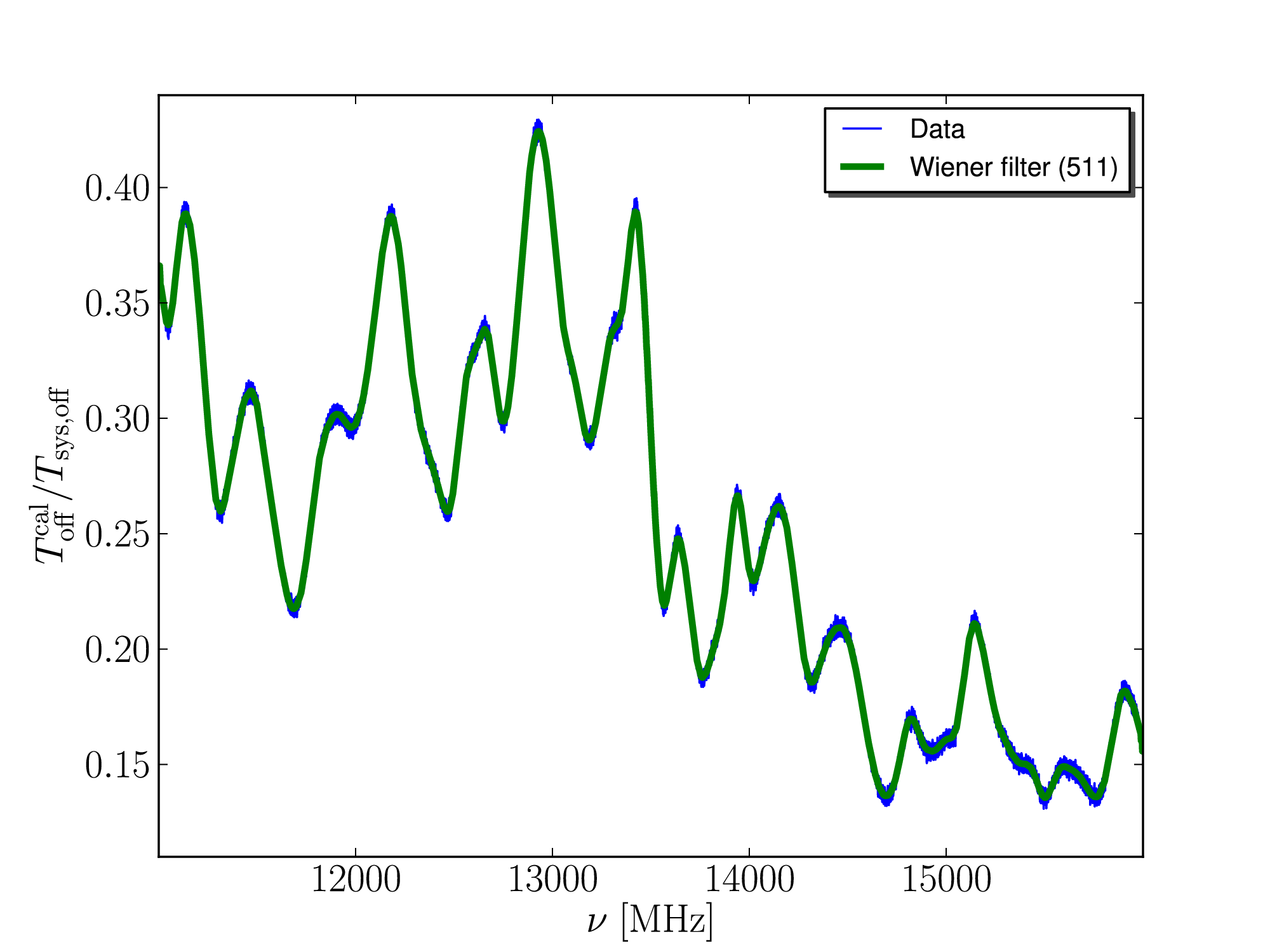}
\caption{A plot to show $\kappa_\mathrm{off}^{-1}$ and a low-noise model obtained from applying a Wiener filter \citep[window size: 511; see][]{wiener49}.}%
\label{fig:pswitchcomplicatedkappa}%
\end{figure}

In this instance, the first method as described in Section\,\ref{subsec:pswitchmethod1} is applied using spectral filtering instead of polynomial models to describe $\kappa_\mathrm{off}^{-1}$. The result is shown in Fig.\,\ref{fig:pswitchcomplicatedkappa}. The second approach  requires spectral windows to be set around the (expected) emission lines such that a simple filtering approach would not work. It is therefore neglected here.

\begin{figure}[!t]
\centering%
\includegraphics[width=0.48\textwidth,bb=20 1 522 392,clip=]{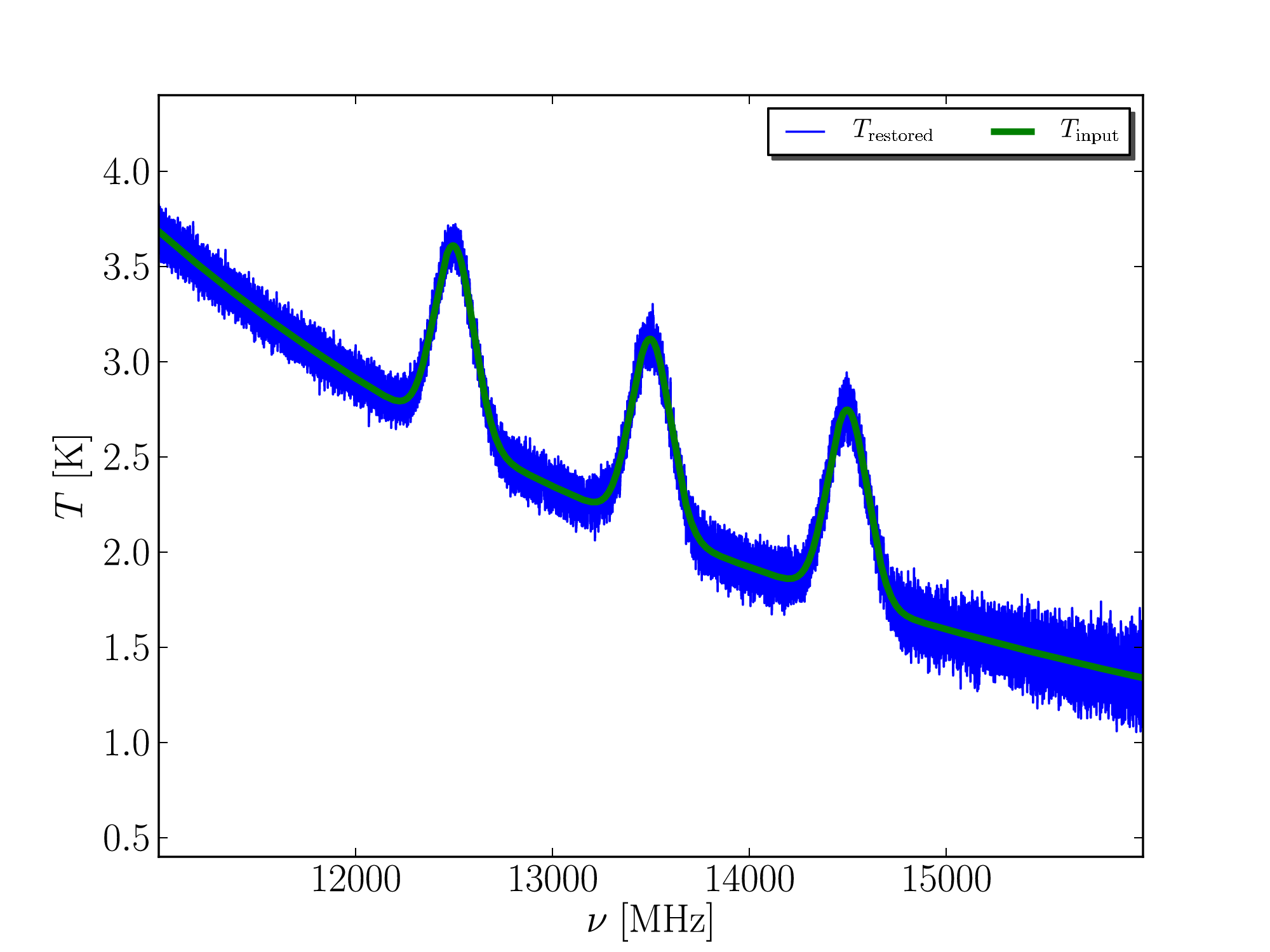}
\caption{The resulting spectrum in comparison with the input flux.}%
\label{fig:pswitchcomplicatedresult}%
\end{figure}

\begin{figure}[!t]
\centering%
\includegraphics[width=0.48\textwidth,bb=20 1 522 392,clip=]{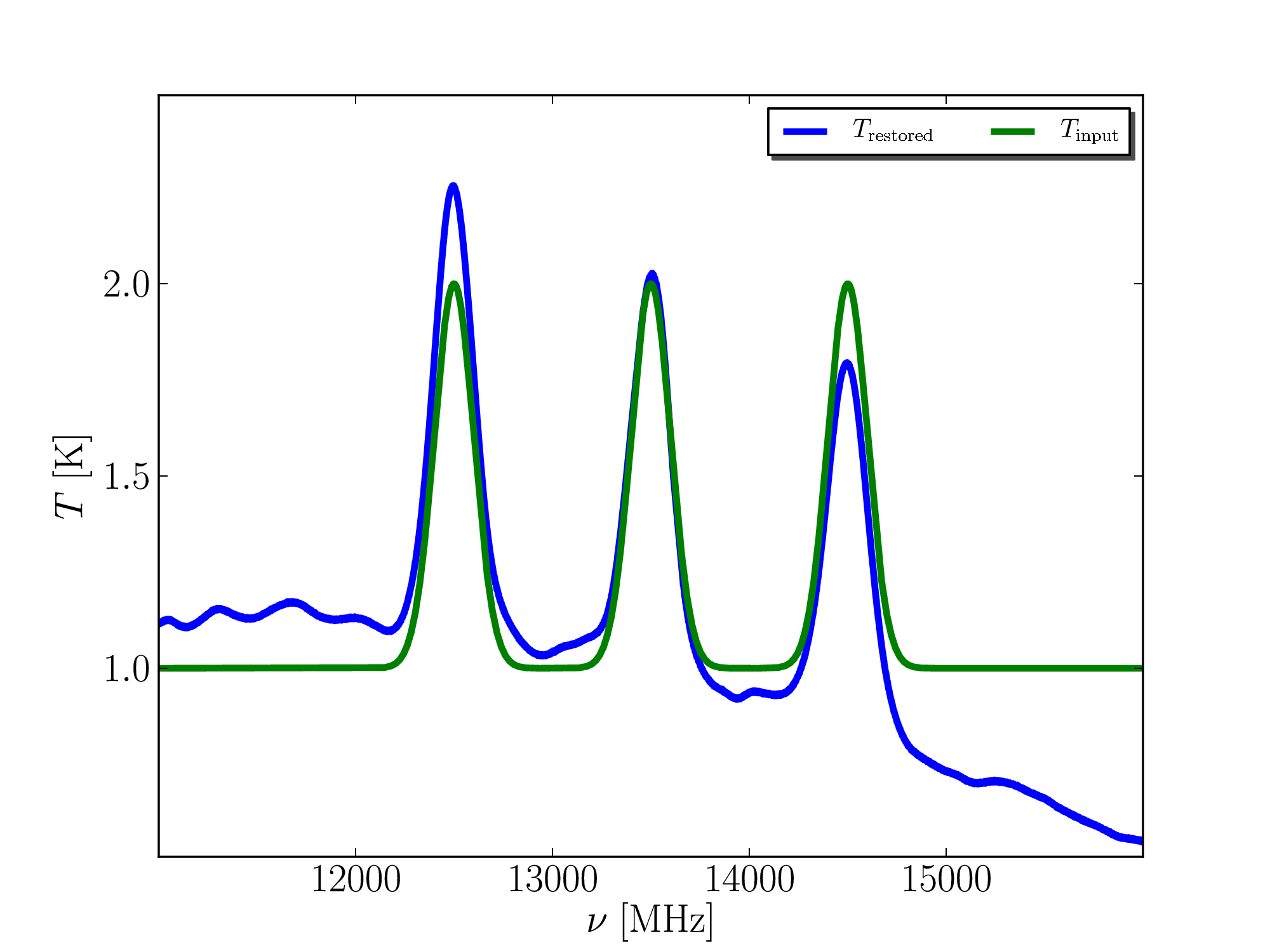}
\caption{The classic calibration scheme leads to highly biased results. Note that for visualisation purposes a flat continuum contribution (1\,K) of the source was used and no noise was added to the input temperatures.}%
\label{fig:pswitchcomplicatedclassic}%
\end{figure}

Fig.\,\ref{fig:pswitchcomplicatedresult} displays the resulting spectrum. Despite the very complex set-up, the calibration worked well and the underlying continuum flux of the observed source could be reconstructed. Fig.\,\ref{fig:pswitchcomplicatedclassic} shows the result of the classic approach. For purposes of comparison, a constant continuum contribution was chosen. The resulting baseline is very complex and is in fact the incorrectly calibrated continuum flux of the source. In a worst case scenario, this can lead to a misinterpretation of the results depending on the line widths of the expected astronomical features. For example, Fig.\,\ref{fig:pswitchcomplicatedclassic} suggests that the middle spectral line has a blue-shifted wing.

\section{Frequency switching}\label{sec:fswitch}
In frequency switching, one uses two different local oscillator (LO) frequencies to provide a reference spectrum. The shifted spectrum serves as a reference to remove the bandpass dependence in the non-shifted spectrum. In practice, a symmetric LO shifting pattern is often used, such that Eq.\,(\ref{eqbasiceq}) may be written as
\begin{align}
P_\mathrm{sig}^\mathrm{[cal]}(\nu)=&G_\mathrm{IF}(\nu)G_\mathrm{RF}(\nu-\Delta \nu)\times\nonumber\\
&\left[ T_\mathrm{sou}(\nu-\Delta \nu)+T_\mathrm{sys,sig}^\mathrm{[cal]}(\nu-\Delta \nu)\right],\\
P_\mathrm{ref}^\mathrm{[cal]}(\nu)=&G_\mathrm{IF}(\nu)G_\mathrm{RF}(\nu+\Delta \nu)\times\nonumber\\
&\left[ T_\mathrm{sou}(\nu+\Delta \nu)+T_\mathrm{sys,ref}^\mathrm{[cal]}(\nu+\Delta \nu)\right].
\end{align}
Together with the \textit{cal} and \textit{non-cal} phases, the LO switching leads to a total of four different phases of \textit{sig+cal, ref, sig}, and \textit{ref+cal}, which are usually rapidly stepped through in order to avoid time-dependent instabilities.

Using a shorter notation and assuming $T_\mathrm{sys,sig}^\mathrm{[cal]}(\nu)=T_\mathrm{sys,ref}^\mathrm{[cal]}(\nu)$, which should be fulfilled since switching is performed on short timescales
\begin{align}
P_\mathrm{sig}^\mathrm{[cal]}(\nu)&=G_\mathrm{IF}G_\mathrm{RF,-}\left[ T_\mathrm{sou,-}+T_\mathrm{sys,-}^\mathrm{[cal]}\right],\\
P_\mathrm{ref}^\mathrm{[cal]}(\nu)&=G_\mathrm{IF}G_\mathrm{RF,+}\left[ T_\mathrm{sou,+}+T_\mathrm{sys,+}^\mathrm{[cal]}\right].
\end{align}
We first assume that $G_\mathrm{RF,-}= G_\mathrm{RF,+}$, which is fulfilled only in the very rare cases that $G_\mathrm{RF}(\nu)=\textrm{constant}$. Then
\begin{align}
\frac{P_\mathrm{sig}^\mathrm{[cal]} - P_\mathrm{ref}^\mathrm{[cal]}}{ P_\mathrm{ref}^\mathrm{[cal]}}&=\frac{T_\mathrm{sou,-}+T_\mathrm{sys,-}^\mathrm{[cal]}-T_\mathrm{sou,+}-T_\mathrm{sys,+}^\mathrm{[cal]}}{T_\mathrm{sou,+}+T_\mathrm{sys,+}^\mathrm{[cal]}}\label{eqsigref}\\
&=\frac{T_\mathrm{line,-}^\mathrm{sou}+T_\mathrm{cont,-}^\mathrm{sou}-T_\mathrm{line,+}^\mathrm{sou}-T_\mathrm{cont,+}^\mathrm{sou}+\Delta T_\mathrm{sys,\pm}^\mathrm{[cal]}}{T_\mathrm{line,+}^\mathrm{sou}+T_\mathrm{cont,+}^\mathrm{sou}+T_\mathrm{sys,+}^\mathrm{[cal]}}\nonumber,
\end{align}
where $\Delta T_\mathrm{sys,\pm}^\mathrm{[cal]}\equiv T_\mathrm{sys,-}^\mathrm{[cal]}-T_\mathrm{sys,+}^\mathrm{[cal]}=\Delta T_\mathrm{sys,\pm}\left[+\Delta T_\mathrm{cal,\pm}\right]$. 
We note that $\Delta T_\mathrm{sys,\pm}$  is equal in the \textit{cal} and \textit{non-cal} phases, while $\Delta T_\mathrm{cal,\pm}$ is different ($\Delta T_\mathrm{cal,\pm}=0$ in the \textit{non-cal} phase and $\Delta T_\mathrm{cal,\pm}\neq0$ in the \textit{cal} phase). Both quantities depend  only on the slope of $T_\mathrm{sys}$ and $T_\mathrm{cal}$ because all contributors to $T_\mathrm{sys}$ remain approximately equal during two adjacent switching phases.
Using Eq.\,(\ref{eqsigref}), it follows that 
\begin{align}
T_\mathrm{sou,-}-T_\mathrm{sou,+}+\Delta T_\mathrm{sys,\pm}^\mathrm{[cal]} &=\left(T_\mathrm{sou,+}+T_\mathrm{sys,+}^\mathrm{[cal]}\right)\frac{P_\mathrm{sig}^\mathrm{[cal]} - P_\mathrm{ref}^\mathrm{[cal]}}{ P_\mathrm{ref}^\mathrm{[cal]}}\nonumber\\&\equiv \tilde T_\mathrm{sig}^\mathrm{[cal]}\label{eqfswitchbaseeq1},\\
T_\mathrm{sou,+}-T_\mathrm{sou,-}-\Delta T_\mathrm{sys,\pm}^\mathrm{[cal]} &=\left(T_\mathrm{sou,-}+T_\mathrm{sys,-}^\mathrm{[cal]}\right)\frac{P_\mathrm{ref}^\mathrm{[cal]} - P_\mathrm{sig}^\mathrm{[cal]}}{ P_\mathrm{sig}^\mathrm{[cal]}}\nonumber\\&\equiv \tilde T_\mathrm{ref}^\mathrm{[cal]}.\label{eqfswitchbaseeq2}
\end{align}
In Sections\,\ref{subsec:fswitchmethod1} and \ref{subsec:fswitchmethod2}, we discuss how to perform the calibration for frequency switching, i.e. how to determine $T_\mathrm{sou,\pm}^\mathrm{cont}+T_\mathrm{sys,\pm}^\mathrm{[cal]}$.

To obtain an optimal S/N, one should not only average \textit{cal} and \textit{non-cal} phases but also shift (to match RF frequencies) and average $\tilde T_\mathrm{sig}$ and $\tilde T_\mathrm{ref}$
\begin{equation}
\begin{split}
&\frac{1}{2}\left[\tilde T_\mathrm{sig}^\mathrm{[cal]}(\nu+\Delta \nu)+\tilde T_\mathrm{ref}^\mathrm{[cal]}(\nu-\Delta \nu)\right]\\
&=T_\mathrm{sou}(\nu)-\frac{1}{2}T_\mathrm{sou}(\nu+2\Delta \nu)-\frac{1}{2}T_\mathrm{sou}(\nu-2\Delta \nu)\\
&+T_\mathrm{sys}^\mathrm{[cal]}(\nu)-\frac{1}{2}T_\mathrm{sys}^\mathrm{[cal]}(\nu+2\Delta \nu)-\frac{1}{2}T_\mathrm{sys}^\mathrm{[cal]}(\nu-2\Delta \nu)\label{eqfswitchfinalequation}.
\end{split}
\end{equation}
This is clearly a much more complicated procedure than for the position switching case.

\subsection{Bandpass ghosts}\label{subsec:fswitchghosts}

In Eq.\,(\ref{eqfswitchfinalequation}), apart from the desired signal, $T_\mathrm{sou}(\nu)$, so-called `spectral-line ghosts' (sometimes also referred to as bandpass ghosts), $-\frac{1}{2}T_\mathrm{sou}(\nu\pm2\Delta \nu)$, appear in the resulting equation. They have half of the amplitude of the original signal. Furthermore, there is a residual \textit{additive}\footnote{This is an important point, as it allows for a simple subtraction of a baseline fit, without destroying the calibration.} baseline depending on the relative changes in $T_\mathrm{sys}$ and $T_\mathrm{cal}$ with frequency.  These can make the final baseline fitting a challenging task, especially in the presence of standing waves (SW).

\begin{figure}[!t]
\centering
\includegraphics[width=0.44\textwidth,bb=8 20 523 200,clip=]{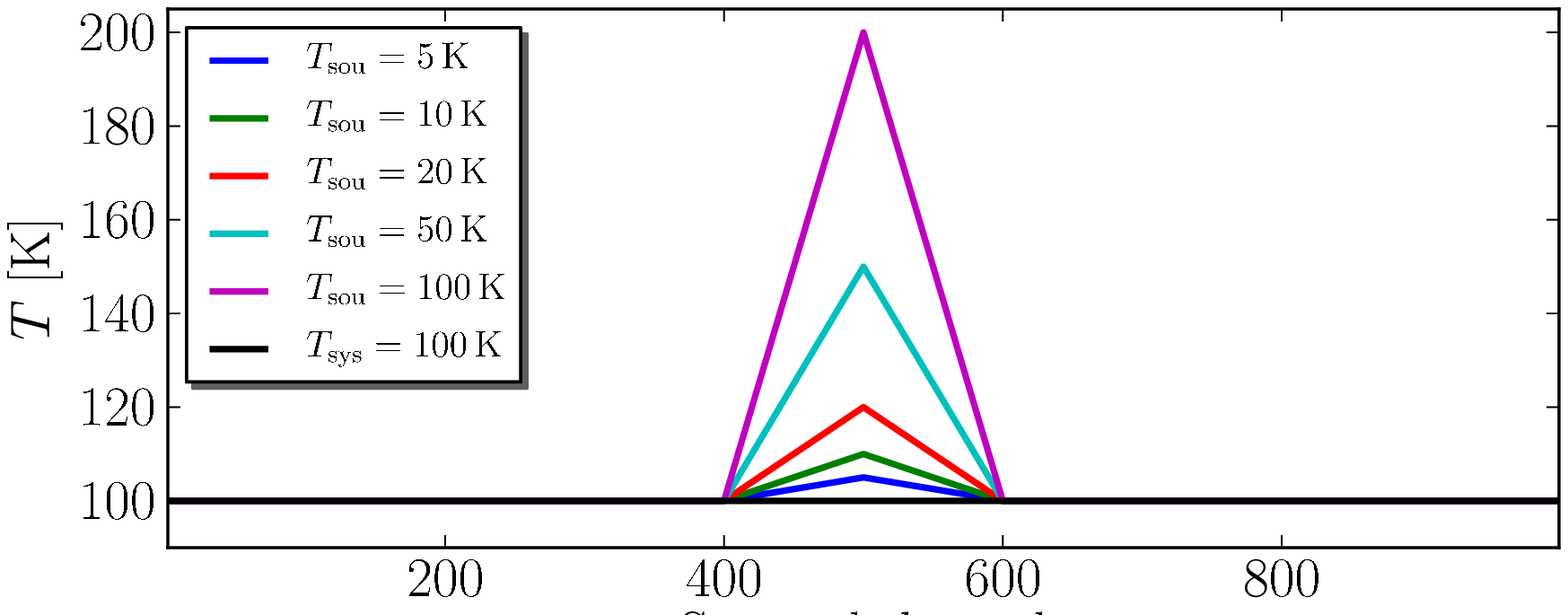}\\[0ex]
\includegraphics[width=0.44\textwidth,bb=8 27 523 263,clip=]{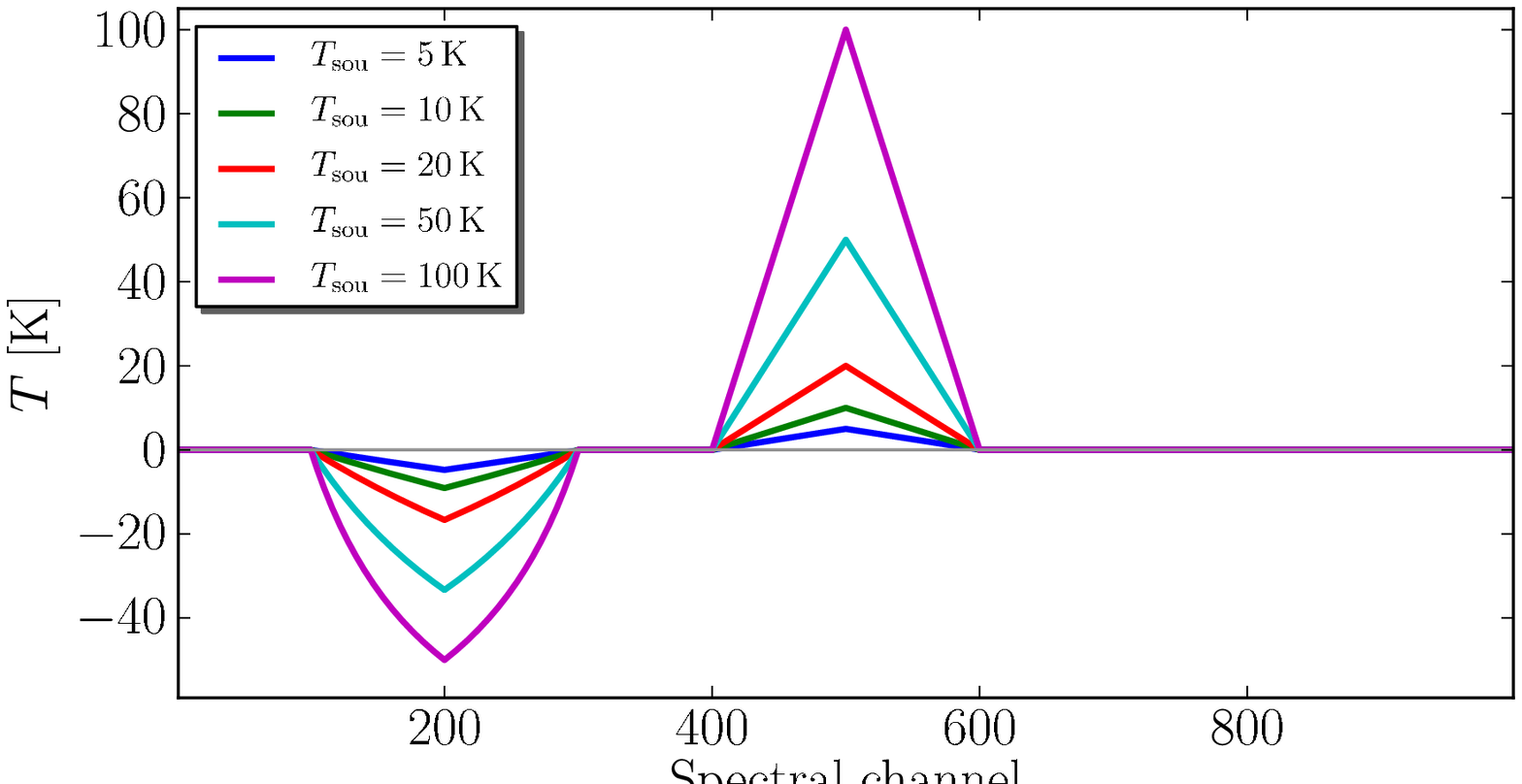}\\[0ex]
\includegraphics[width=0.44\textwidth,bb=8 0 523 328,clip=]{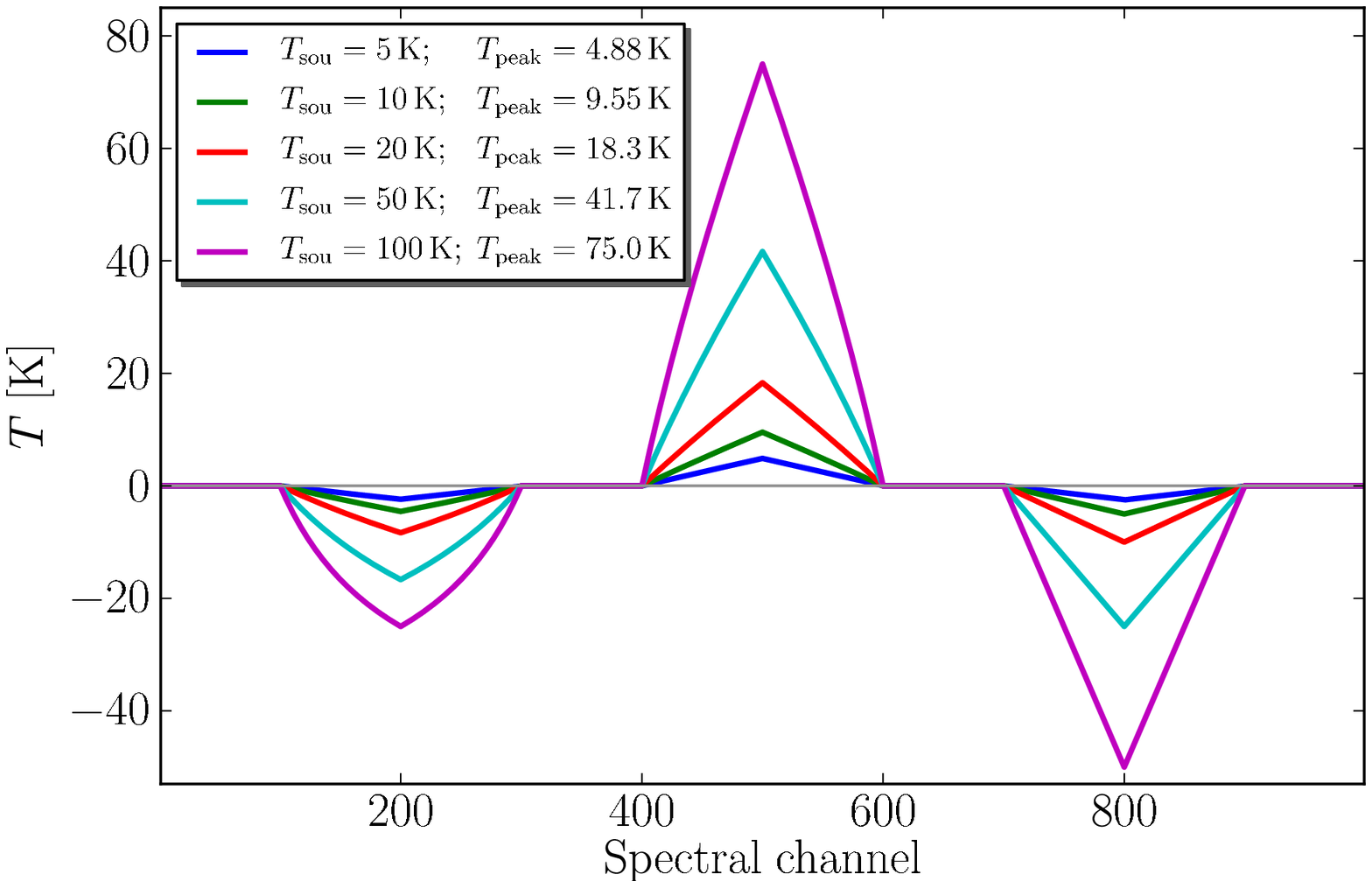}
\caption{Applying the frequency-switching base equation, Eq.\,(\ref{eqfswitchfinalequation}), one would expect the spectral-line ghosts to have half the amplitude of the original line. However, in practice one is usually unable to determine $T_\mathrm{sou,\pm}+T_\mathrm{sys,\pm}^\mathrm{[cal]}=T_\mathrm{sou,\pm}^\mathrm{line}+T_\mathrm{sou,\pm}^\mathrm{cont}+T_\mathrm{sys,\pm}^\mathrm{[cal]}$ --- required for Eq.\,(\ref{eqfswitchbaseeq1}) and (\ref{eqfswitchbaseeq2}) --- but only $T_\mathrm{sou,\pm}^\mathrm{cont}+T_\mathrm{sys,\pm}^\mathrm{[cal]}$. As a consequence the inferred calibration is wrong at frequencies where the spectral-line ghosts appear. This does not only affect the amplitude of the ghosts, but even their shape gets distorted. To show the effect of this, we use a simple model in which  different source intensities ($T_\mathrm{sou}=5,10,20,50,100\,\mathrm{K}$) were added to a hypothetical system temperature, $T_\mathrm{sys}=100\,\mathrm{K}$ \textbf{(top panel)}. The bandpass was assumed to be flat. For visualisation purposes, the spectral lines are triangular-shaped. Increasing the input source temperature with respect to $T_\mathrm{sys}$ decreases the line ratios of $T_\mathrm{ghost}$ to $T_\mathrm{sou}$ after applying frequency shifting \textbf{(middle panel)}. In addition the distortion becomes more and more pronounced. Finally, in the bottom panel the result of a simple shift--flip-sign--averaging (so-called \textit{folding}; see also Eq.\,(\ref{eqclassfolding})) is presented. The legend contains the reconstructed peak temperatures, which are systematically deficient for larger ratios of $T_\mathrm{sou}/T_\mathrm{sys}$. This \textit{folding}  algorithm should clearly be avoided for cases where $T_\mathrm{sou}\nll T_\mathrm{sys}$.}%
\label{fig:bpghosts}%
\end{figure}

One thing to note is that usually we are unable to determine $T_\mathrm{sou,\pm}+T_\mathrm{sys,\pm}^\mathrm{[cal]}=T_\mathrm{sou,\pm}^\mathrm{line}+T_\mathrm{sou,\pm}^\mathrm{cont}+T_\mathrm{sys,\pm}^\mathrm{[cal]}$ and can only compute $T_\mathrm{sou,\pm}^\mathrm{cont}+T_\mathrm{sys,\pm}^\mathrm{[cal]}$ which then is substituted into the right-hand side of  Eq.\,(\ref{eqfswitchbaseeq1}) and (\ref{eqfswitchbaseeq2}). This is an essential drawback. Consequently, the estimated $\tilde T_\mathrm{sig,ref}$ are wrong for all frequencies where $T^\mathrm{line}_\mathrm{sou}(\nu\pm\Delta \nu)\ll T_\mathrm{sys}$ is not fulfilled (see Fig.\,\ref{fig:bpghosts}). 

Luckily, this usually only happens for frequencies at which the spectral-line ghosts appear (except for complicated cases; see Section\,\ref{subsec:fswitchrealisticexample}). If one uses both Eq.\,(\ref{eqfswitchbaseeq1}) and (\ref{eqfswitchbaseeq2}) and applies shift-and-averaging as described above, this is not a problem. \textit{However, if just one of these two equations is used and instead a so-called `fold' procedure}
(e.g., the \textsc{Class} task \texttt{fold} from the \textsc{Gildas} package\footnote{\texttt{http://www.iram.fr/IRAMFR/GILDAS}}) \textit{is applied, one will achieve incorrect results if $T_\mathrm{sou}\nll T_\mathrm{sys}$} (Fig.\,\ref{fig:bpghosts}, bottom panel). This is not a rare case, especially for observations in the decimetre-wavelength regime. The term \textit{folding} denotes the use of spectral-line ghosts themselves to add to the positive emission line, by changing the sign of and both shifting and averaging the signals
\begin{equation}
\frac{1}{2}\left[\tilde T_\mathrm{sig}^\mathrm{[cal]}(\nu+\Delta \nu)-\tilde T_\mathrm{sig}^\mathrm{[cal]}(\nu-\Delta \nu)\right]\label{eqclassfolding}\,.
\end{equation}

\subsection{Setting up simulations}
\begin{figure}[!t]
\centering
\includegraphics[width=0.44\textwidth,bb=23 42 521 392,clip=]{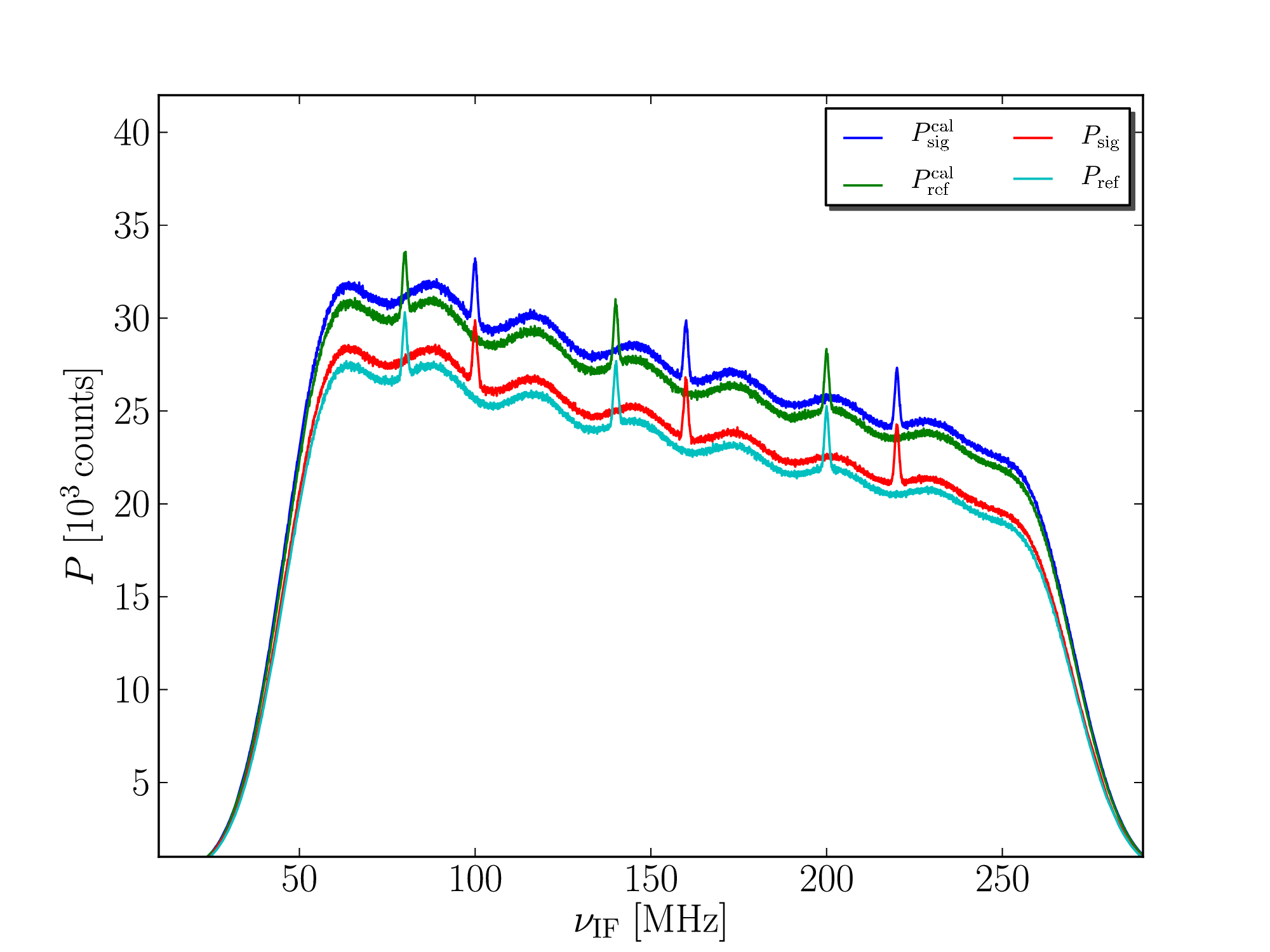}\\[0ex]
\includegraphics[width=0.44\textwidth,bb=23 1 521 392,clip=]{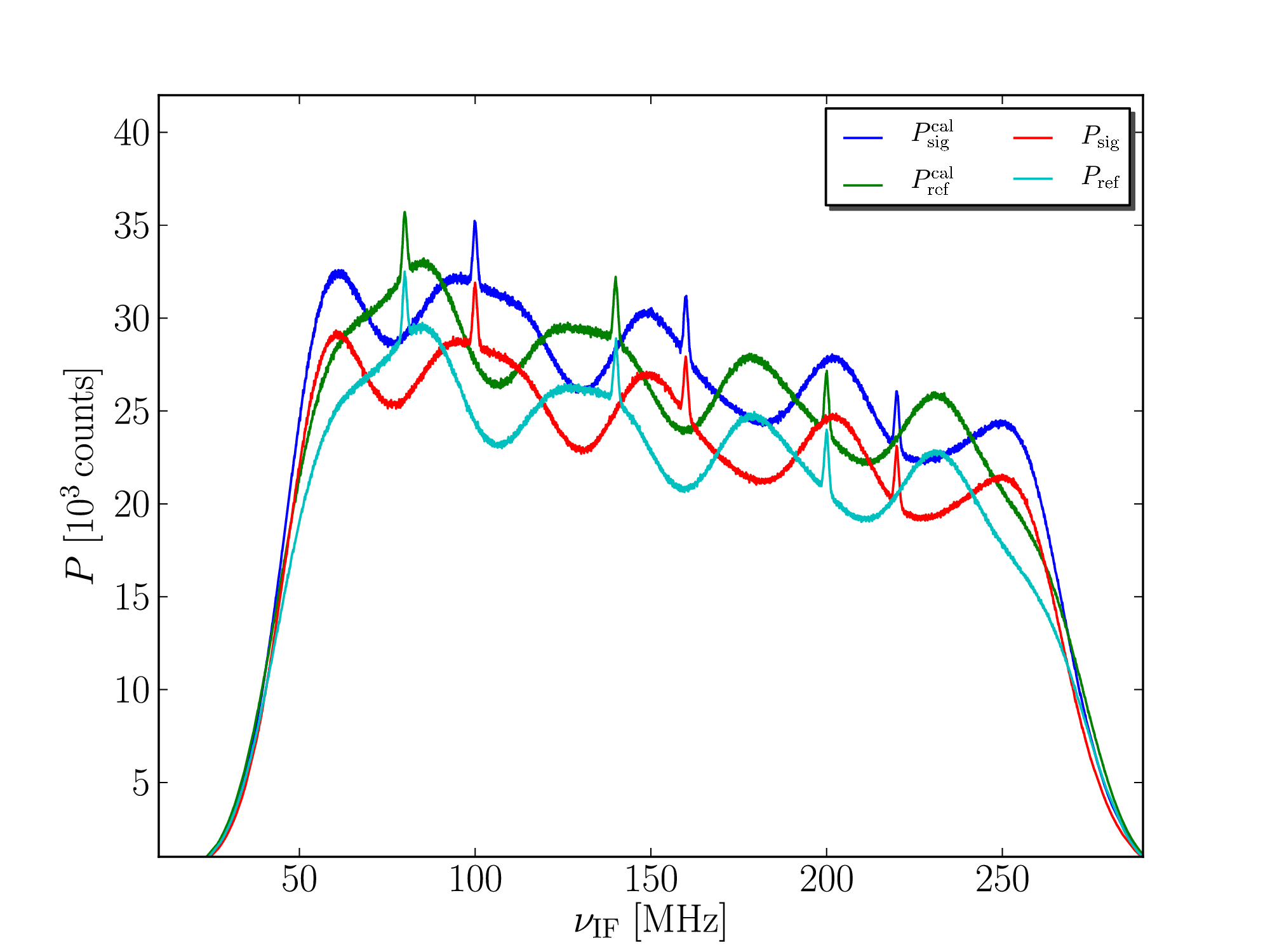}
\caption{Shifting the input spectral densities with $\pm\Delta\nu_\mathrm{LO}=\pm10\,\mathrm{MHz}$ and multiplying by the (IF) bandpass shape (see Fig.\,\ref{fig:syntheticbandpass}) returns the `measured' spectra $P_\mathrm{sig,ref}$. The \textbf{upper panel} shows the results for the simpler case, while the \textbf{lower} panel includes a standing wave contribution ($\nu_\mathrm{sw}=50\,\mathrm{MHz}$). Note that for the method described in Section\,\ref{subsec:fswitchmethod2}  we used a smaller standing wave frequency ($\nu_\mathrm{sw}=\nu_\mathrm{LO}=10\,\mathrm{MHz}$) in contrast to the other examples in order to avoid resonances (see also Section\,\ref{subsec:fswitchresonances}).}%
\label{fig:fswitchsyntheticspectraP}%
\end{figure}

\begin{figure*}[!t]
\centering
\includegraphics[width=0.45\textwidth,bb=10 1 521 392,clip=]{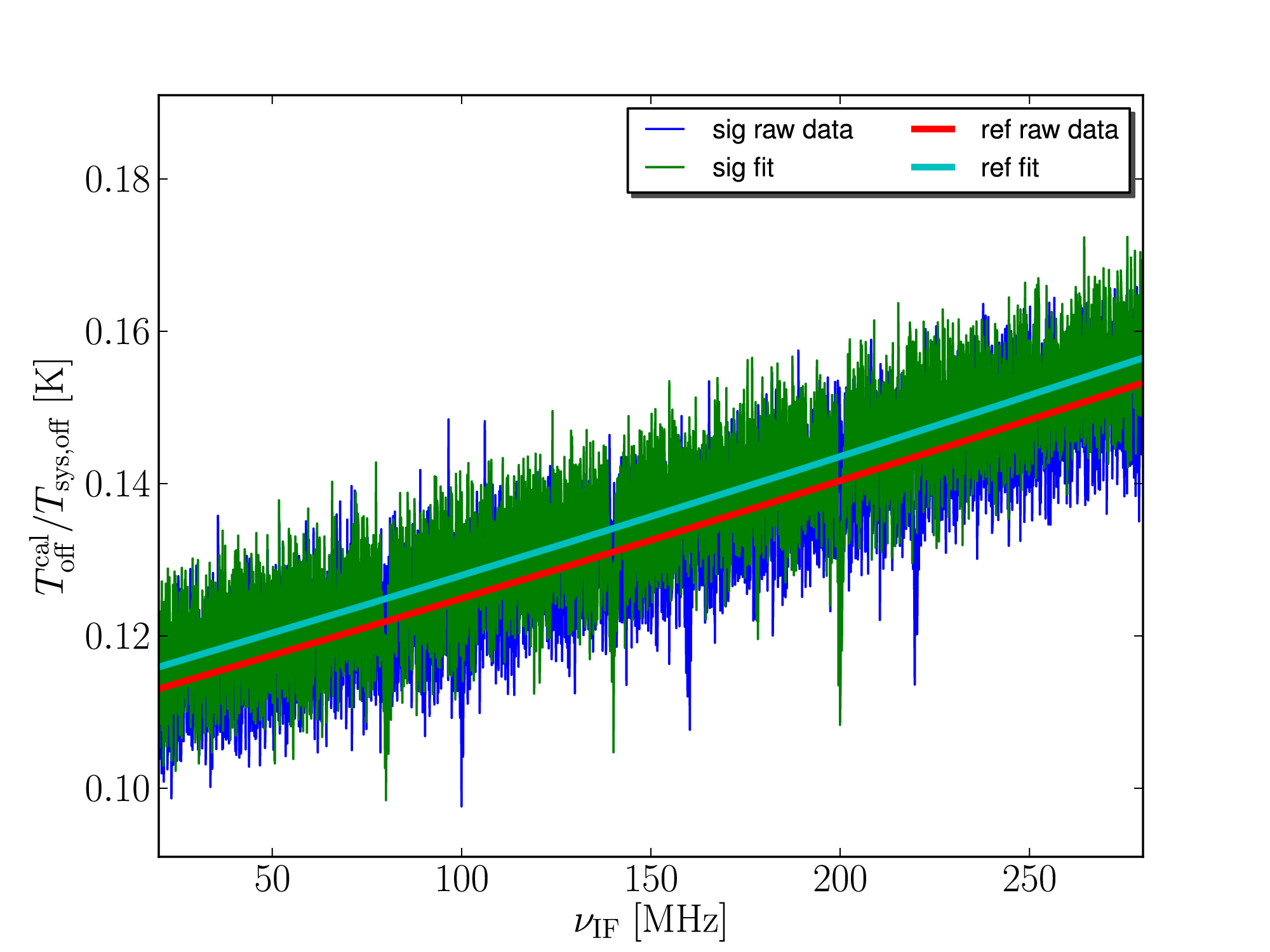}\quad
\includegraphics[width=0.45\textwidth,bb=10 1 521 392,clip=]{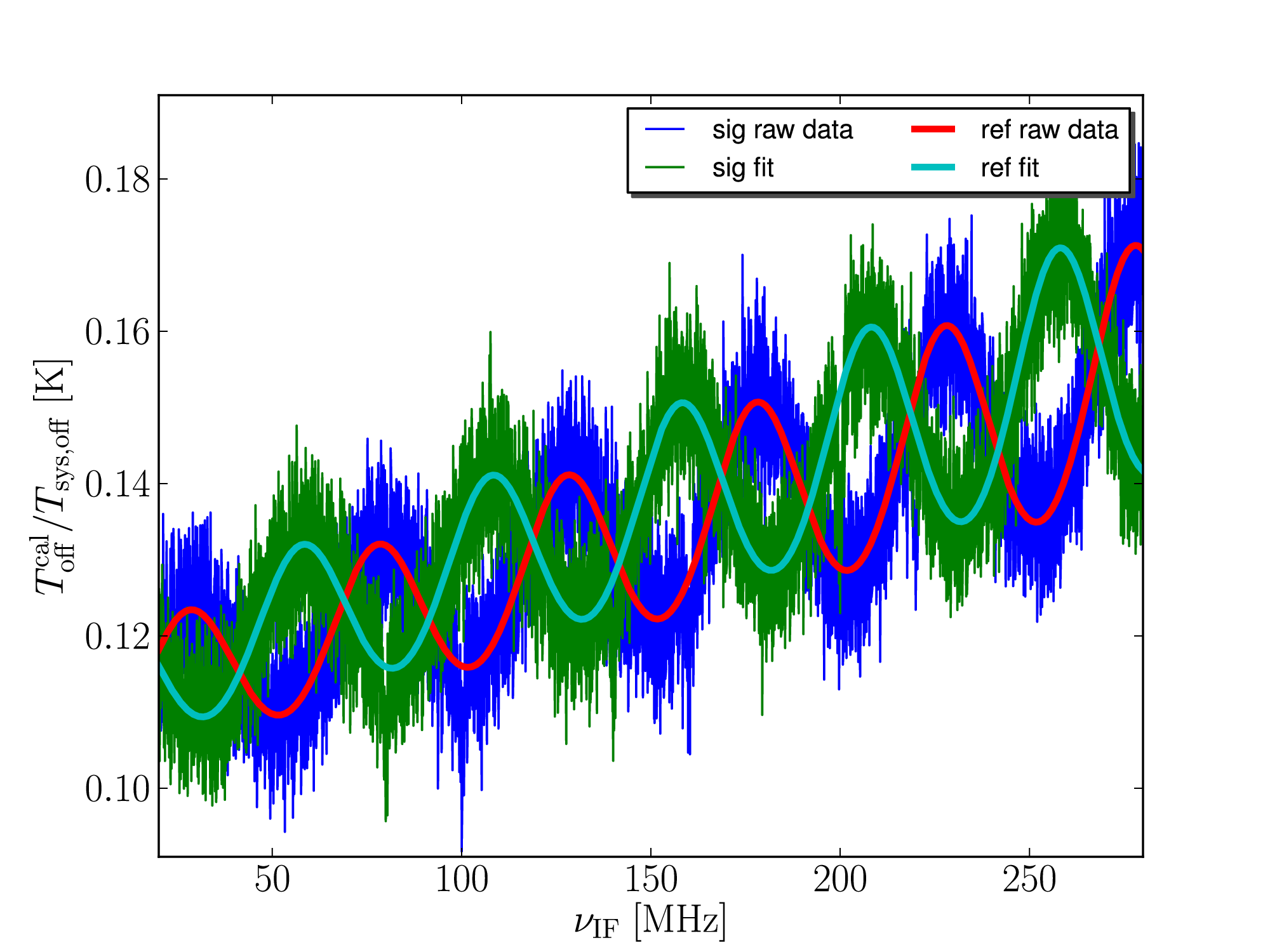}
\caption{The spectrum of $\kappa_\mathrm{sig,ref}^{-1}$ for the simpler (\textbf{left panel}) and SW (\textbf{right panel}) cases and the inferred fitting models. Spectral windows are set around parts containing the spectral lines (not clearly visible in the plots due to noise), otherwise the models would have been influenced to some extent.}%
\label{fig:fswitch_method1_modelfits}%
\end{figure*}

The input spectra for frequency switching are similar to those shown in Fig.\,\ref{fig:syntheticspectraT}, except that each measurement phase contains the $T_\mathrm{sou}$ contribution. They are also shifted according to an LO frequency of $\pm\Delta\nu_\mathrm{LO}=\pm10\,\mathrm{MHz}$ before multiplying by the bandpass $G_\mathrm{IF}$ ($G_\mathrm{RF}\equiv1$). A different standing wave frequency, $\nu_\mathrm{sw}=10\,\mathrm{MHz}$, is also used for the method that we describe in Section\,\ref{subsec:fswitchmethod2} (since $\nu_\mathrm{LO}=10\,\mathrm{MHz}$ must be a multiple of $\nu_\mathrm{sw}$ for this second method). Fig.\,\ref{fig:fswitchsyntheticspectraP} shows the resulting input spectra as they would have been measured by the backend. We note that for frequency switching the measured \textit{sig} and \textit{ref} phases belong to different RF frequencies. To clarify whether an operation is performed in either the IF or RF domain, we assume a hypothetical IF centre frequency of 150\,MHz and plot figures accordingly.

\subsection{Using the two switching phases individually to obtain $T_\mathrm{sys}$}\label{subsec:fswitchmethod1}

As in the previous sections, the signal of the calibration diode is used. Computing
\begin{align}
\left.\frac{P_\mathrm{sig}^\mathrm{cal}}{P_\mathrm{sig}}\right|_\mathrm{cont}-1&=\frac{T_\mathrm{cal,-}}{T_\mathrm{cont,-}^\mathrm{sou}+T_\mathrm{sys,-}}\equiv \kappa_\mathrm{sig}^{-1}\label{eqfswitchtsystcal1},\\
\left.\frac{P_\mathrm{ref}^\mathrm{cal}}{P_\mathrm{ref}}\right|_\mathrm{cont}-1&=\frac{T_\mathrm{cal,+}}{T_\mathrm{cont,+}^\mathrm{sou}+T_\mathrm{sys,+}}\equiv \kappa_\mathrm{ref}^{-1}\label{eqfswitchtsystcal2}
\end{align}
and inserting these expressions into Eq.\,(\ref{eqfswitchbaseeq1}) and Eq.\,(\ref{eqfswitchbaseeq2}) leads to
\begin{align}
\tilde T_\mathrm{sig}^\mathrm{[cal]} &=\left(T_\mathrm{line,+}^\mathrm{sou}+\kappa_\mathrm{ref}T_\mathrm{cal,+} + \left[T_\mathrm{cal,+}\right] \right)\frac{P_\mathrm{sig}^\mathrm{[cal]} - P_\mathrm{ref}^\mathrm{[cal]}}{ P_\mathrm{ref}^\mathrm{[cal]}}\label{eqfswitchmethod1eq1},\\
\tilde T_\mathrm{ref}^\mathrm{[cal]}&=\left(T_\mathrm{line,-}^\mathrm{sou}+\kappa_\mathrm{sig}T_\mathrm{cal,-} + \left[T_\mathrm{cal,-}\right]\right)\frac{P_\mathrm{ref}^\mathrm{[cal]} - P_\mathrm{sig}^\mathrm{[cal]}}{ P_\mathrm{sig}^\mathrm{[cal]}}.\label{eqfswitchmethod1eq2}
\end{align}
One needs to neglect the line emission contribution, $T_\mathrm{sou}$, on the right hand side of the equations, but this has no serious impact, except that spectral-line ghosts have unexpected amplitudes.

\begin{figure*}[!t]
\centering
\includegraphics[width=0.45\textwidth,bb=20 42 521 392,clip=]{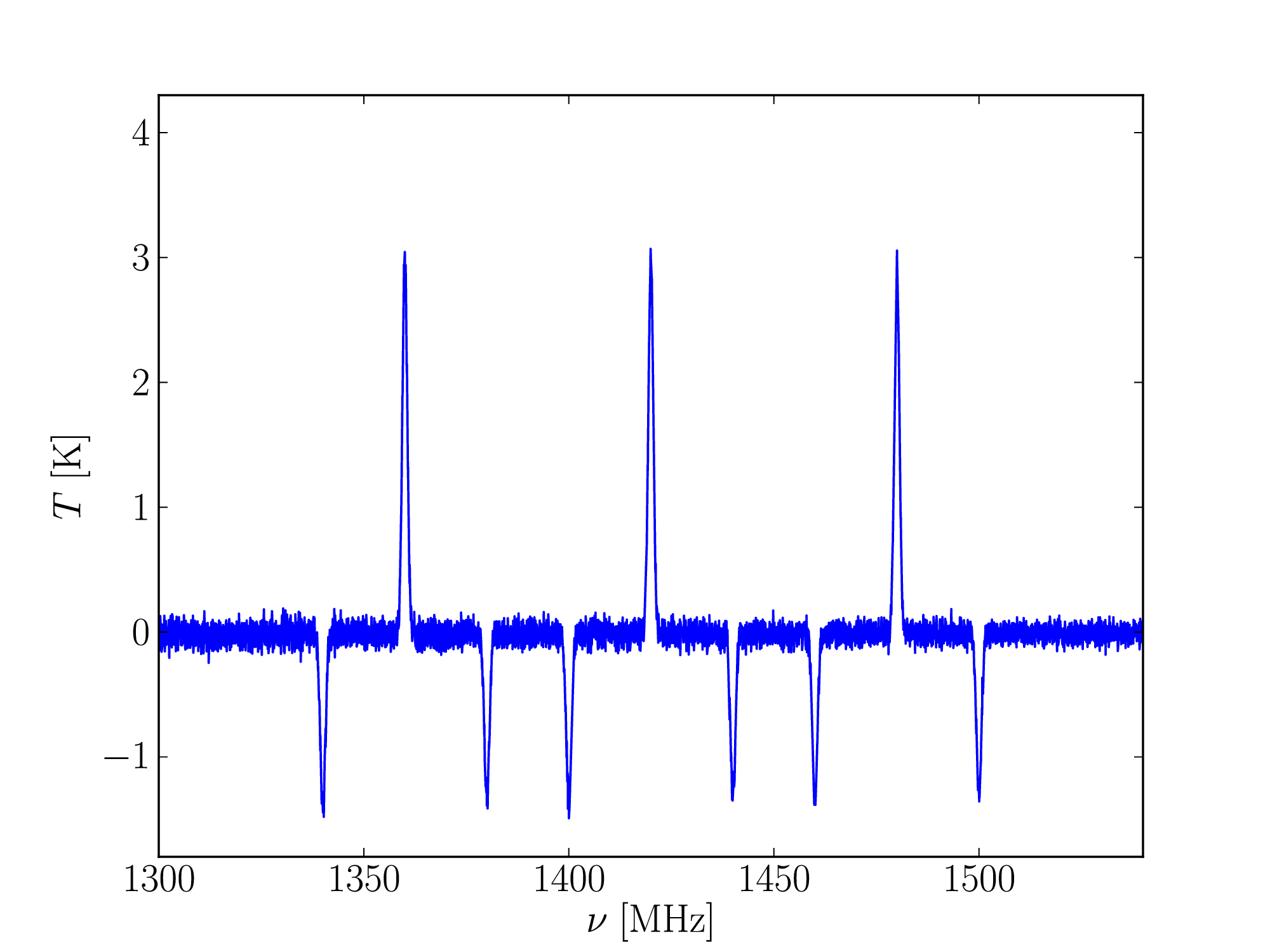}\quad
\includegraphics[width=0.45\textwidth,bb=20 42 521 392,clip=]{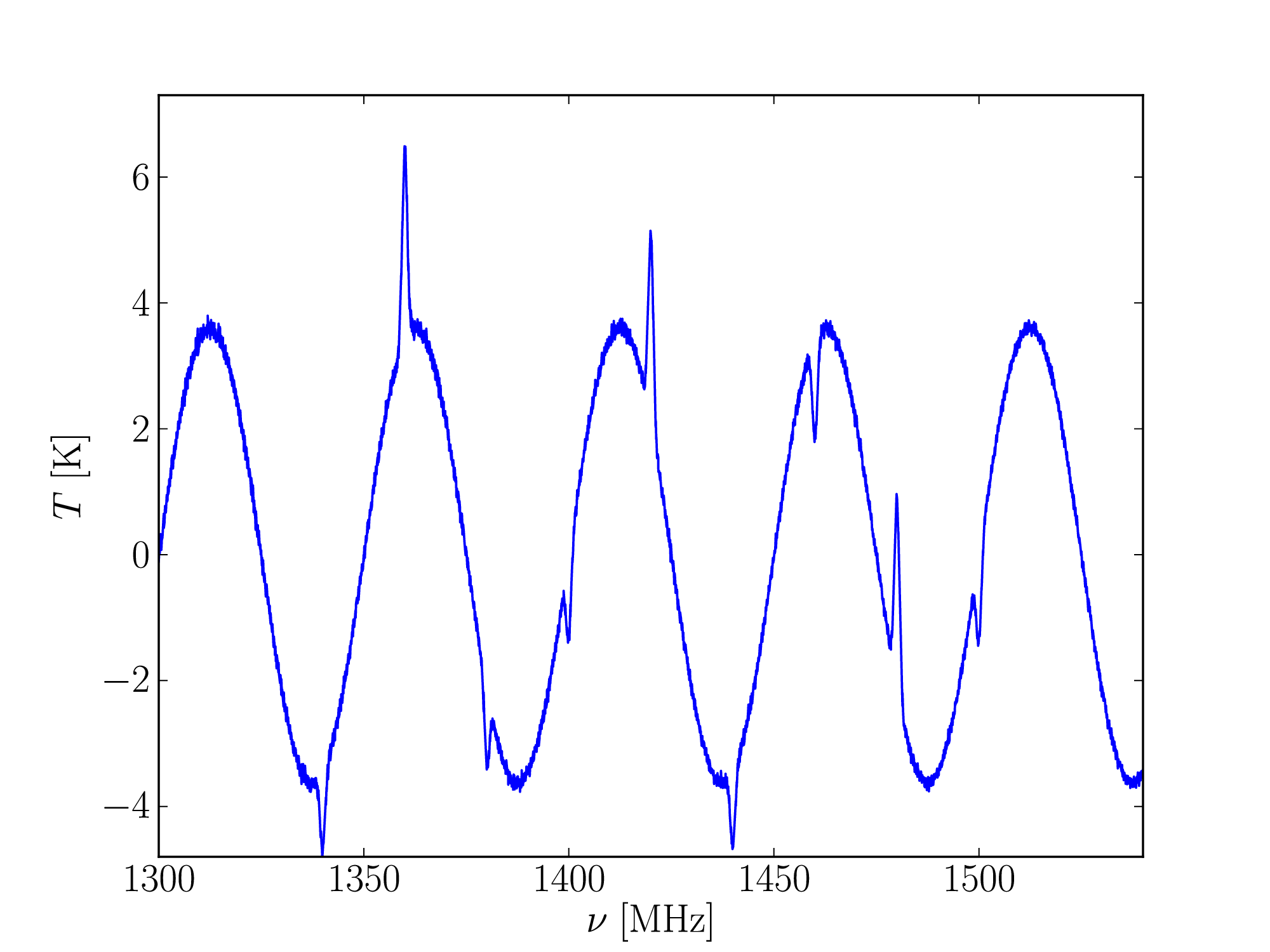}\\[0ex]
\includegraphics[width=0.45\textwidth,bb=20 1 521 392,clip=]{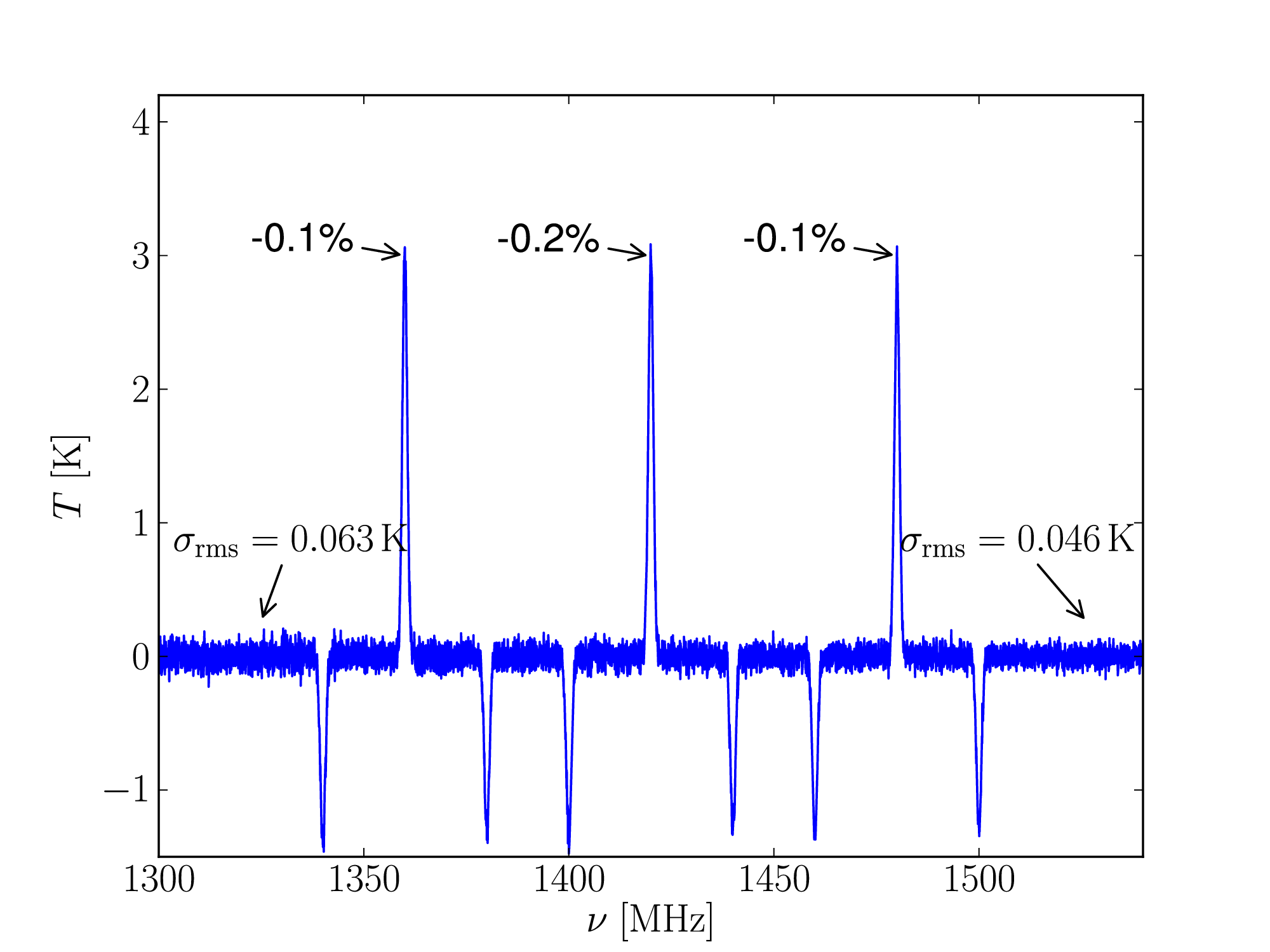}\quad
\includegraphics[width=0.45\textwidth,bb=20 1 521 392,clip=]{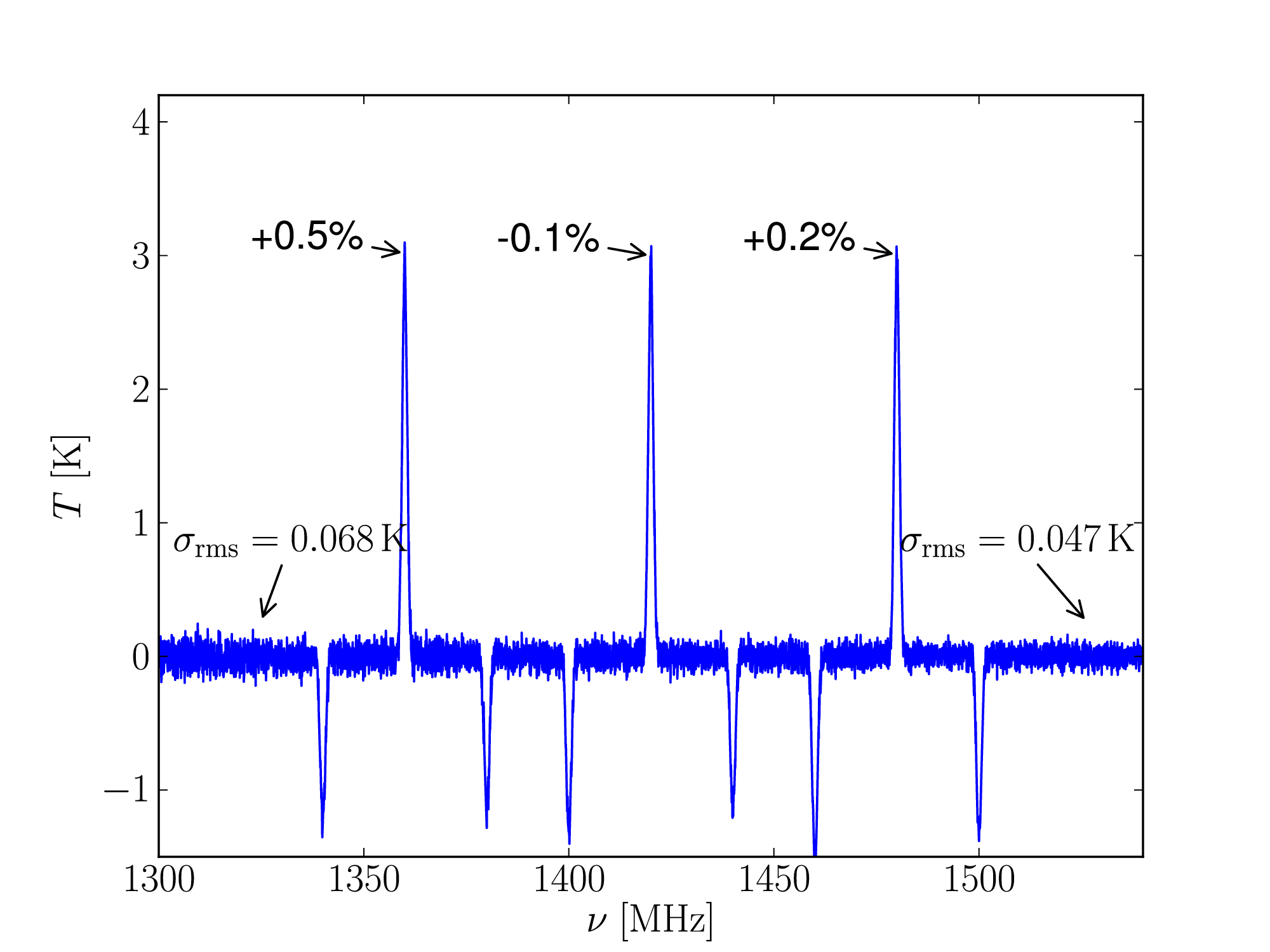}
\caption{Using Eq.\,(\ref{eqfswitchfinalequation}) and the fitting models from Fig.\,\ref{fig:fswitch_method1_modelfits}, one obtains the calibrated spectra. The \textbf{upper left panel} shows the result for the non-SW case, and the \textbf{upper right panel} for the SW case. A baseline was subtracted for both in the \textbf{lower panels}. In the SW case, the residual baseline is very complicated. The continuum of the source is \textit{not} reconstructed.}%
\label{fig:fswitch_method1_result}%
\end{figure*}

Fig.\,\ref{fig:fswitch_method1_modelfits} displays the $\kappa_\mathrm{sig,ref}^{-1}$ spectra, along with appropriate fitting models. Using these models leads to the calibrated spectra shown in Fig.\,\ref{fig:fswitch_method1_result} (top panels). Whilst the more simple case without SW shows no residual baseline owing to the specific form of Eq.\,(\ref{eqfswitchfinalequation}) in the resulting spectrum, the SW case displays a strong residual pattern. For the lower right panel in Fig.\,\ref{fig:fswitch_method1_result}, we subtracted a sine-wave signal modulated with a polynomial.

\subsection{Using both switching phases together to obtain $T_\mathrm{sys}$}\label{subsec:fswitchmethod2}
\begin{figure*}[!t]
   \centering%
   \includegraphics[width=0.45\textwidth,bb=20 1 521 392,clip=]{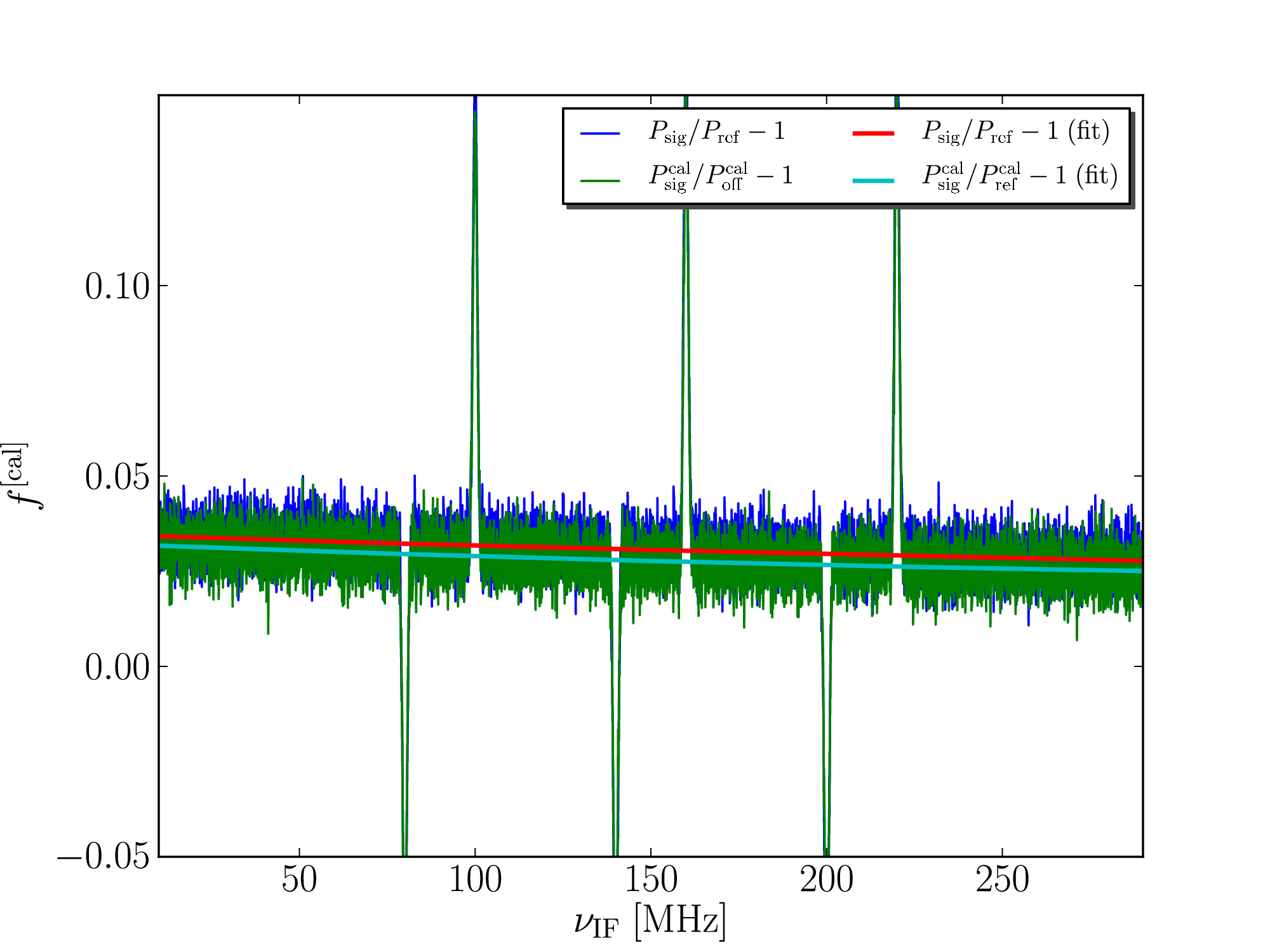}\quad
   \includegraphics[width=0.45\textwidth,bb=20 1 521 392,clip=]{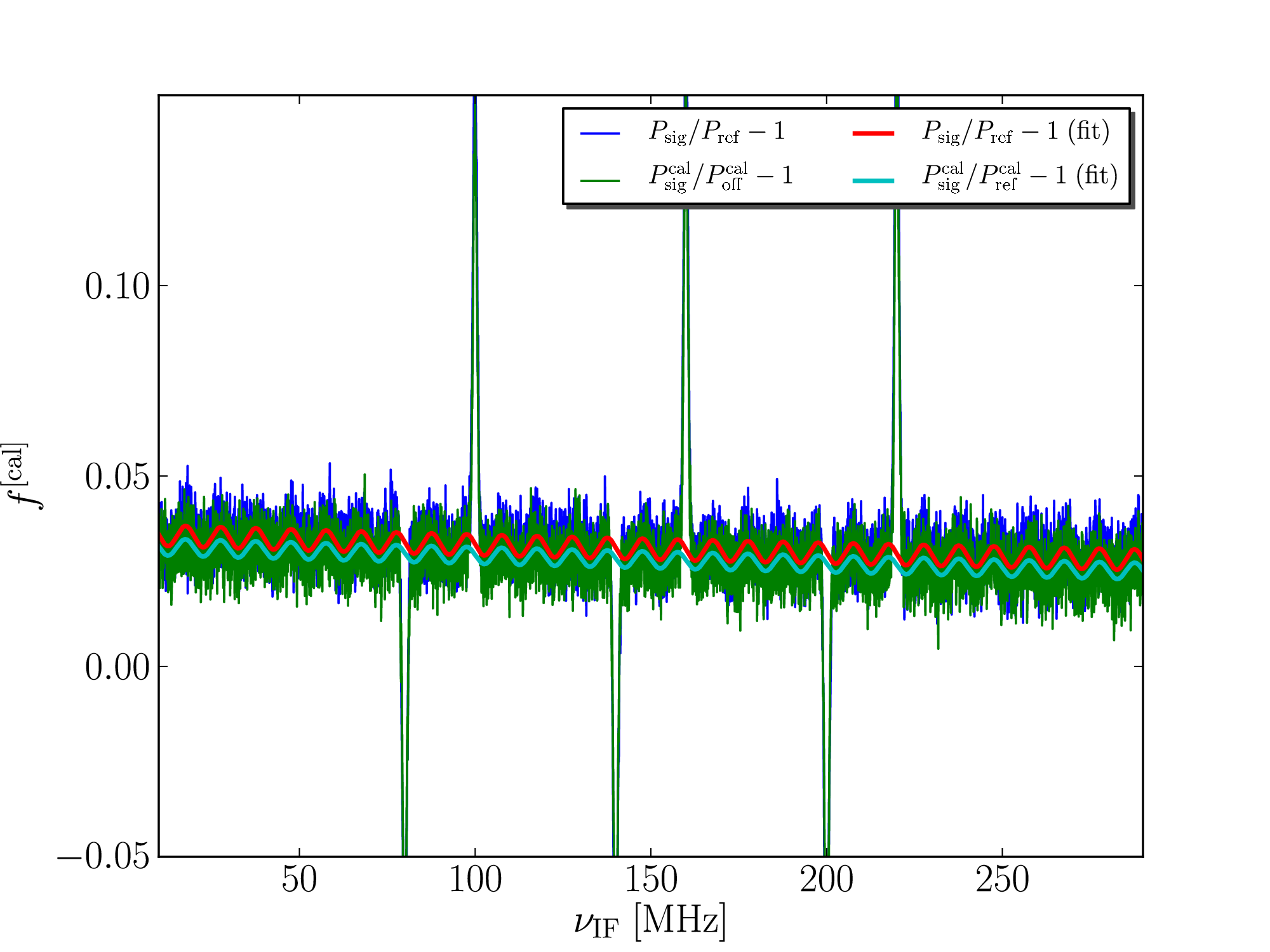}
  \caption{To correctly reconstruct the original fluxes, our proposed method utilises models of the intermediate spectra $(P_\mathrm{sig}^\mathrm{[cal]}-P_\mathrm{ref}^\mathrm{[cal]})/P_\mathrm{ref}^\mathrm{[cal]}$ (see Eq.\,(\ref{eqfswitchf}) and (\ref{eqfswitchfcal})). Appropriate windows around the spectral lines must be set as one is interested in the continuum contribution only. The \textbf{left panel} shows the results for the simpler case using second-order polynomials that are able to describe the baseline. For the standing-wave case (\textbf{right panel}), a more complicated fitting model $m(\nu)$ had to be applied (see Eq.\,(\ref{eqfswitchfittingmodel})).}%
   \label{fig:fswitch_method2_fit_through_div}%
\end{figure*}

We parametrise
\begin{align}
f(\nu)&\equiv \left.\frac{P_\mathrm{sig} - P_\mathrm{ref}}{ P_\mathrm{ref}}\right|_\mathrm{cont}=\frac{T_\mathrm{cont,-}^\mathrm{sou}-T_\mathrm{cont,+}^\mathrm{sou}+\Delta T_\mathrm{sys,\pm}}{T_\mathrm{cont,+}^\mathrm{sou}+T_\mathrm{sys,+}}\label{eqfswitchf},\\
f^\mathrm{cal}(\nu)&\equiv \left.\frac{P_\mathrm{sig}^\mathrm{cal} - P_\mathrm{ref}^\mathrm{cal}}{ P_\mathrm{ref}^\mathrm{cal}}\right|_\mathrm{cont}=
\frac{T_\mathrm{cont,-}^\mathrm{sou}-T_\mathrm{cont,+}^\mathrm{sou}+\Delta T_\mathrm{sys,\pm}+\Delta T_\mathrm{cal,\pm}}{T_\mathrm{cont,+}^\mathrm{sou}+T_\mathrm{sys,+}+T_\mathrm{cal,+}}\label{eqfswitchfcal}
\end{align}
using only the continuum contributions to compute a model (see Fig.\,\ref{fig:fswitch_method2_fit_through_div}). 
The SW frequency was set to 10\,MHz, for reasons discussed in Section\,\ref{subsec:fswitchresonances}.

The SW case requires a relatively complex model of the form
\begin{equation}
\begin{split}
m(\nu)\equiv&\left[L(n,\nu-\Delta\nu)-L(n,\nu+\Delta\nu)+\right.\\
&\left.A\sin(a(\nu-\Delta\nu)+b)-A\sin(a(\nu+\Delta\nu)+b)\right]\times\\
&\left[P(n,\nu+\Delta\nu)+A\sin(a(\nu+\Delta\nu)+b)\right]^{-1},\label{eqfswitchfittingmodel}
\end{split}
\end{equation}
where $L(n,\nu)$ is a polynomial function of degree $n$.  Fig.\,\ref{fig:fswitch_method2_fit_through_div} shows the results of these fits. In the presence of SW (right panel), one must be sure to have a good initial estimate of $\nu_\mathrm{sw}$ in order to obtain a meaningful result. This may impose difficulties in practice, though it is likely that using a particularly long integration time (or using several data sets) will aid the measurement of $\nu_\mathrm{sw}$.

From Eq.\,(\ref{eqfswitchf}) and Eq.\,(\ref{eqfswitchfcal}), we obtain
\begin{align}
T_\mathrm{cont,+}^\mathrm{sou}+T_\mathrm{sys,+}&=T_\mathrm{cal,+}\frac{f^\mathrm{cal}}{f-f^\mathrm{cal}} -\frac{ \Delta T_\mathrm{cal,\pm}}{ f-f^\mathrm{cal}} \equiv \alpha(\nu)  \label{eqalpha},\\
T_\mathrm{cont,+}^\mathrm{sou}+T_\mathrm{sys,+}^\mathrm{cal}&=T_\mathrm{cal,+}\frac{f}{f-f^\mathrm{cal}} -\frac{ \Delta T_\mathrm{cal,\pm}}{ f-f^\mathrm{cal}} \equiv \beta(\nu). \label{eqbeta}
\end{align}
The term $\Delta T_\mathrm{cal,\pm}/(f-f^\mathrm{cal})$ contributes significantly to $\alpha$ and $\beta$ and must not be neglected. This correction term is the only fundamental difference between the frequency and position switching equations, which otherwise are very similar. This is because the \textit{ref} phase is not independent of the \textit{sig} phase, but contains only frequency-shifted quantities, introducing a degeneracy. As for position switching, a problematic situation occurs if $f\approx f^\mathrm{cal}$ since the denominator in Eq.\,(\ref{eqalpha}) and (\ref{eqbeta}) becomes singular. While in the case of SW this is a frequent phenomenon (see Section\,\ref{subsec:fswitchresonances}), it may also occur when $\Delta T_\mathrm{cont,\pm}^\mathrm{sou}\approx0$, $\Delta T_\mathrm{sys,\pm}\approx0$, and $\Delta T_\mathrm{cal,\pm}\approx0$. At first glance, this might appear to be rather unlikely, but it can indeed arise if the continuum and $T_\mathrm{cal}$ slopes are not very steep in relation to the LO shift.

Using the substitution $\nu\rightarrow\nu-2\Delta\nu$ in Eq.\,(\ref{eqalpha}) and (\ref{eqbeta}), one can also infer 
 \begin{align}
T_\mathrm{cont,-}^\mathrm{sou}+T_\mathrm{sys,-}&=\alpha(\nu-2\Delta \nu)\label{eqalpha2}, \\
T_\mathrm{cont,-}^\mathrm{sou}+T_\mathrm{sys,-}^\mathrm{cal}&=\beta(\nu-2\Delta \nu)\label{eqbeta2}.
\end{align}
Subtracting Eq.\,(\ref{eqalpha}) from (\ref{eqalpha2}) and  Eq.\,(\ref{eqbeta}) from (\ref{eqbeta2}) results in
 \begin{align}
\alpha(\nu-2\Delta \nu)-\alpha(\nu)&=T_\mathrm{cont,-}^\mathrm{sou}  -T_\mathrm{cont,+}^\mathrm{sou} + \Delta T_\mathrm{sys,\pm}\label{eqalphadiff},\\
\beta(\nu-2\Delta \nu)-\beta(\nu)&=T_\mathrm{cont,-}^\mathrm{sou}  -T_\mathrm{cont,+}^\mathrm{sou} + \Delta T_\mathrm{sys,\pm}+ \Delta T_\mathrm{cal,\pm}.\label{eqbetadiff}
\end{align}
Inserting Eq.\,(\ref{eqalphadiff}) and (\ref{eqalpha}) into Eq.\,(\ref{eqsigref}) leads to
\begin{equation}
T_\mathrm{line,-}^\mathrm{sou}-T_\mathrm{line,+}^\mathrm{sou}+\alpha(\nu-2\Delta \nu)-\alpha(\nu)=\frac{P_\mathrm{sig} - P_\mathrm{ref}}{ P_\mathrm{ref}}\left[T_\mathrm{line,+}^\mathrm{sou}+\alpha(\nu)\right]
\end{equation}
and
\begin{align}
T_\mathrm{line,-}^\mathrm{sou}-T_\mathrm{line,+}^\mathrm{sou}&=\frac{P_\mathrm{sig}^\mathrm{\phantom{cal}} - P_\mathrm{ref}^\mathrm{\phantom{cal}}}{ P_\mathrm{ref}}\left[T_\mathrm{line,+}^\mathrm{sou}+\alpha(\nu)\right]+\alpha(\nu)-\alpha(\nu-2\Delta \nu)\label{eqfswitchfinal}\\
&=\frac{P_\mathrm{sig}^\mathrm{cal} - P_\mathrm{ref}^\mathrm{cal}}{ P_\mathrm{ref}^\mathrm{cal}}\left[T_\mathrm{line,+}^\mathrm{sou}+\beta(\nu)\right]+\beta(\nu)-\beta(\nu-2\Delta \nu).
\end{align}
Similarly,
\begin{align}
T_\mathrm{line,+}^\mathrm{sou}-T_\mathrm{line,-}^\mathrm{sou}=&\frac{P_\mathrm{ref}^\mathrm{\phantom{cal}} - P_\mathrm{sig}^\mathrm{\phantom{cal}}}{ P_\mathrm{sig}}\left[T_\mathrm{line,-}^\mathrm{sou}+\alpha(\nu-2\Delta \nu)\right]\nonumber\\
&+\alpha(\nu-2\Delta \nu)-\alpha(\nu) \\
=&\frac{P_\mathrm{ref}^\mathrm{cal} - P_\mathrm{sig}^\mathrm{cal}}{ P_\mathrm{sig}^\mathrm{cal}}\left[T_\mathrm{line,-}^\mathrm{sou}+\beta(\nu-2\Delta \nu)\right]\nonumber\\
&+\beta(\nu-2\Delta \nu)-\beta(\nu).\label{eqfswitchfinal2}
\end{align}

Finally, to average all four phases to form the final spectrum,
each resulting spectrum must be shifted to achieve that the emission lines lie at the same position in the spectrum. 

We note that, in contrast to position switching, one cannot reconstruct $T_\mathrm{cont}^\mathrm{sou}$ and $T_\mathrm{sys}$ as separate quantities, only $T_\mathrm{cont}^\mathrm{sou}+T_\mathrm{sys}$ using Eq.\,(\ref{eqalpha}), even if $T_\mathrm{cal}(\nu)$ is known. This is due to the degeneracy described above. 

One advantage of this method, despite being more complex, is that on the left hand side of Eq.\,(\ref{eqfswitchfinal}) to Eq.\,(\ref{eqfswitchfinal2}), there is no continuum contribution (in contrast to the method presented in Section\,\ref{subsec:fswitchmethod1}). This means that SW do not propagate into the final spectrum producing complicated baselines as in Fig.\,\ref{fig:fswitch_method1_result}. However, SW can still have a negative influence on the model fitting such that the calibration could be affected. This means that if one has to reduce frequency switched spectra having a SW contribution, this method may be the most suited in terms of residual baselines, however the calibration accuracy is lowered.

\begin{figure*}[!t]
\centering
   \includegraphics[width=0.45\textwidth,bb= 20 42 521 392,clip=]{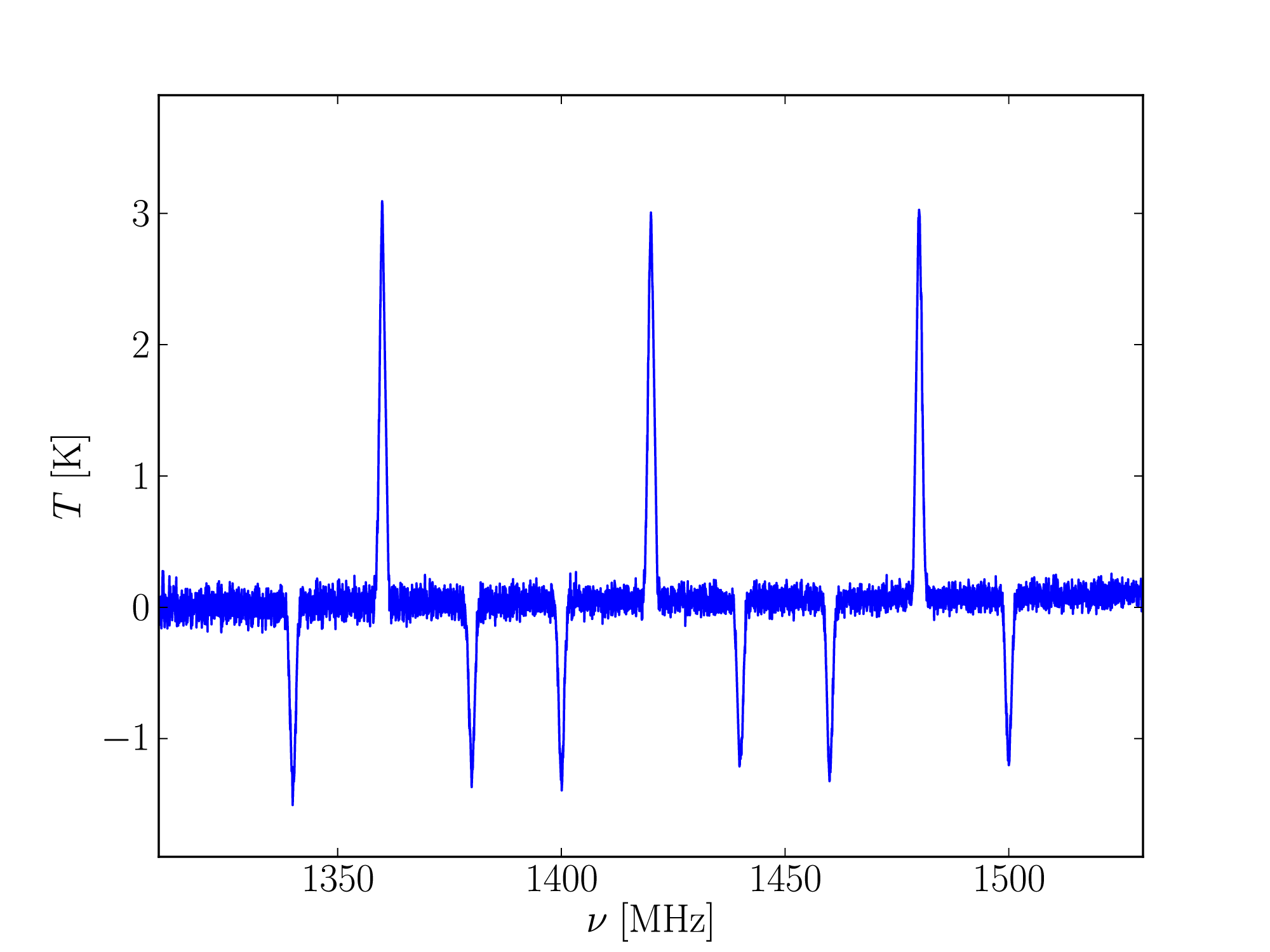}\quad
   \includegraphics[width=0.45\textwidth,bb= 20 42 521 392,clip=]{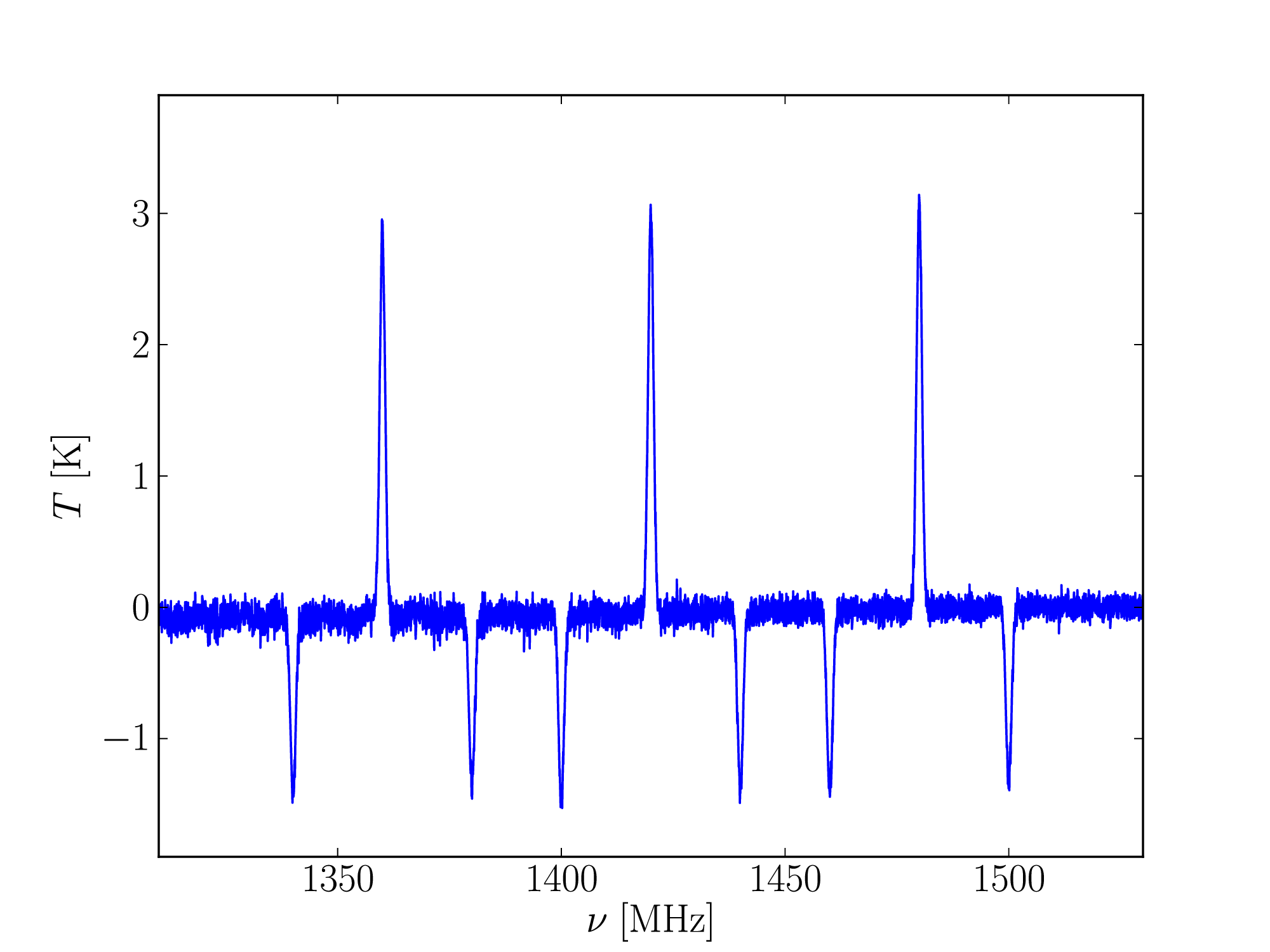}\\[0ex]
   \includegraphics[width=0.45\textwidth,bb= 20 1 521 392,clip=]{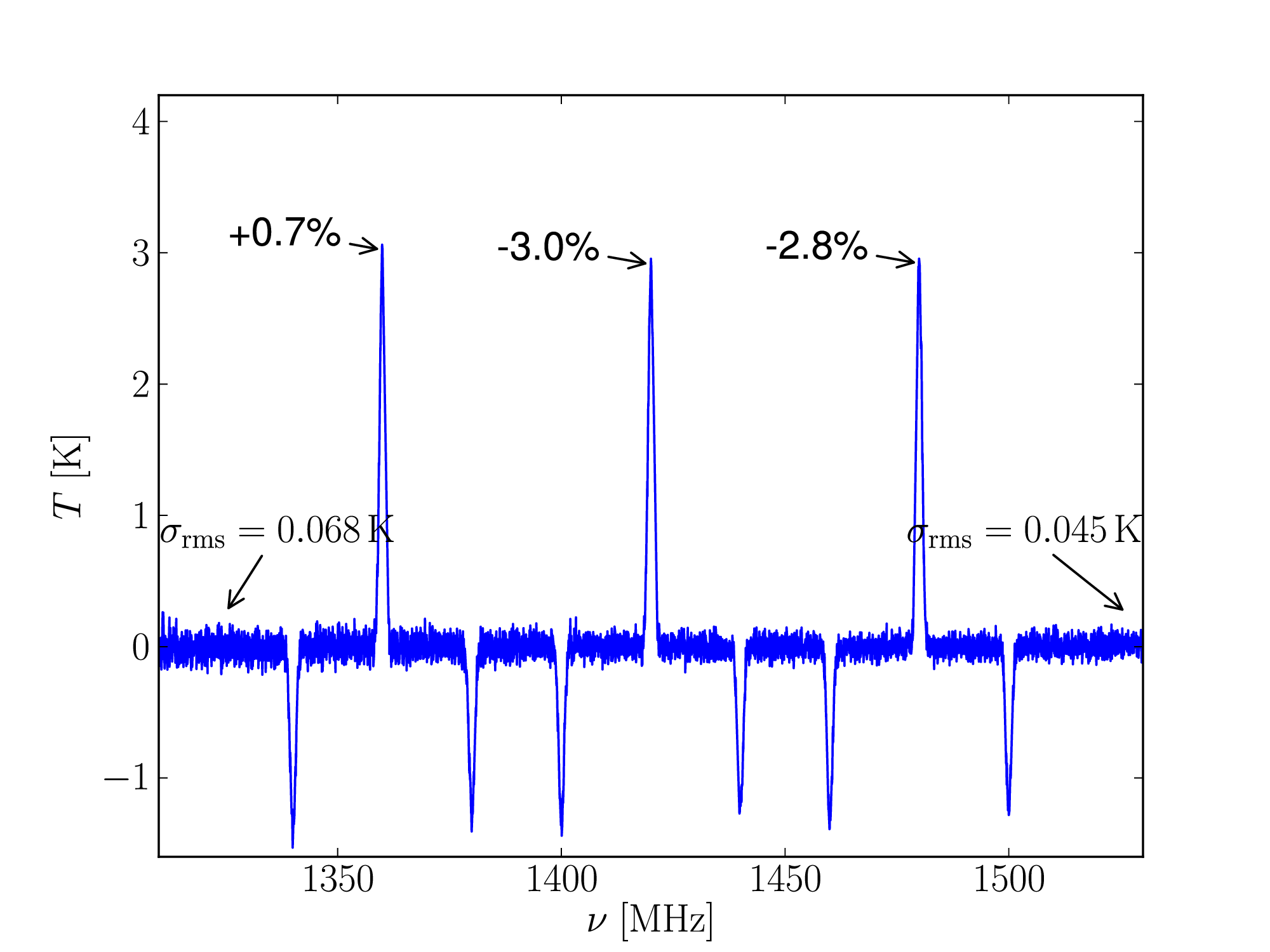}\quad
   \includegraphics[width=0.45\textwidth,bb= 20 1 521 392,clip=]{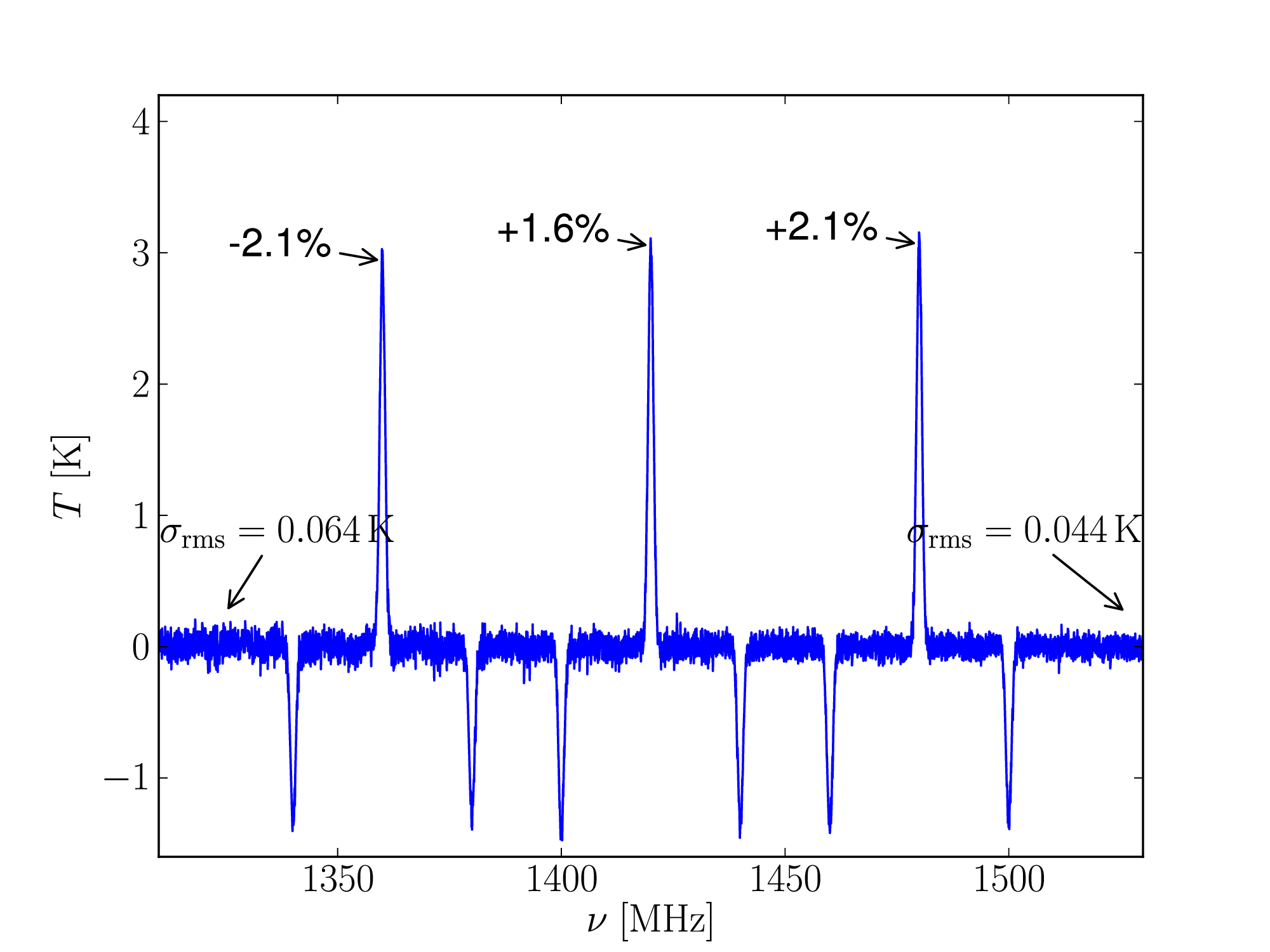}
  \caption{Inserting the models $f^\mathrm{[cal]}(\nu)$ into Eq.\,(\ref{eqfswitchfinal}) results in a unbiased flux calibration, as shown in Section\,\ref{subsec:simulationfswitch}. The figure shows the outcome for the non-SW (\textbf{left panels}) and SW (\textbf{right panels}) scenarios with (lower row) and without (upper row) baseline subtraction. It appears that the method is quite significantly affected by small errors in the fitting models of $f$ and $f^\mathrm{cal}$ (compare with Section\,\ref{subsec:simulationfswitch}). Furthermore, to obtain acceptable fits, we had to choose very good prior model parameters, e.g. from a low-noise run. Nonetheless, compared to the method in Section\,\ref{subsec:fswitchmethod1} the SW do not directly convert into a sine-wave imprint in the final spectrum (compare upper right panel with Fig.\,\ref{fig:fswitch_method1_result}), although relatively often higher order effects are caused when the model fits are inaccurate.}%
   \label{fig:fswitch_method2_result}%
\end{figure*}

Results for the non-SW and SW cases are displayed in Fig.\,\ref{fig:fswitch_method2_result} (upper left and upper right panels respectively). The results are less accurate than for the method presented in Section\,\ref{subsec:fswitchmethod1}. Nonetheless, further investigation, e.g. studying the distribution of the flux calibration error using several hundreds of simulated spectra (see Section\,\ref{subsec:simulationfswitch}) shows that the calibration is still unbiased.  The error distribution is however wider than for the other methods, owing to the use of two independent fitting models and the relative errors between both fits then having a rather large impact on the final results. The method is even less robust for the SW case.

\subsubsection{Problems caused by standing wave contributions}\label{subsec:fswitchresonances}
\begin{figure*}[!tp]
   \centering%
   \includegraphics[width=0.45\textwidth,bb=20 1 521 392,clip=]{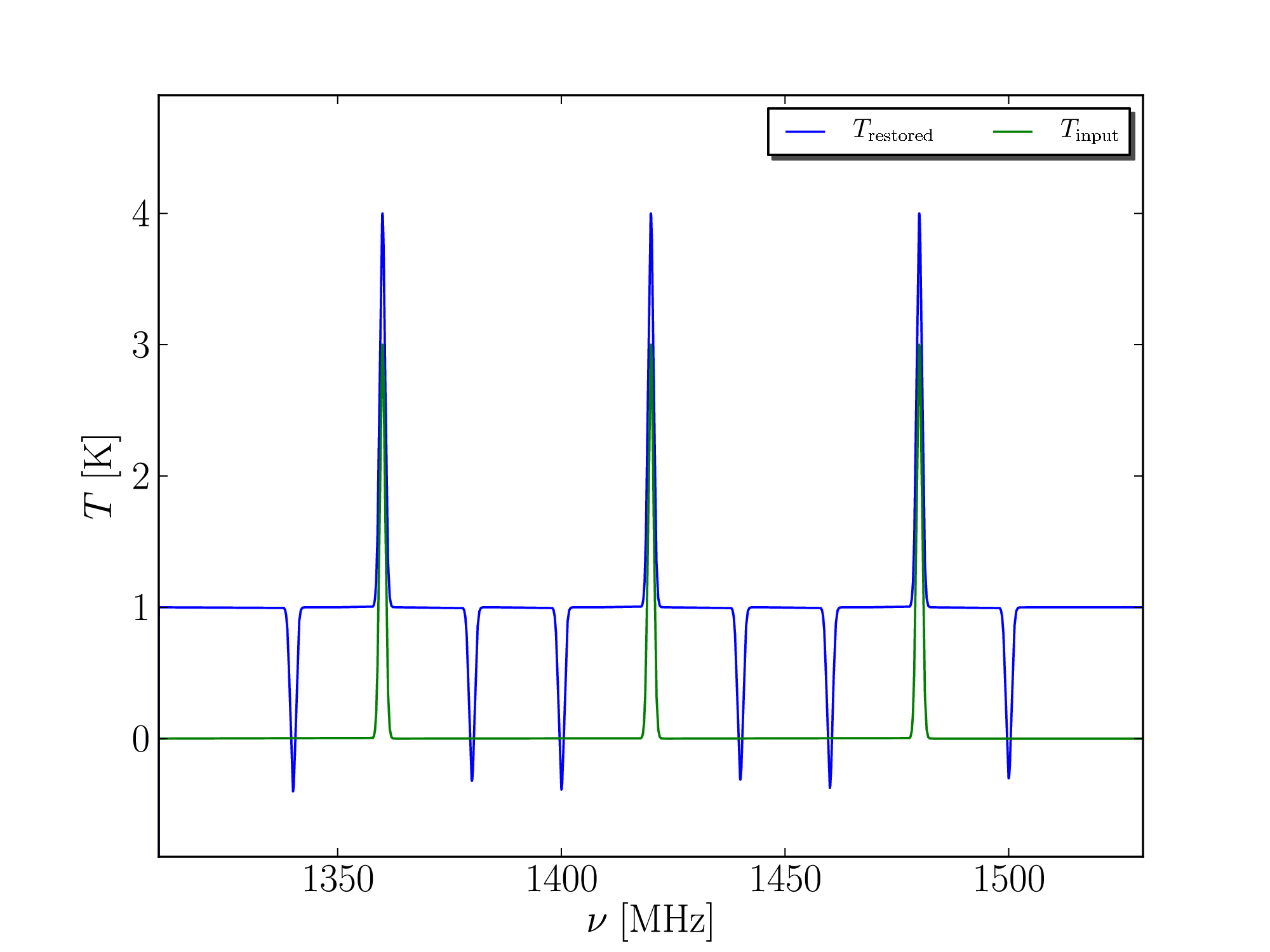}\quad
   \includegraphics[width=0.45\textwidth,bb=20 1 521 392,clip=]{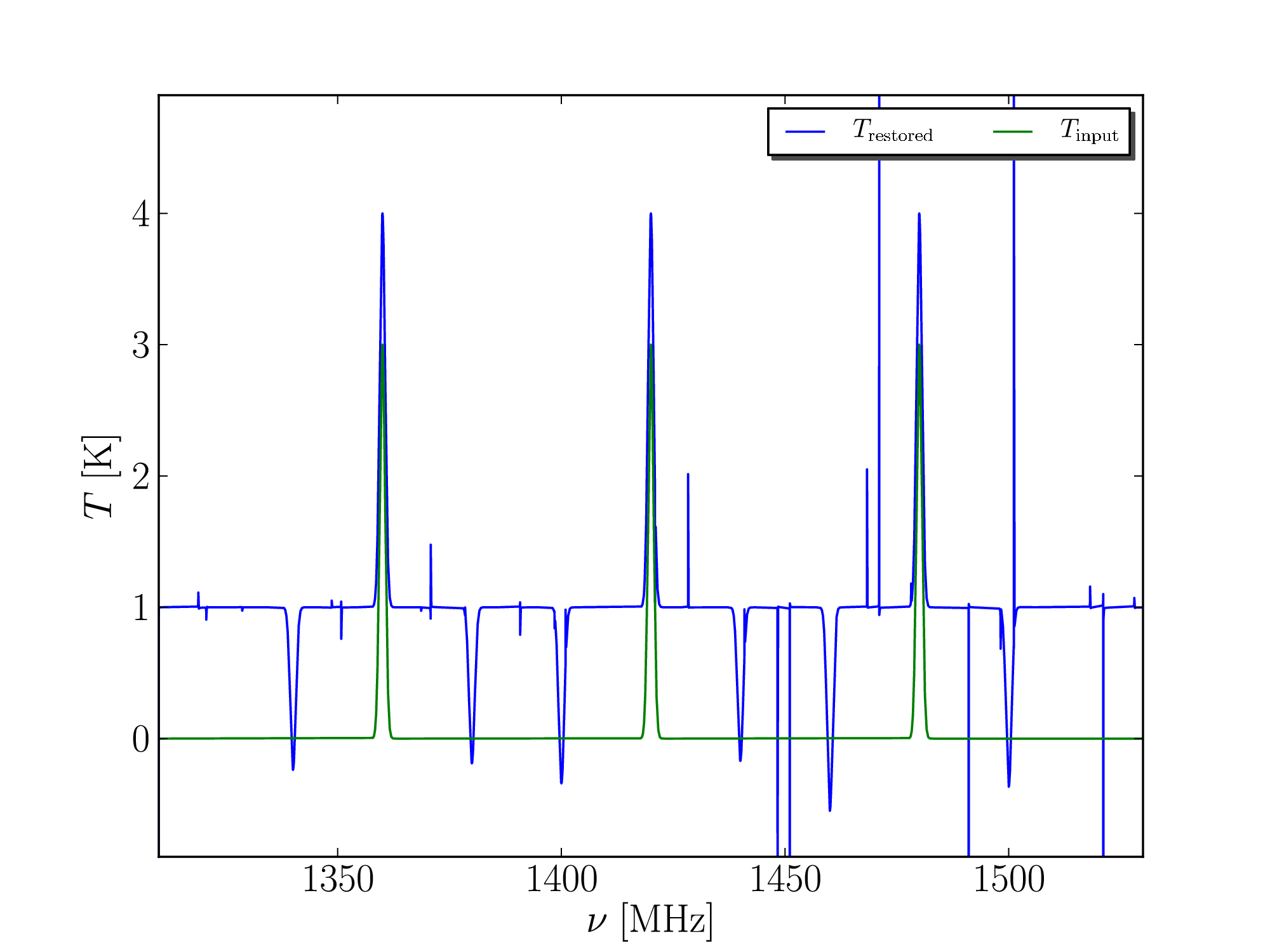}\\[0ex]
   \includegraphics[width=0.45\textwidth,bb=20 42 521 392,clip=]{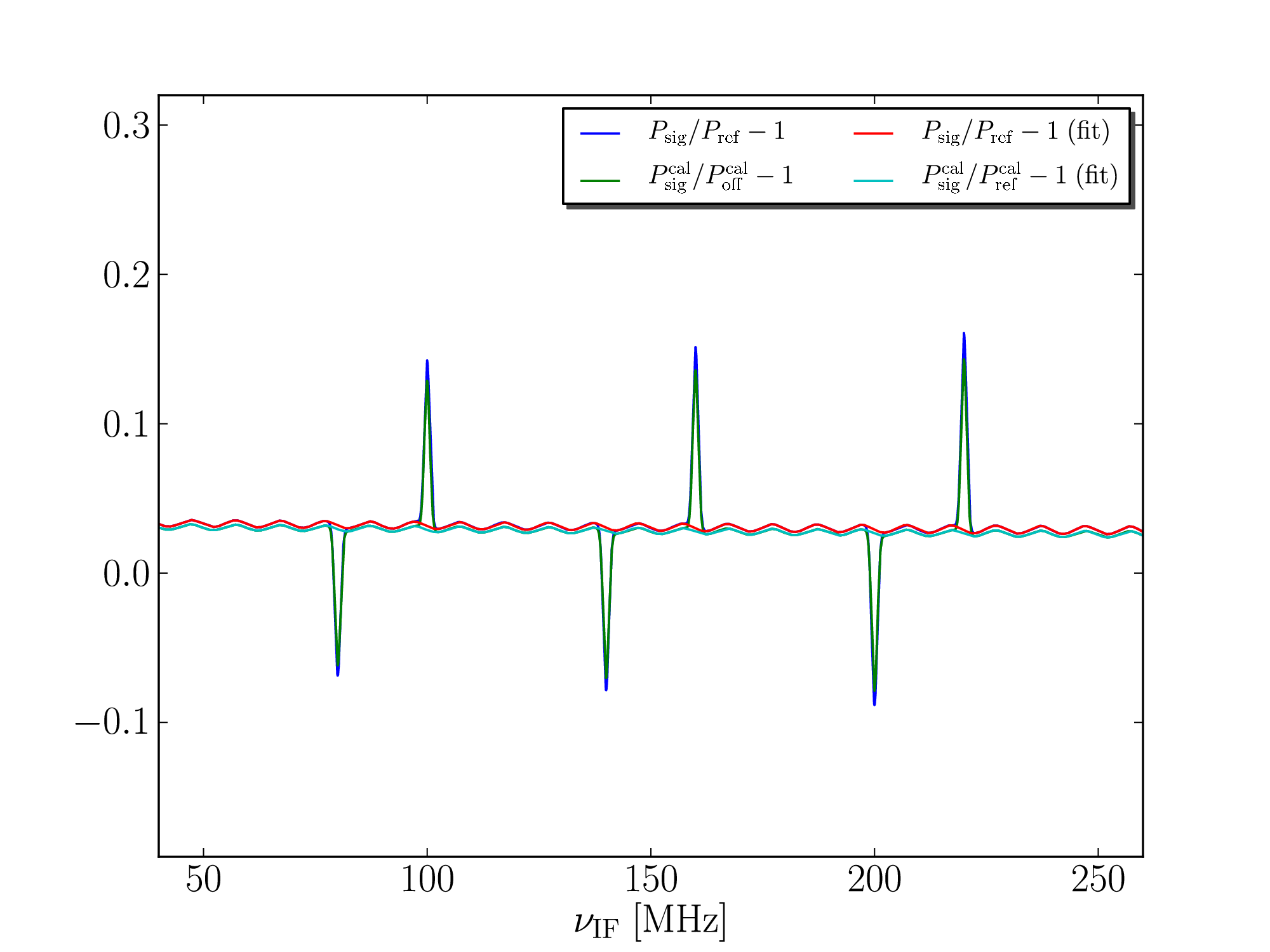}\quad
   \includegraphics[width=0.45\textwidth,bb=20 42 521 392,clip=]{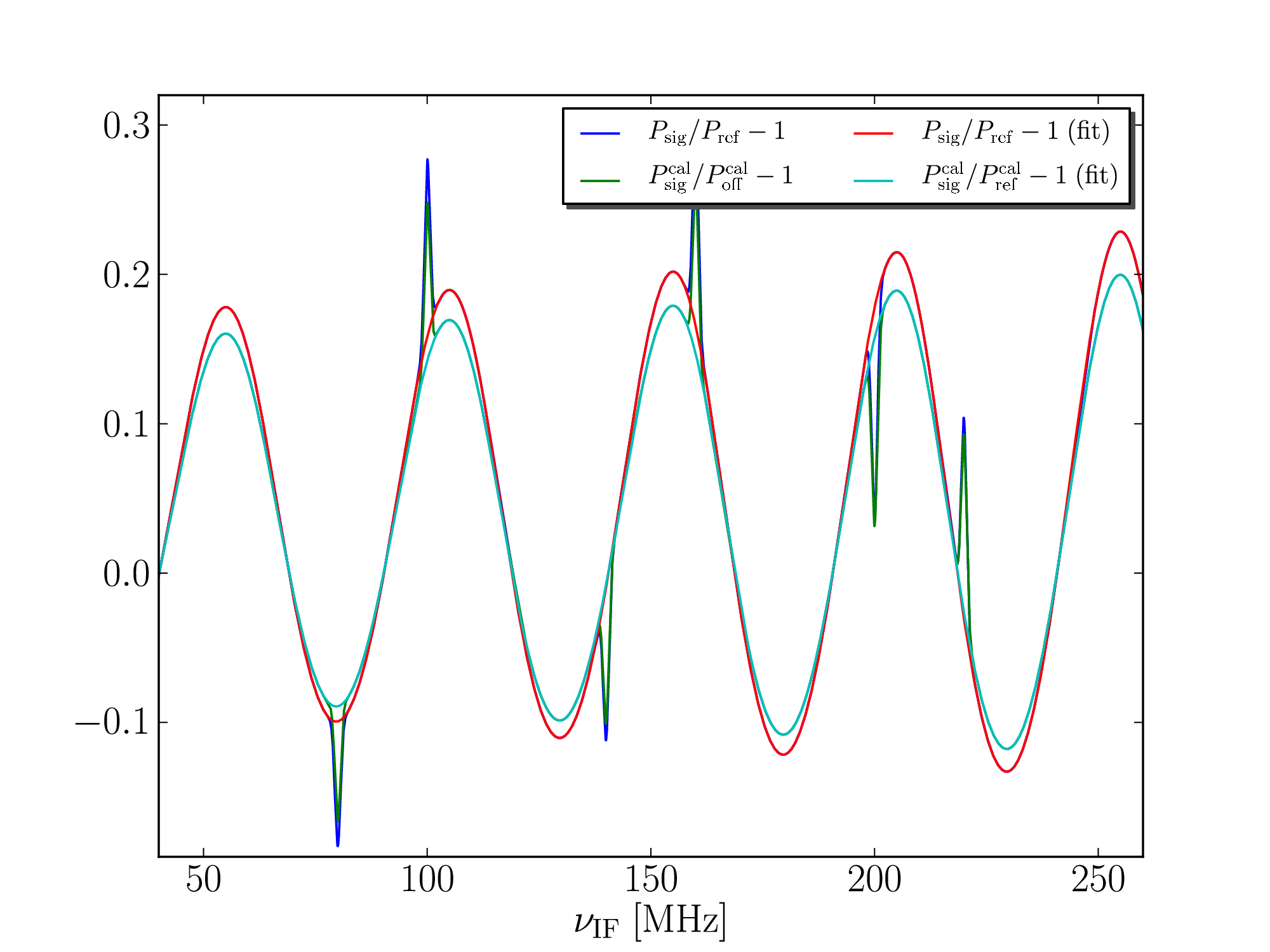}\\[0ex]
   \includegraphics[width=0.45\textwidth,bb=20 1 521 392,clip=]{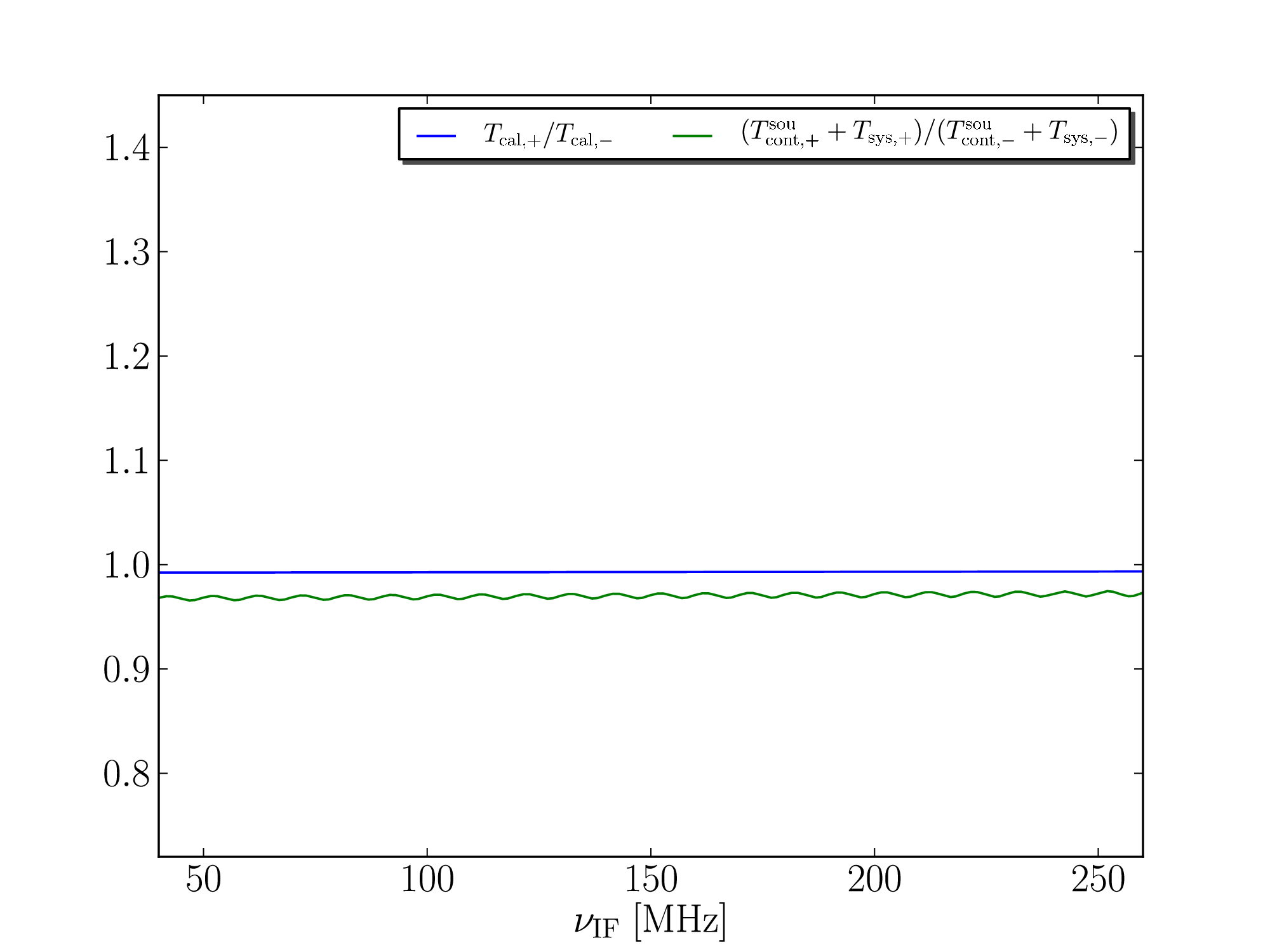}\quad
   \includegraphics[width=0.45\textwidth,bb=20 1 521 392,clip=]{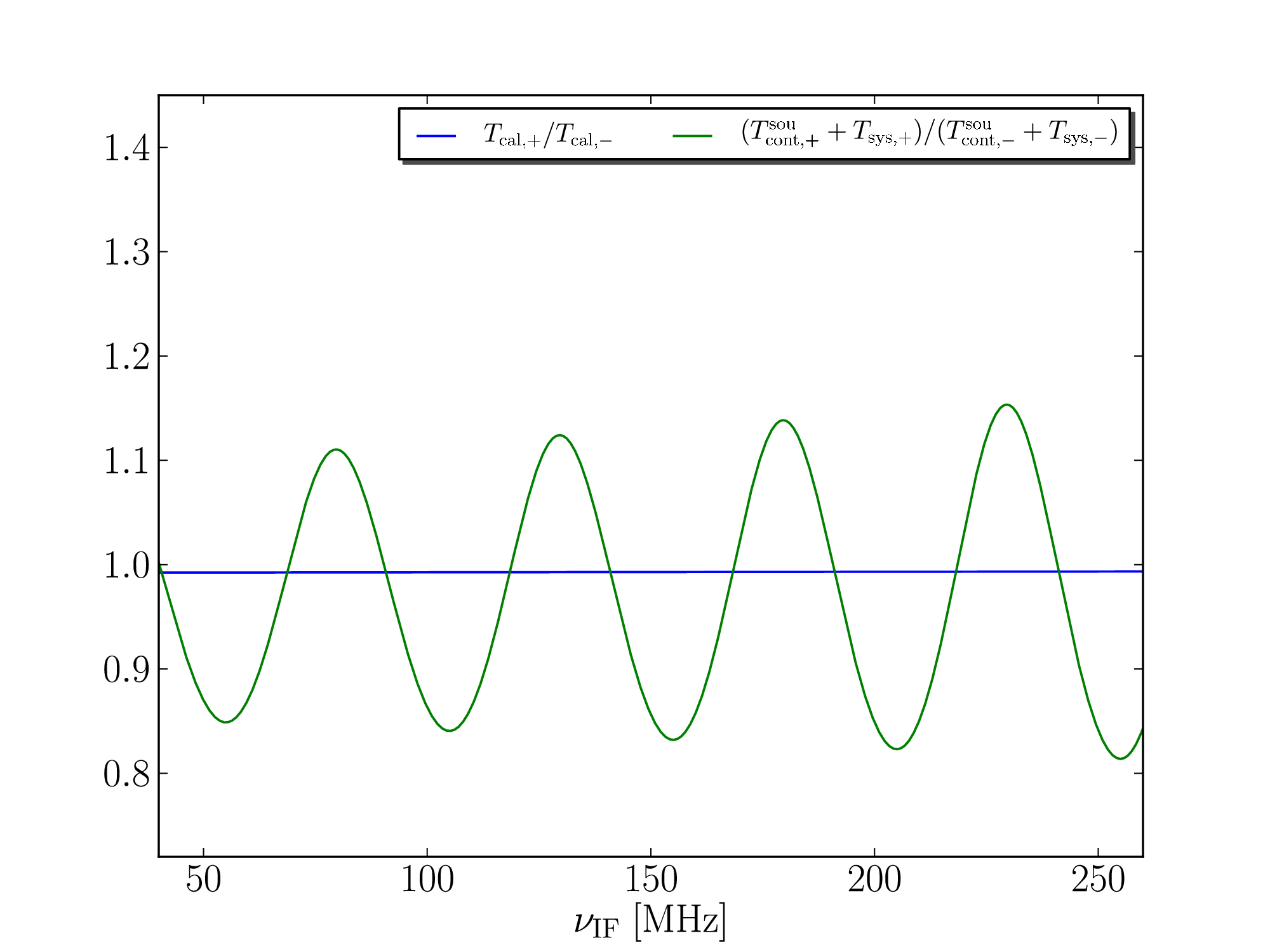}
  \caption{If the LO shift $\nu_\mathrm{LO}$ is not a multiple of the standing wave frequency $\nu_\mathrm{sw}$, the quantity  $f-f^\mathrm{cal}$ can have several roots causing `resonances' in the final spectrum. The \textbf{left panels} have $\nu_\mathrm{sw}=10\,\mathrm{MHz}$ matching $\nu_\mathrm{LO}=10\,\mathrm{MHz}$, while for the \textbf{right panels} we chose $\nu_\mathrm{sw}=50\,\mathrm{MHz}$. The \textbf{upper panels} show the resulting spectra (simulations were performed without a noise contribution for a clearer visualisation).  In the upper right panel, spikes do appear. The \textbf{middle panels} show the model fitting through $(P_\mathrm{sig}^\mathrm{[cal]}-P_\mathrm{ref}^\mathrm{[cal]})/P_\mathrm{ref}^\mathrm{[cal]}$. While in the left middle panel $f^\mathrm{cal}>f$ for all frequencies, in the right middle panel several crossings occur. The \textbf{lower panels} depict Eq.\,(\ref{eqresonance}), which represents the condition for resonances to appear.}%
   \label{fig:fswitchresonances}%
\end{figure*}
During our simulations, an important issue emerged. Inspection of Eq.\,(\ref{eqalpha}) reveals that $\alpha(\nu)$ can become singular, if $f=f^\mathrm{cal}$ or
\begin{equation}
\frac{T_\mathrm{cal,+}}{T_\mathrm{cal,-}}=\frac{T_\mathrm{cont,+}^\mathrm{sou}+T_\mathrm{sys,+}}{T_\mathrm{cont,-}^\mathrm{sou}+T_\mathrm{sys,-}}.\label{eqresonance}
\end{equation}
Under normal circumstances, this condition will not be fulfilled. However, in the presence of a standing wave contribution, $T_\mathrm{sw,+/-}$ to $T_\mathrm{sys,+/-}$, there will be several frequencies in the spectrum where Eq.\,(\ref{eqresonance}) is true. This results in `resonance' spikes in the reduced spectra, shifted according to $\nu_\mathrm{LO}$) (see Fig.\,\ref{fig:fswitchresonances}, upper right panel). These occur if the LO frequency is not (close to) a multiple of the standing wave frequency. The middle panels of Fig.\,\ref{fig:fswitchresonances} show $(P_\mathrm{sig}^\mathrm{[cal]}-P_\mathrm{ref}^\mathrm{[cal]})/P_\mathrm{ref}^\mathrm{[cal]}$ for the case where the LO shift matches $\nu_\mathrm{sw}$ (left) and does not match $\nu_\mathrm{sw}$ (right). The lower panels depict Eq.\,(\ref{eqresonance}), i.e. the resonance condition in terms of temperatures. Consequently, to avoid this problem one must ensure that $T_\mathrm{sw,+}\approx T_\mathrm{sw,-} $ by matching the LO shift to the frequency of the standing wave, i.e. $\nu_\mathrm{LO}\approx k\nu_\mathrm{sw},~k\in \mathds{N}_+$.  

\subsection{Determination of $T_\mathrm{sys}$ neglecting frequency dependencies --- the `classical' approach}\label{subsec:fswitchclassicalmethod}
\begin{figure*}[!t]
   \centering%
   \includegraphics[width=0.45\textwidth,bb=20 42 521 392,clip=]{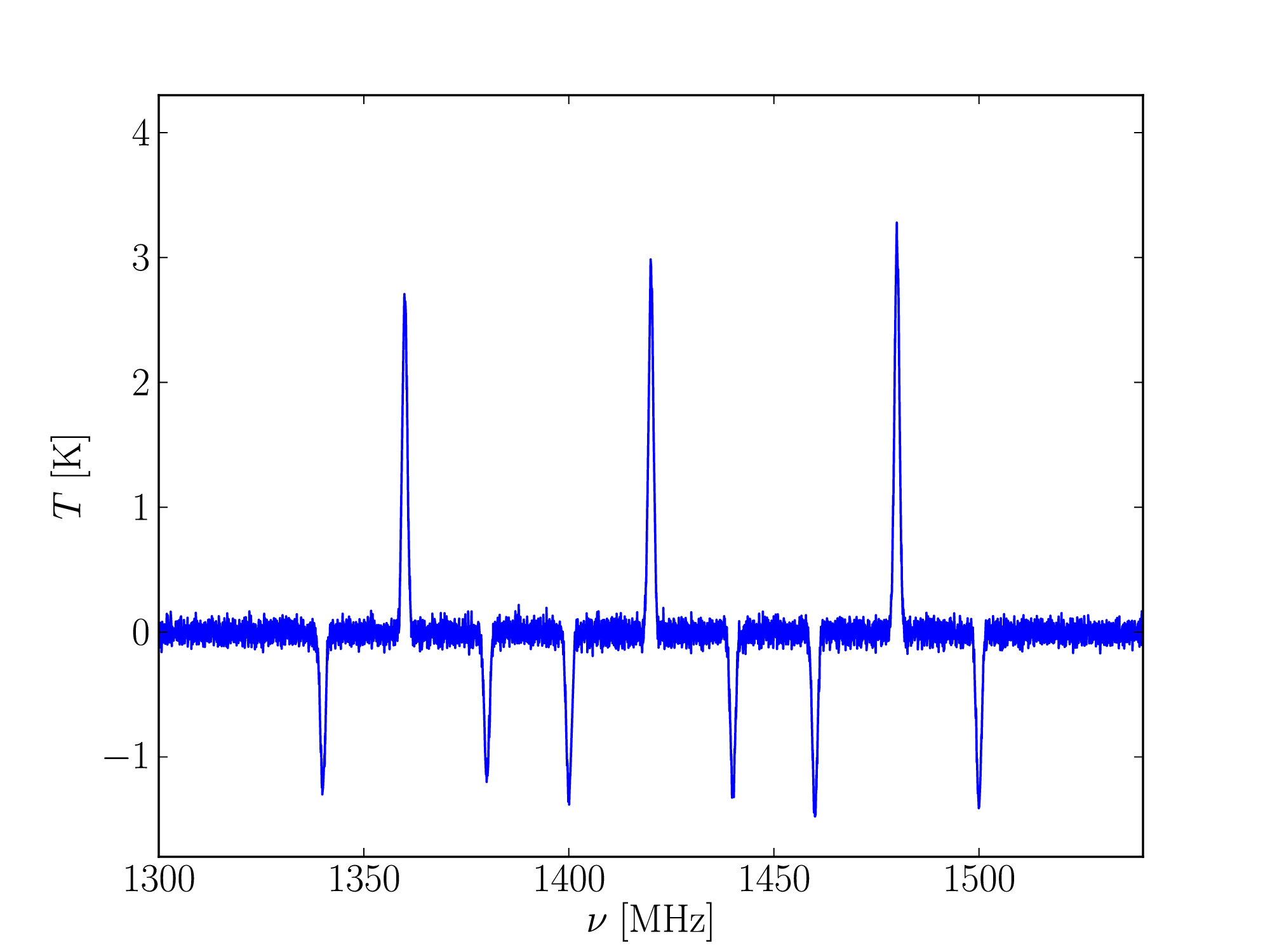}\quad
  \includegraphics[width=0.45\textwidth,bb=20 42 521 392,clip=]{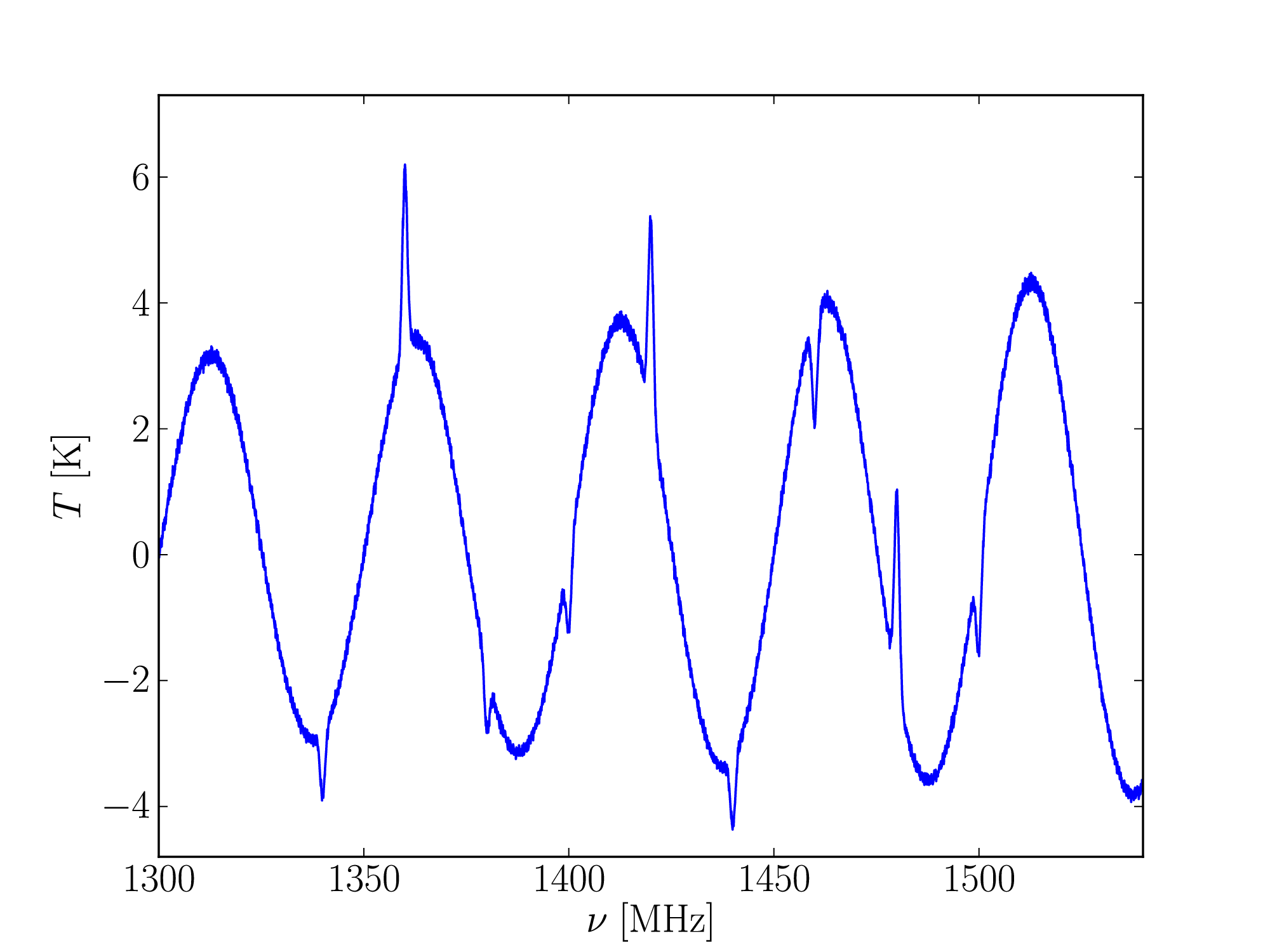}\\[0ex]
   \includegraphics[width=0.45\textwidth,bb=20 1 521 392,clip=]{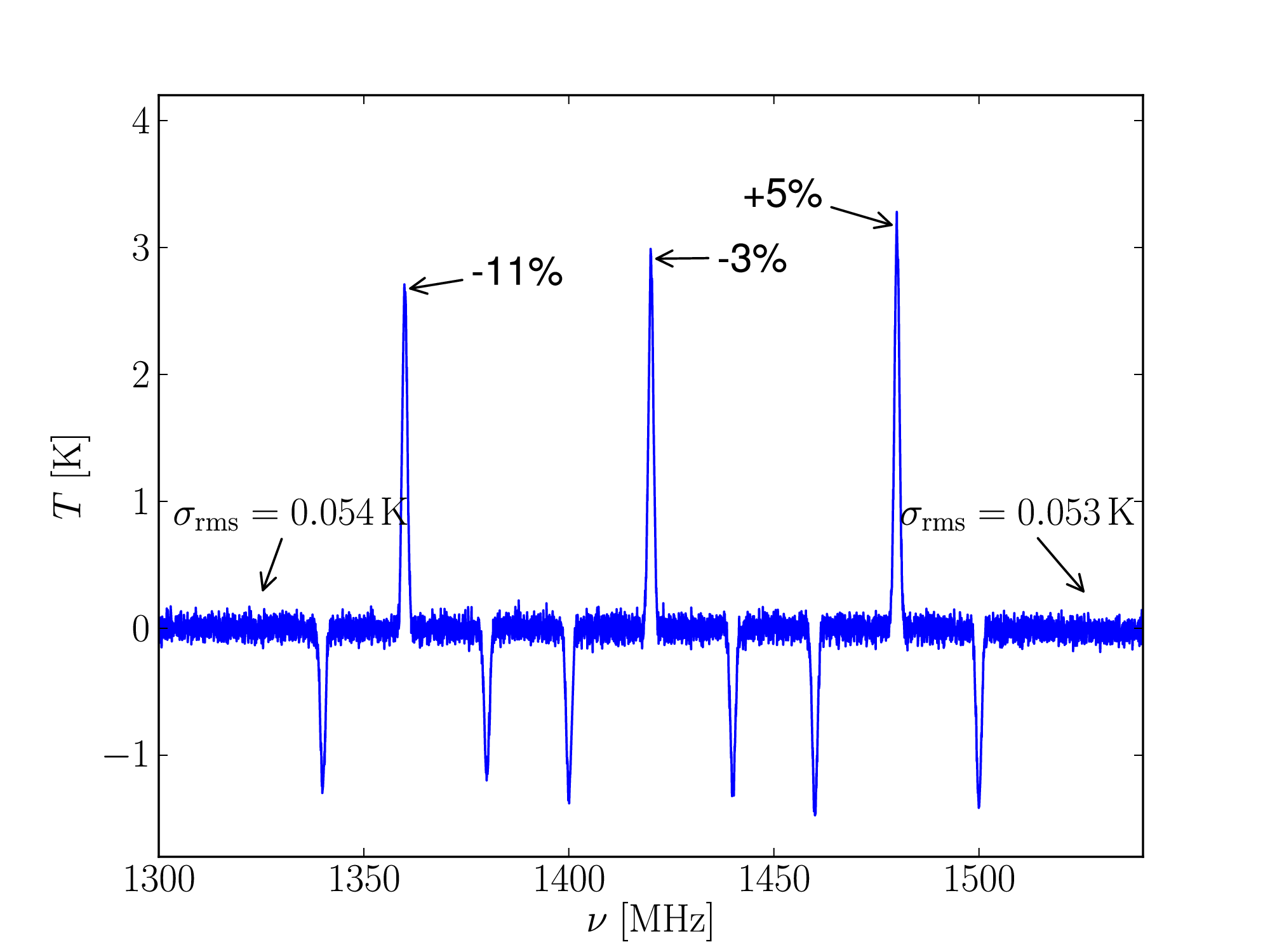}\quad
  \includegraphics[width=0.45\textwidth,bb=20 1 521 392,clip=]{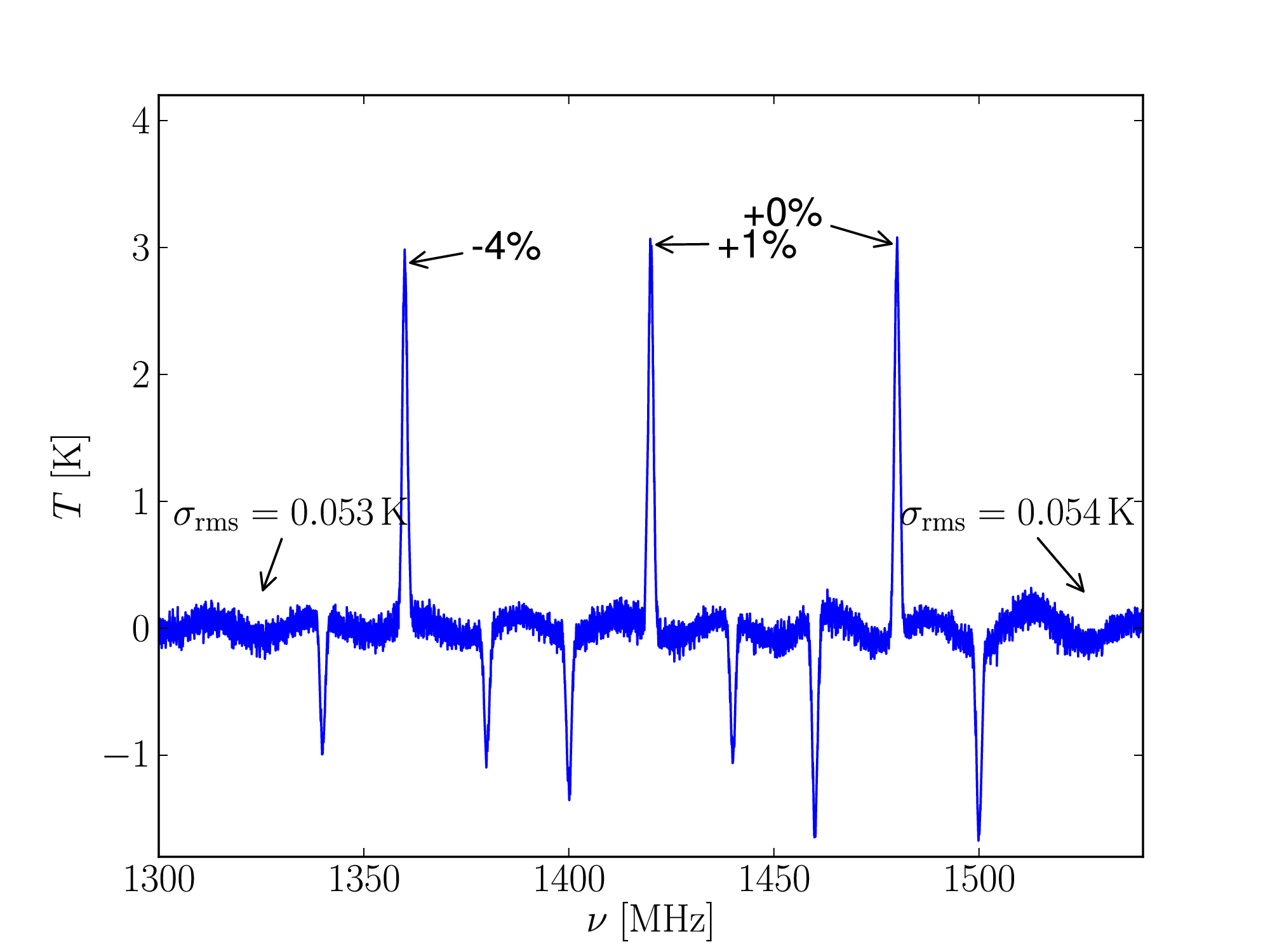}
  \caption{These plots show the results of the classical scheme before (upper rows) and after (lower rows) baseline subtraction (\textbf{left panels}: simple case; \textbf{right panels}: standing wave case). There is obviously a frequency-dependent flux error. As in Fig.\,\ref{fig:fswitch_method1_result}, the SW case has a complicated baseline that cannot be easily fitted.  }%
   \label{fig:fswitch_classic_result_base}%
\end{figure*}

On the basis of Eq.\,(\ref{eqfswitchtsystcal1}) and Eq.\,(\ref{eqfswitchtsystcal2}), but explicitly not restricting ourselves to continuum contributions, we find that
\begin{align}
\frac{T_\mathrm{line,-}^\mathrm{sou}+T_\mathrm{cont,-}^\mathrm{sou}+T_\mathrm{sys,-}}{T_\mathrm{cal,-}}&=\frac{(P_\mathrm{sig}^\mathrm{cal}+P_\mathrm{sig})-(P_\mathrm{sig}^\mathrm{cal}-P_\mathrm{sig})}{2(P_\mathrm{sig}^\mathrm{cal}-P_\mathrm{sig})},\\
\frac{T_\mathrm{line,+}^\mathrm{sou}+T_\mathrm{cont,+}^\mathrm{sou}+T_\mathrm{sys,+}}{T_\mathrm{cal,+}}&=\frac{(P_\mathrm{ref}^\mathrm{cal}+P_\mathrm{ref})-(P_\mathrm{ref}^\mathrm{cal}-P_\mathrm{ref})}{2(P_\mathrm{ref}^\mathrm{cal}-P_\mathrm{ref})}.
\end{align}
Neglecting line-emission contributions completely and treating  $T_\mathrm{sys}$ and $T_\mathrm{cal}$ as constants, one can write 
\begin{align}
\overline{T}_\mathrm{cont,-}^\mathrm{sou}+ \overline{T}_\mathrm{sys,-}&=\langle T_\mathrm{cont,-}^\mathrm{sou}+T_\mathrm{sys,-}\rangle_\nu\nonumber\\
&= T_\mathrm{cal} \left\langle\frac{(P_\mathrm{sig}^\mathrm{cal}+P_\mathrm{sig})-\langle P_\mathrm{sig}^\mathrm{cal}-P_\mathrm{sig}\rangle_\nu}{2\langle P_\mathrm{sig}^\mathrm{cal}-P_\mathrm{sig}\rangle_\nu} \right\rangle_\nu, \\
\overline{T}_\mathrm{cont,+}^\mathrm{sou}+\overline{T}_\mathrm{sys,+}&=\langle T_\mathrm{cont,+}^\mathrm{sou}+T_\mathrm{sys,+}\rangle_\nu\nonumber\\
&= T_\mathrm{cal} \left\langle\frac{(P_\mathrm{ref}^\mathrm{cal}+P_\mathrm{ref})-\langle P_\mathrm{ref}^\mathrm{cal}-P_\mathrm{ref}\rangle_\nu}{2\langle P_\mathrm{ref}^\mathrm{cal}-P_\mathrm{ref}\rangle_\nu} \right\rangle_\nu.
\end{align}
Under the assumption of frequency independence, Eq.\,(\ref{eqfswitchbaseeq1}) and Eq.\,(\ref{eqfswitchbaseeq2}) simplify to
\begin{align}
T_\mathrm{sou,-}-T_\mathrm{sou,+} &=\left(T_\mathrm{line,+}^\mathrm{sou}+\overline{ T}_\mathrm{cont,+}^\mathrm{sou}+\overline{T}_\mathrm{sys,+} \left[+T_\mathrm{cal}\right]\right)\frac{P_\mathrm{sig}^\mathrm{[cal]} - P_\mathrm{ref}^\mathrm{[cal]}}{ P_\mathrm{ref}^\mathrm{[cal]}}\label{eqfswitcheffelsbergpipeline1},\\
T_\mathrm{sou,+}-T_\mathrm{sou,-} &=\left(T_\mathrm{line,-}^\mathrm{sou}+\overline{ T}_\mathrm{cont,-}^\mathrm{sou}+\overline{T}_\mathrm{sys,-} \left[+T_\mathrm{cal}\right]\right)\frac{P_\mathrm{ref}^\mathrm{[cal]} - P_\mathrm{sig}^\mathrm{[cal]}}{ P_\mathrm{sig}^\mathrm{[cal]}}\label{eqfswitcheffelsbergpipeline2}.
\end{align}
Figure~\ref{fig:fswitch_classic_result_base} contains the resulting spectra after applying the `classical' method, shifting, averaging, and subtracting a baseline. The recovered spectral line intensities are systematically wrong. In the SW case, the baseline has a complicated pattern, which can hardly be described by a simple model.

We note that simple spectral averaging to calculate $T_\mathrm{sys}$ is affected by spectral lines and spectral-line ghosts, introducing even more inaccuracies.

\subsection{The realistic case --- limits of frequency switching}\label{subsec:fswitchrealisticexample}
\begin{figure}[!t]
\centering%
\includegraphics[width=0.45\textwidth,bb=20 1 522 392,clip=]{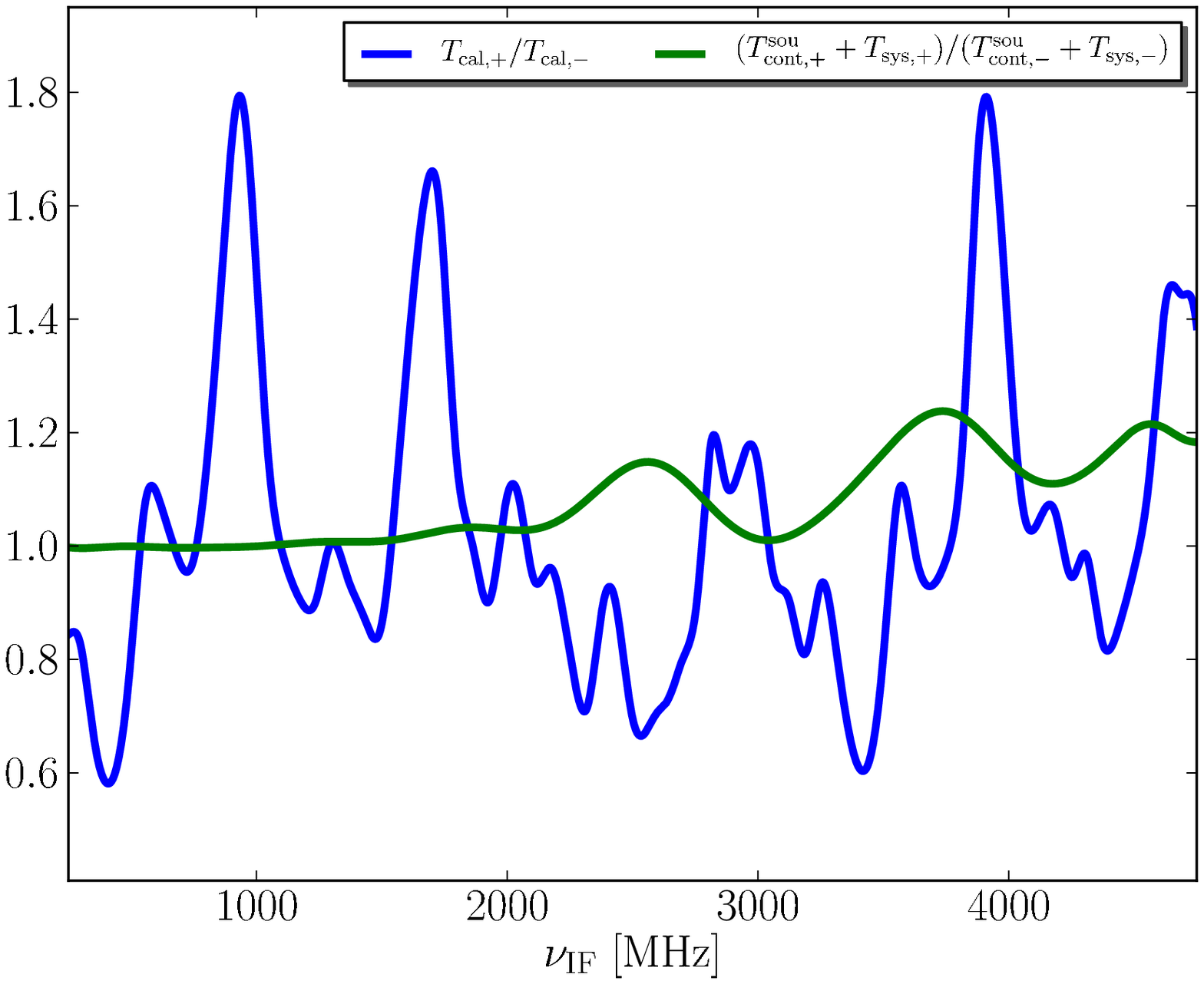}
\caption{Applying the second calibration method is impossible for more complicated system temperatures and/or calibration signals. Using the values from Fig.\,\ref{fig:pswitchcomplicatedtcalinput}, the two functions $T_\mathrm{cal,+}/T_\mathrm{cal,-}$ and $(T_\mathrm{cont,+}^\mathrm{sou}+T_\mathrm{sys,+})/(T_\mathrm{cont,-}^\mathrm{sou}+T_\mathrm{sys,-})$ have several intersections, which would lead to singularities in the result.}%
\label{fig:fswitchcomplicatedmethod2problem}%
\end{figure}

In Section\,\ref{subsec:pswitchrealisticexample}, we have shown that position switching also works well for more complicated examples, where the system temperature and calibration signal are fluctuating. The question now is whether one can also apply the frequency switching schemes to this present example.

For the second calibration model, it is usually impossible for the reasons discussed in Section\,\ref{subsec:fswitchresonances}. Fig.\,\ref{fig:fswitchcomplicatedmethod2problem} illustrates $T_\mathrm{cal,+}/T_\mathrm{cal,-}$ and $(T_\mathrm{cont,+}^\mathrm{sou}+T_\mathrm{sys,+})/(T_\mathrm{cont,-}^\mathrm{sou}+T_\mathrm{sys,-})$. Each intersection of the two functions would lead to a singularity in the final spectrum rendering a fair fraction of the spectral range unusable. Furthermore, as we have seen in Section\,\ref{subsec:fswitchmethod2}, the accuracy of the second method is generally lower than that of the first. 

\begin{figure}[!t]
\centering%
\includegraphics[width=0.45\textwidth,bb=20 1 522 392,clip=]{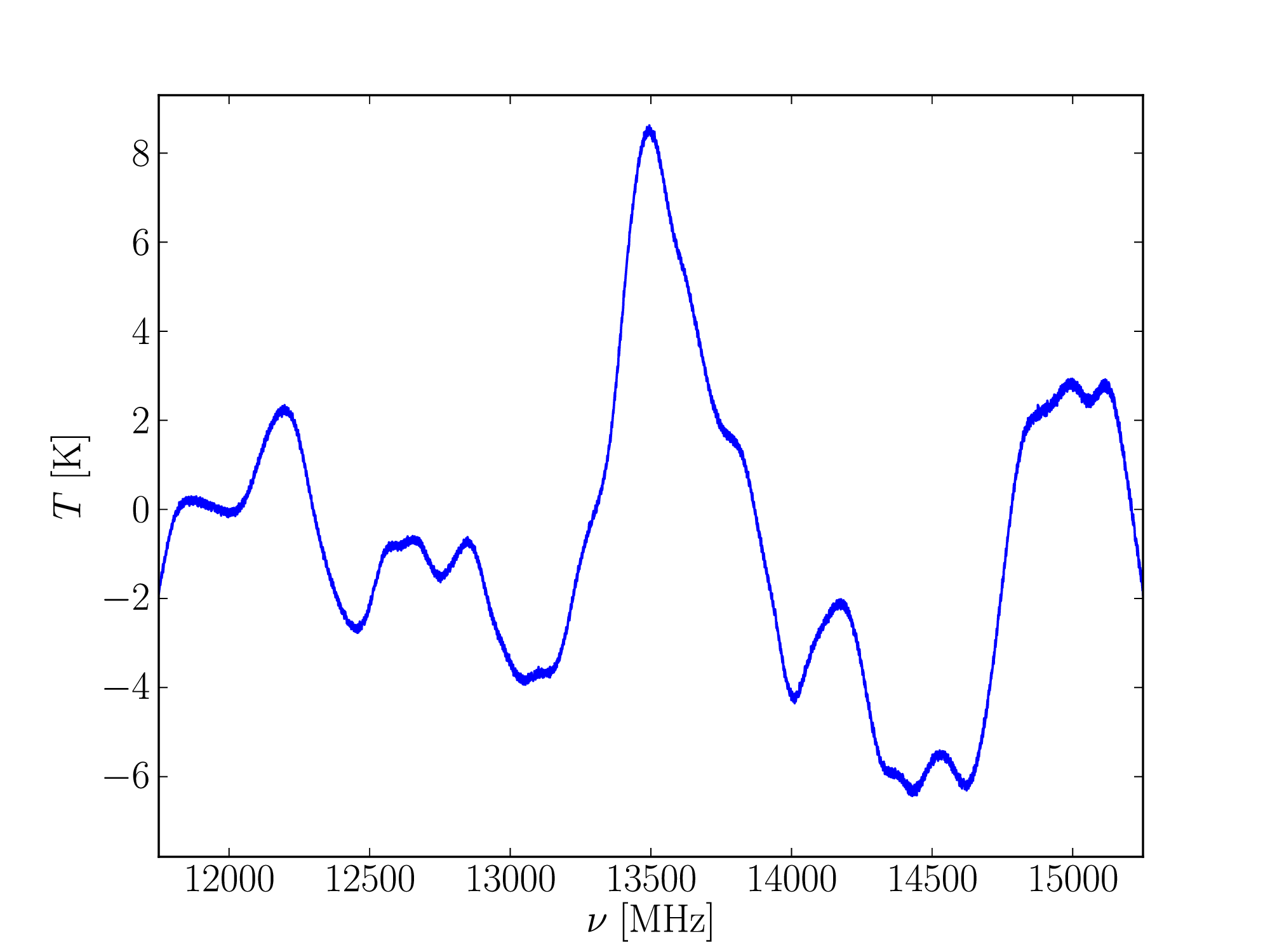}
\caption{The resulting spectrum for a real-world example using frequency switching and the first calibration method (see Fig.\,\ref{fig:pswitchcomplicatedtcalinput} for temperature inputs). The only difference in comparison to the position switching example is the use of a single Gaussian with an amplitude of 10\,K. Despite the larger line amplitude, the emission peak is not very pronounced in the result.}%
\label{fig:fswitchcomplicatedresult}%
\end{figure}
Does the first method (see Section\,\ref{subsec:fswitchmethod1}) work? In short, the answer is yes, but the result is still unusable (see Fig.\,\ref{fig:fswitchcomplicatedresult}). While the reconstructed spectrum is correctly calibrated, the residual baseline (which is also correct) has such a complex shape that one can hardly distinguish the spectral line from the structures in the baseline. This has nothing to do with our calibration scheme but is an intrinsic problem with frequency switching. Inspecting Eq.\,(\ref{eqfswitchfinalequation}) reveals that not only does the source produce spectral-line ghosts, but there is the term
\begin{equation}
T_\mathrm{sys}^\mathrm{[cal]}(\nu)-\frac{1}{2}T_\mathrm{sys}^\mathrm{[cal]}(\nu+2\Delta \nu)-\frac{1}{2}T_\mathrm{sys}^\mathrm{[cal]}(\nu-2\Delta \nu)\,.
\end{equation}
Hence, as soon as the system temperature (and $T_\mathrm{cal}$) has spectral features on the same scale as the astronomical lines, frequency switching will fail (or rather the astronomer who has to extract useful information out of the spectrum). This is an important aspect to be considered for future broadband instruments. 

\subsection{Combining position and frequency switching to handle $G_\mathrm{RF,-}\neq G_\mathrm{RF,+}$ cases}\label{subsec:fswitchcombination}
At the beginning of Section\,\ref{sec:fswitch}, we discussed how frequency switching is limited to cases where $G_\mathrm{RF,-}= G_\mathrm{RF,+}$. This is unfortunate because it substantially limits the number of possible applications as almost no real receiving system will provide a constant RF gain, at least if moderate or large bandwidths are involved.

One may however combine a frequency- and position-switching measurement, preferably of a calibration source, to infer the term $G_\mathrm{RF,-}/G_\mathrm{RF,+}$.  This is sufficient to open the frequency switching method to a broader range of applications where $G_\mathrm{RF,-}\neq G_\mathrm{RF,+}$, as we show below. Without any restriction with respect to $G_\mathrm{RF}$, Eq.\,(\ref{eqsigref}) reads
\begin{equation}
\frac{P_\mathrm{sig}^\mathrm{[cal]} - P_\mathrm{ref}^\mathrm{[cal]}}{ P_\mathrm{ref}^\mathrm{[cal]}}= \frac{G_\mathrm{RF,-}}{G_\mathrm{RF,+}} \frac{T_\mathrm{sou,-}+T_\mathrm{sys,-}^\mathrm{[cal]}-T_\mathrm{sou,+}-T_\mathrm{sys,+}^\mathrm{[cal]}}{T_\mathrm{sou,+}+T_\mathrm{sys,+}^\mathrm{[cal]}}\,.
\end{equation}
With two frequency-switching measurements, one \textit{on} source the other \textit{off} source, i.e.
\begin{align}
P_\mathrm{sig,on}^\mathrm{[cal]}&=G_\mathrm{IF}G_\mathrm{RF,-}\left[ T_\mathrm{sou,-}+T_\mathrm{sys,-}^\mathrm{[cal]}\right],\\
P_\mathrm{ref,on}^\mathrm{[cal]}&=G_\mathrm{IF}G_\mathrm{RF,+}\left[ T_\mathrm{sou,+}+T_\mathrm{sys,+}^\mathrm{[cal]}\right],\\
P_\mathrm{sig,off}^\mathrm{[cal]}&=G_\mathrm{IF}G_\mathrm{RF,-} T_\mathrm{sys,-}^\mathrm{[cal]},\\
P_\mathrm{ref,off}^\mathrm{[cal]}&=G_\mathrm{IF}G_\mathrm{RF,+} T_\mathrm{sys,+}^\mathrm{[cal]},
\end{align}
one can determine
\begin{equation}
\frac{G_\mathrm{RF,-}}{G_\mathrm{RF,+}}=\frac{P_\mathrm{sig,on}^\mathrm{[cal]}-P_\mathrm{sig,off}^\mathrm{[cal]}}{P_\mathrm{ref,on}^\mathrm{[cal]}-P_\mathrm{ref,off}^\mathrm{[cal]}} \frac{T_\mathrm{sou,+}}{T_\mathrm{sou,-}},
\end{equation}
where the term $T_\mathrm{sou,+}/T_\mathrm{sou,-}$ is known if a calibration source is used. A simple rescaling
\begin{equation}
\frac{P_\mathrm{sig}^\mathrm{[cal]} - P_\mathrm{ref}^\mathrm{[cal]}}{ P_\mathrm{ref}^\mathrm{[cal]}} \rightarrow \frac{G_\mathrm{RF,+}}{G_\mathrm{RF,-}} \frac{P_\mathrm{sig}^\mathrm{[cal]} - P_\mathrm{ref}^\mathrm{[cal]}}{ P_\mathrm{ref}^\mathrm{[cal]}}
\end{equation}
of all occurrences of the term $(P_\mathrm{sig}^\mathrm{[cal]} - P_\mathrm{ref}^\mathrm{[cal]})/ P_\mathrm{ref}^\mathrm{[cal]}$ in all of the frequency-switching equations is sufficient to fully incorporate the $G_\mathrm{RF,-}\neq G_\mathrm{RF,+}$ cases.

The \textsc{On--Off} measurement of the calibration source has of course to be repeated on appropriate timescales to account for time-dependent instabilities in the RF gain. However, in most cases this will allow a much longer effective integration time \textit{on} source than a position-switching measurement, which in the past has been considered the only viable option for `futile' cases.

\subsection{Least squares frequency switching}

Least squares frequency switching (LSFS) is a technique developed by \citet{heiles07}, which makes use of not only two, but several different LO shifts\footnote{This might require changes to the hardware of a receiving system to allow for more than the usual two LO frequencies. We note that LSFS does not deliver calibrated data but only normalised spectra that must be flux-calibrated using other measures (e.g. using $T_\mathrm{cal}$) later on.}. The underlying mathematical problem can be approximated by a linear equation and efficiently solved using maximum likelihood methods. A great property of LSFS is that it does not produce spectral-line ghosts. Hence, this method is the only option among frequency switching techniques for cases where blending of emission lines with spectral-line ghosts would be a problem, or where the system temperature or $T_\mathrm{cal}$ has structure (see also Section\,\ref{subsec:fswitchrealisticexample}). Improvements to the original LSFS method were proposed by \citet{winkel07b} to incorporate RFI treatment and to increase its robustness in the presence of strong emission lines. 

In contrast to classic frequency switching, one is able to determine the IF bandpass curve, as well as the product $G_\mathrm{RF}\left(T_\mathrm{sou}+T_\mathrm{sys}\right)$. The latter however shows that as for normal frequency switching, only the IF gain can be dealt with. Hence, if $G_\mathrm{RF}$ is frequency dependent (which is often the case especially for large bandwidths), frequency switching will lead to calibration errors. A similar method to that discussed in Section\,\ref{subsec:fswitchcombination} may be applied to work around this problem, i.e. combining LSFS with a position switch. \textit{It is then possible to even disentangle $G_\mathrm{RF}$ and $G_\mathrm{IF}$, which can be considered a great advantage over classical frequency switching.} Some overhead would clearly be involved to conduct (repeated) necessary \textsc{On--Off} measurements toward calibration sources.

\section{Direct determination of the bandpass curve}\label{sec:directmethod}
\begin{figure*}[!t]
\centering%
\includegraphics[width=0.44\textwidth,bb=20 42 535 392,clip=]{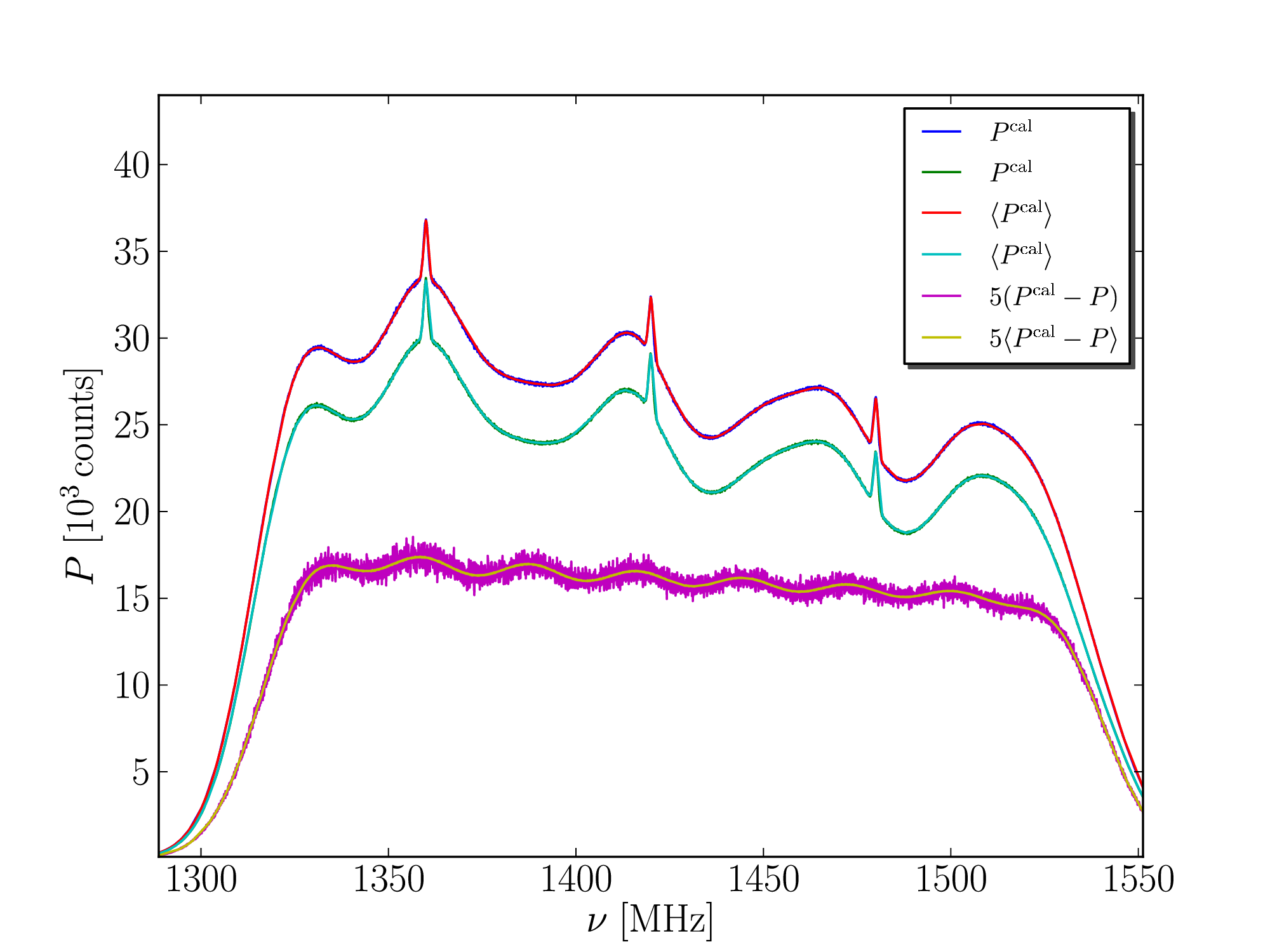}\quad
\includegraphics[width=0.44\textwidth,bb=20 42 535 392,clip=]{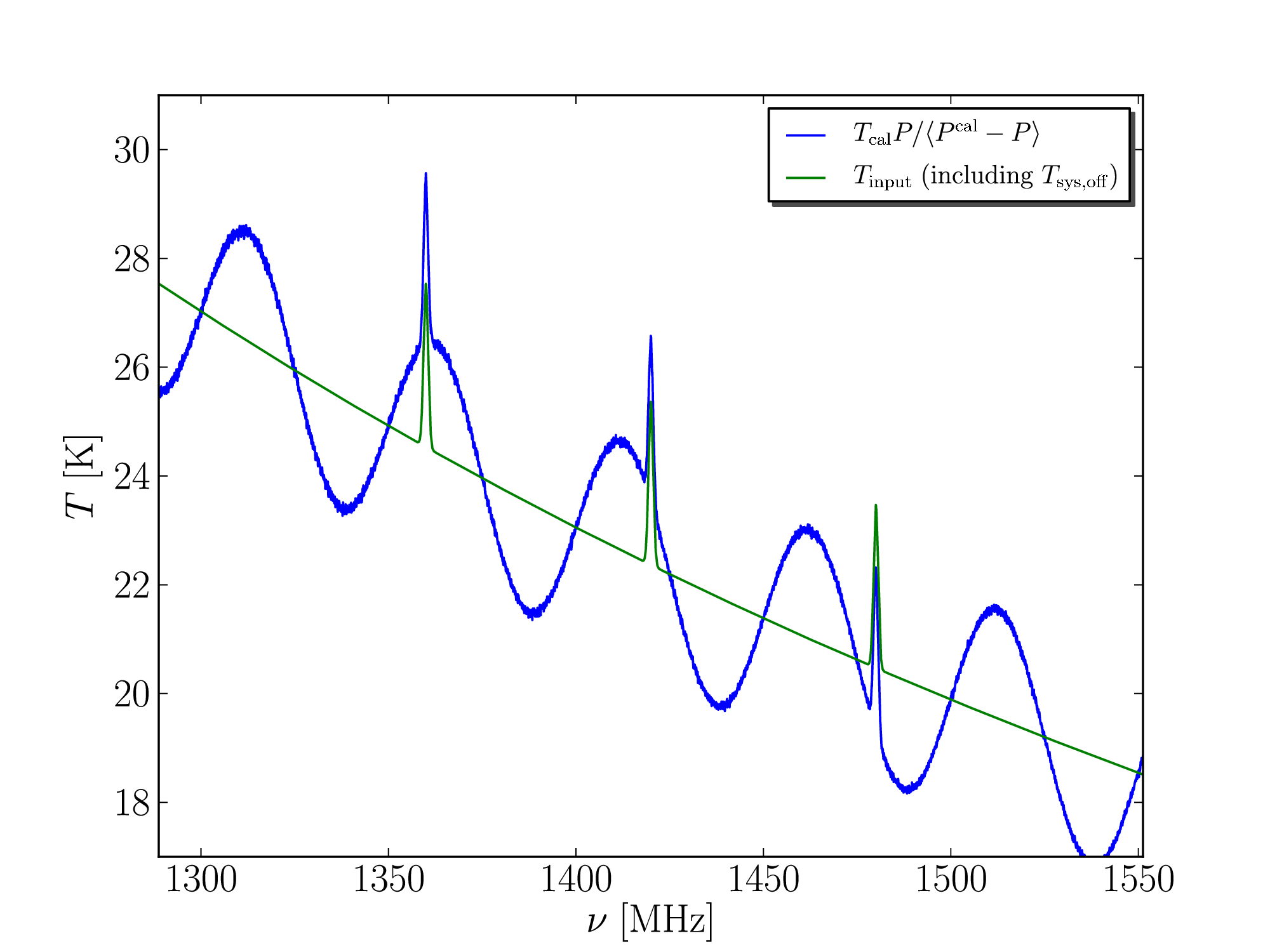}\\[0ex]
\includegraphics[width=0.44\textwidth,bb=20 1 535 392,clip=]{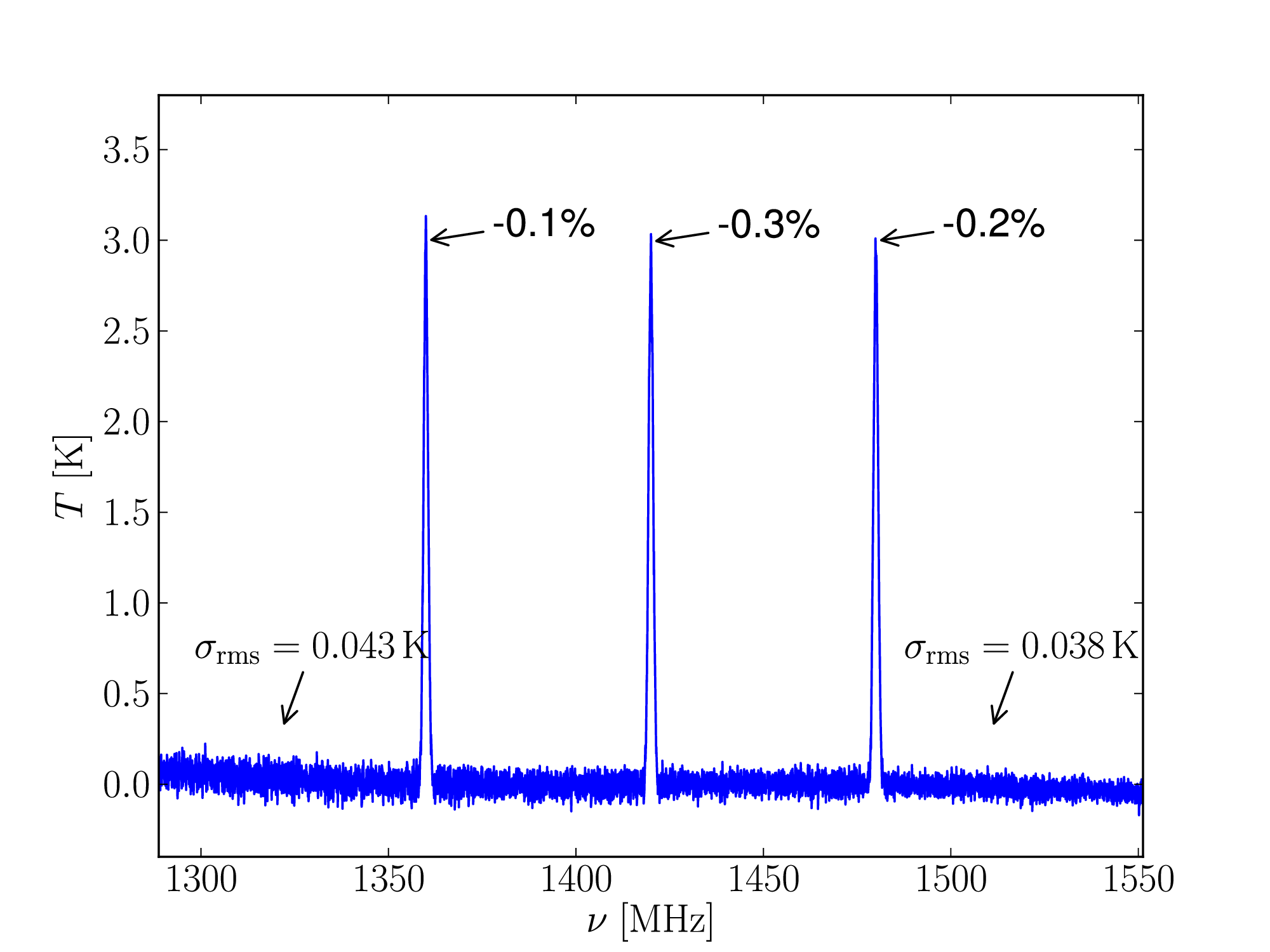}\quad
\includegraphics[width=0.44\textwidth,bb=20 1 535 392,clip=]{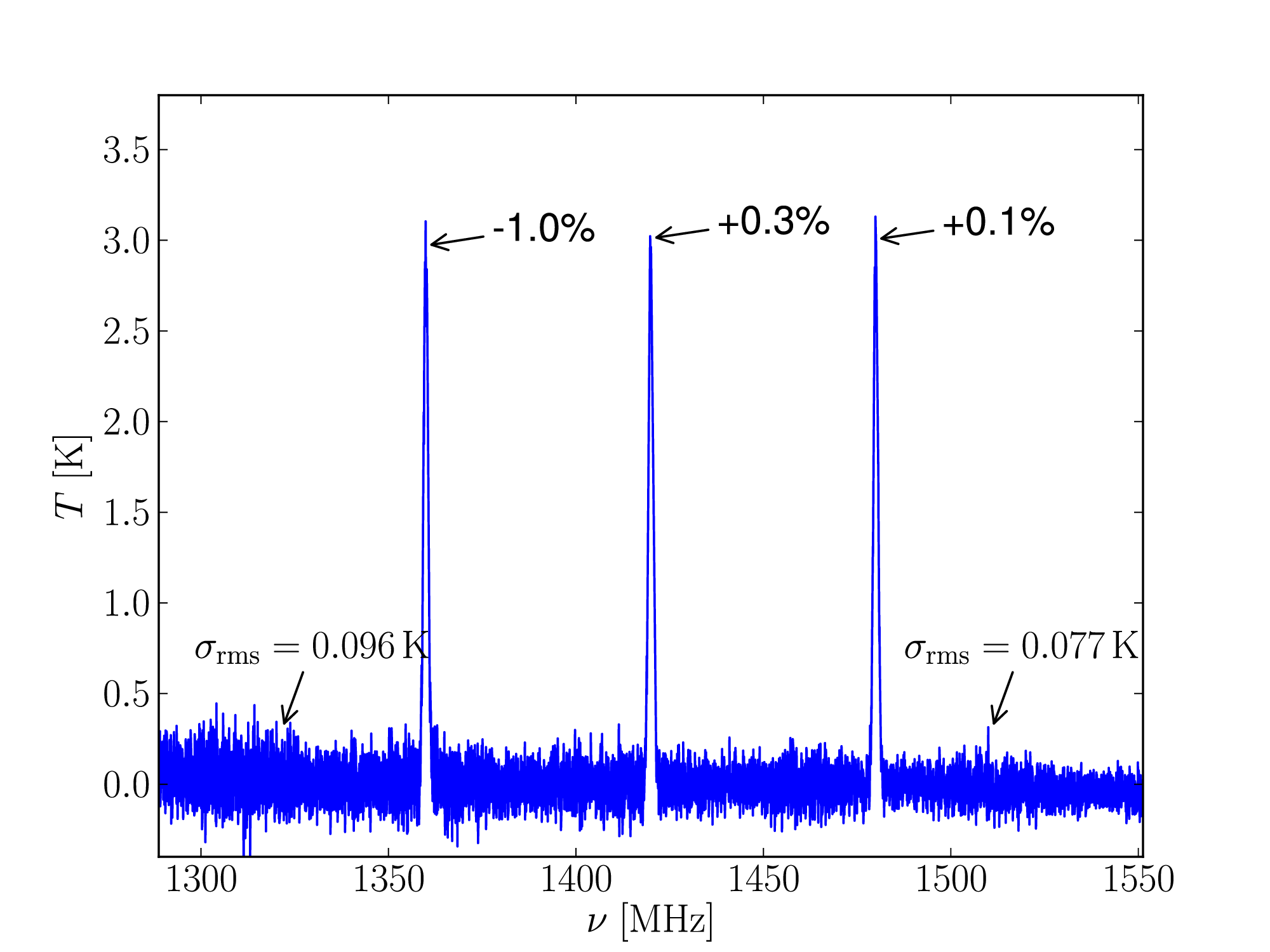}\\[-2ex]
\caption{\textbf{Top left panel:} Input spectra (including SW) used for the direct method to determine the bandpass curve. It also contains $P^\mathrm{cal}-P$ for a single spectral dump and $\langle P^\mathrm{cal}-P\rangle_t$ as computed using the average of 1000 spectral dumps, both scaled by a factor of five for improved visualisation. The \textbf{upper right panel} shows the reduced data after applying Eq.\,(\ref{eqdirectresult}). Note that the standing wave contribution is included in the result. The \textbf{lower panels} show the final spectra after baseline subtraction for $T_\mathrm{cal}/T_\mathrm{sys,off}\approx15\%$ (left panel) and $T_\mathrm{cal}/T_\mathrm{sys,off}\approx2.5\%$ (right panel).}%
\label{fig:directmethod}%
\end{figure*}

For the sake of completeness, we now present a third method that may be suitable for obtaining a sufficient calibration. It uses averaging over a large number of sequential spectral dumps, $P_i$, in order to measure the bandpass directly
\begin{equation}
\begin{split}
P_i^\mathrm{cal}-P_{i}&=G_\mathrm{RF}G_\mathrm{IF}\left( T_i^\mathrm{sou}+T_\mathrm{i,sys}^\mathrm{cal} - T_{i}^\mathrm{sou}-T_\mathrm{i,sys}\right)\\
&=G_\mathrm{RF}G_\mathrm{IF}\left[T_\mathrm{cal}+\Delta T_\mathrm{sou}+\Delta T_\mathrm{sys}\right]\label{eqdirectbasic}.
\end{split}
\end{equation}
The correct $T_\mathrm{cal}(\nu)$ clearly needs to be known here. Since $\Delta T_\mathrm{sou}+\Delta T_\mathrm{sys}\ll T_\mathrm{cal}$, as continuum levels between two subsequent dumps within one scan can be assumed to be approximately equal, then
\begin{equation}
G_\mathrm{RF}G_\mathrm{IF}=\left\langle\frac{ P^\mathrm{cal}-P}{T_\mathrm{cal}+\Delta T_\mathrm{sou}+\Delta T_\mathrm{sys}}\right\rangle_t
\approx  \frac{\left\langle P^\mathrm{cal}-P\right\rangle_t}{T_\mathrm{cal}} \equiv\gamma(\nu)\label{eqdirectbandpass}\,.
\end{equation}
The function $\gamma(\nu)$ is of course subject to noise, which owing to averaging is however much smaller than in the raw data of a single dump. One may even apply spectral averaging (smoothing) to further decrease the noise in $\gamma(\nu)$. It follows that
\begin{equation}
T_\mathrm{sou}+T_\mathrm{sys}=\frac{P}{\gamma}=\frac{P^\mathrm{cal}}{\gamma}-T_\mathrm{cal}\label{eqdirectresult}.
\end{equation}
An important difference from the position switching method is that one can only infer $T_\mathrm{sou}+T_\mathrm{sys}$, hence afterwards one cannot distinguish the background and source continua. The background continuum also varies with time, and thus the baseline fitting must be carried out on appropriate timescales. A second drawback is that the bandpass/gain must not vary over the course of the observation, otherwise the procedure would only return an average bandpass such that the individual spectra are wrongly calibrated. Whether this method is applicable depends on the properties of the receiver. Nevertheless, the performance of the direct bandpass estimation may still be better than the use of frequency switching (one also has full-time on-source, in contrast to position switching).

This method is used for the Effelsberg--Bonn \ion{H}{i} Survey \citep[EBHIS;][]{winkel10}, where neither position nor frequency switching is applicable. For the 21-cm line at galactic radial velocities, there is simply no suitable \textsc{Off} position as there is emission in every direction. The receiver also exhibits a complex multi-modal SW pattern, which would be hard to model using analytic fitting functions. Frequency switching is impossible as the receiver has a significant bandpass ripple in the RF part. For more details, we refer to \citet{winkel10,winkel11techreport}.

A simulation was conducted using the same set-up as for the position switching, with the only difference being that no \textsc{Off} position was needed. In Fig.\,\ref{fig:directmethod} (upper left panel), the input spectra after multiplying by the bandpass curve are shown (including SW).

Figure\,\ref{fig:directmethod} also presents $P^\mathrm{cal}-P$ for a single spectral dump and $\langle P^\mathrm{cal}-P\rangle_t$, as computed using the average of 1000 spectral dumps, the latter showing significantly less noise. The upper right panel presents the reconstructed spectrum. In contrast to position switching, the full continuum contribution $T_\mathrm{sou}+T_\mathrm{sys}$ is obtained (including SW). Therefore, to extract the pure spectral-line contribution one has to compute an appropriate baseline (in our case the superposition of a polynomial and a sine wave). The resulting spectra (after baseline subtraction) are shown in the lower panels. We use $T_\mathrm{cal}/T_\mathrm{sys,off}\approx15\%$ (left panel) and $T_\mathrm{cal}/T_\mathrm{sys,off}\approx2.5\%$ (right panel). In the resulting spectrum, the noise is of course higher in the latter case. It is however much harder to obtain a noise-free description of the bandpass curve (by filtering or model fitting) than in the former case. The robustness of the method therefore also depends on the ratio of $T_\mathrm{cal}$ to $T_\mathrm{sys,off}$.

\section{Flux density calibration error distributions}\label{sec:errordistribution}
We analyse the spread in the recovered flux values using the three different calibration methods. The distribution of flux errors is a good indicator of the robustness of the calibration schemes and also reveals potential biases. The exact values expected depend strongly on the specific choice of the various temperature contributions (especially the slopes), bandwidth, and so on. Our analysis should therefore be treated as a qualitative study and not a quantitative analysis.

To estimate the distribution of flux errors, we used the simulations (of the more simple example presented in Section\,\ref{subsec:simpleexample}) as described in the previous sections, but repeated each of them one thousand times, each time with a different noise realisation but all other parameters constant. In this way, we were able to explore the stability of the various methods. 

\subsection{Position switching}\label{subsec:simulationpswitch}

\begin{figure*}[!tp]
   \centering%
   \includegraphics[width=0.4\textwidth,bb=14 22 521 392,clip=]{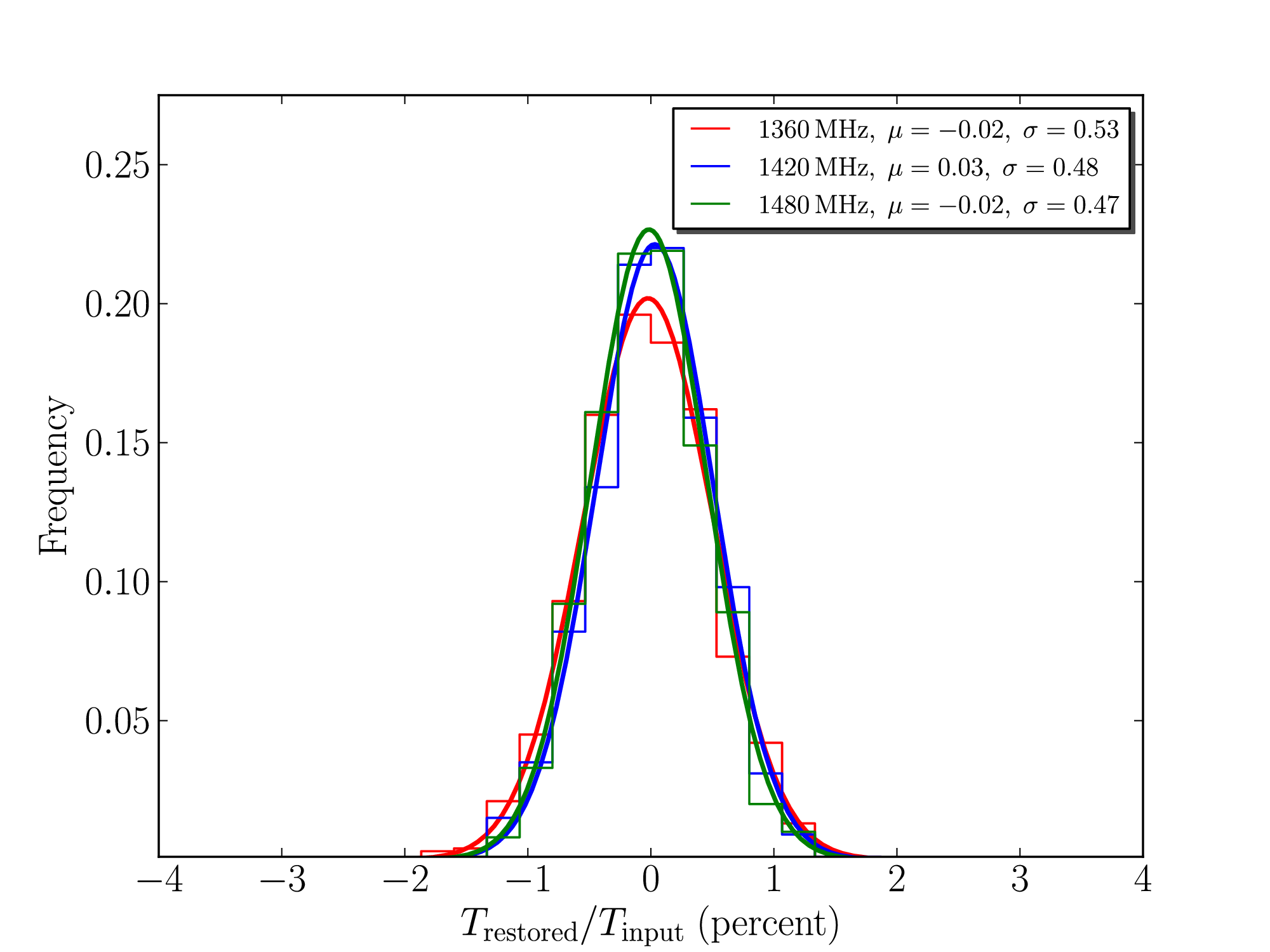}\quad
   \includegraphics[width=0.4\textwidth,bb=14 22 521 392,clip=]{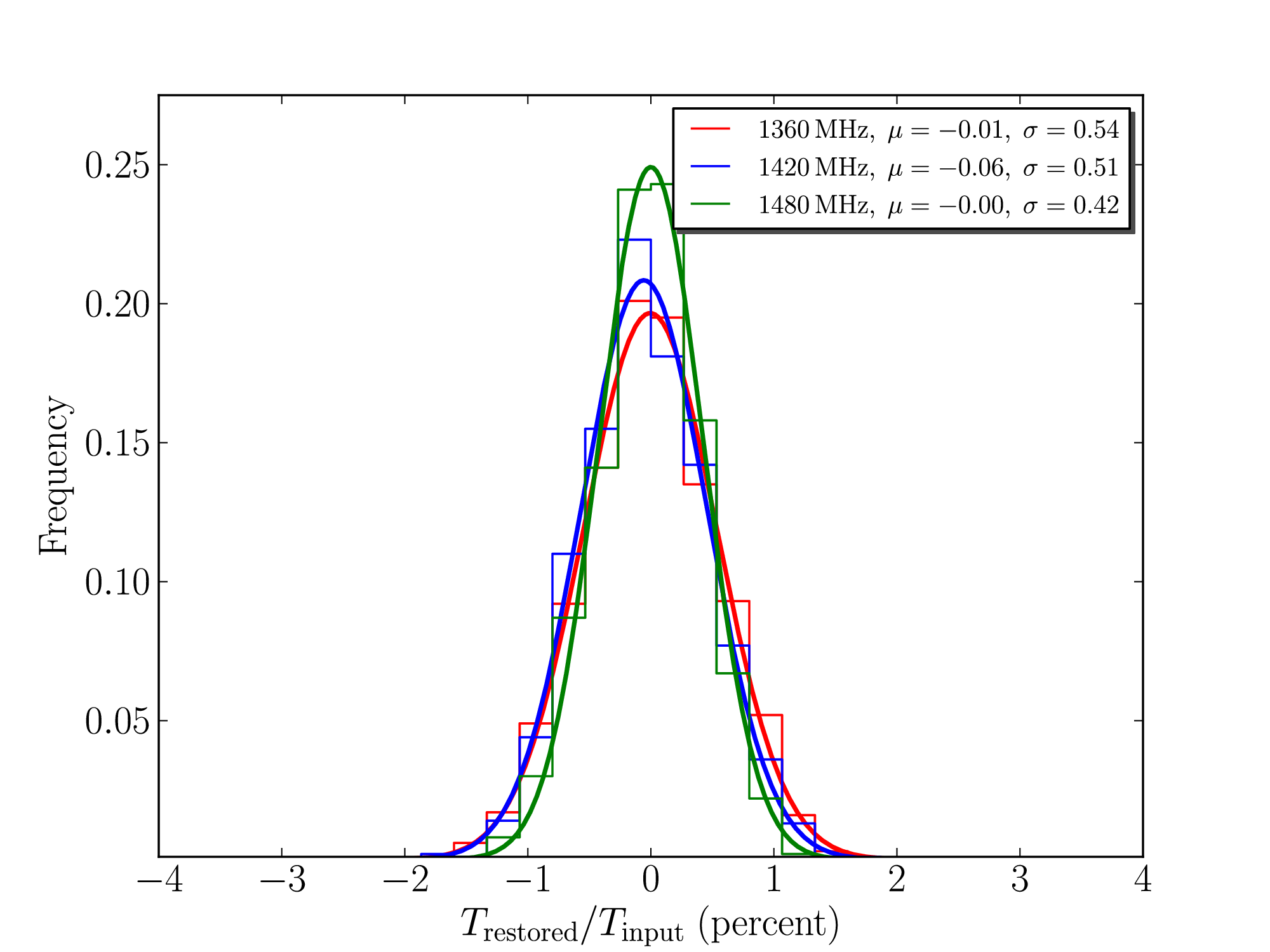}\\[0ex]
   \includegraphics[width=0.4\textwidth,bb=14 23 521 392,clip=]{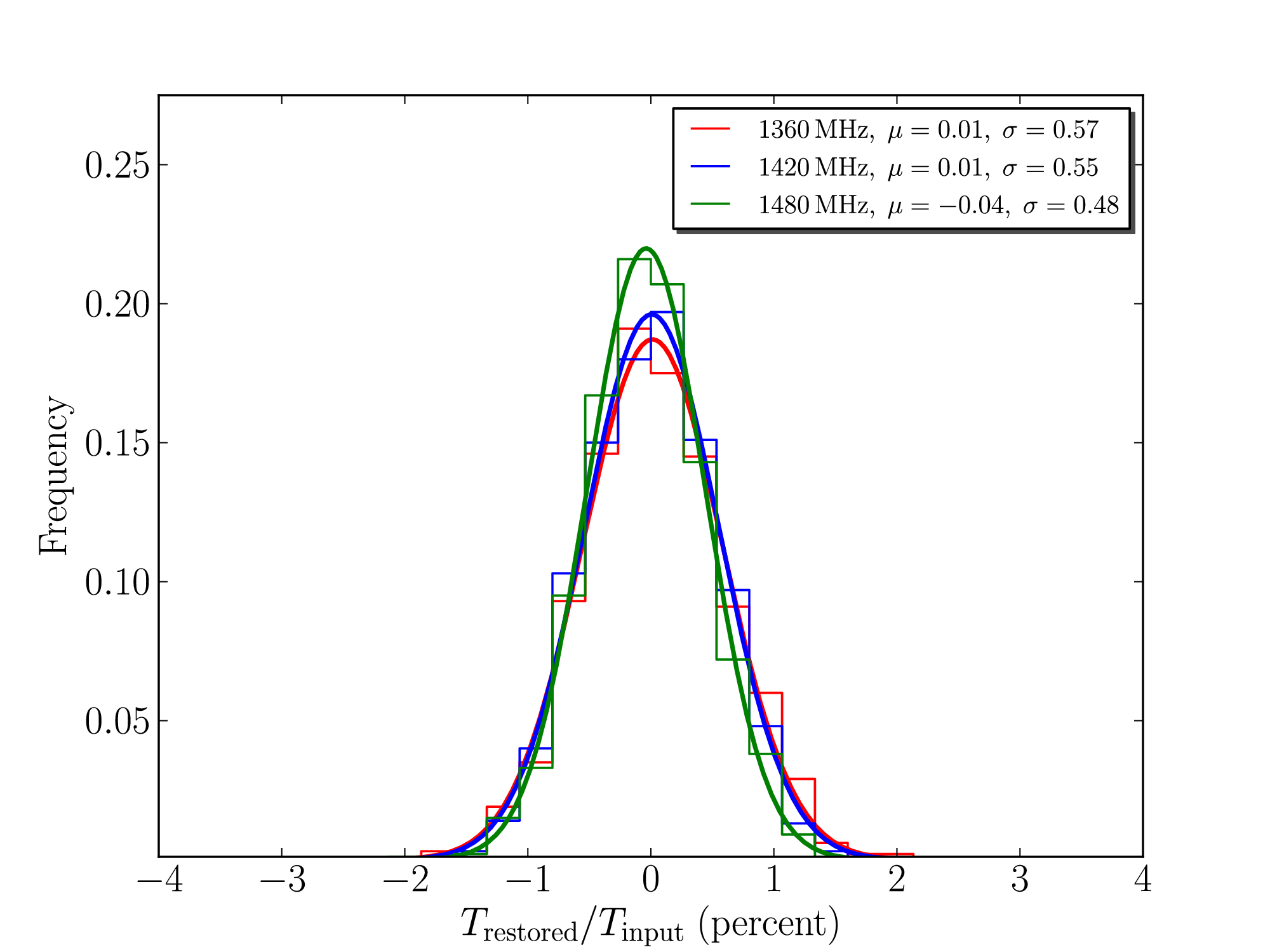}\quad
   \includegraphics[width=0.4\textwidth,bb=14 23 521 392,clip=]{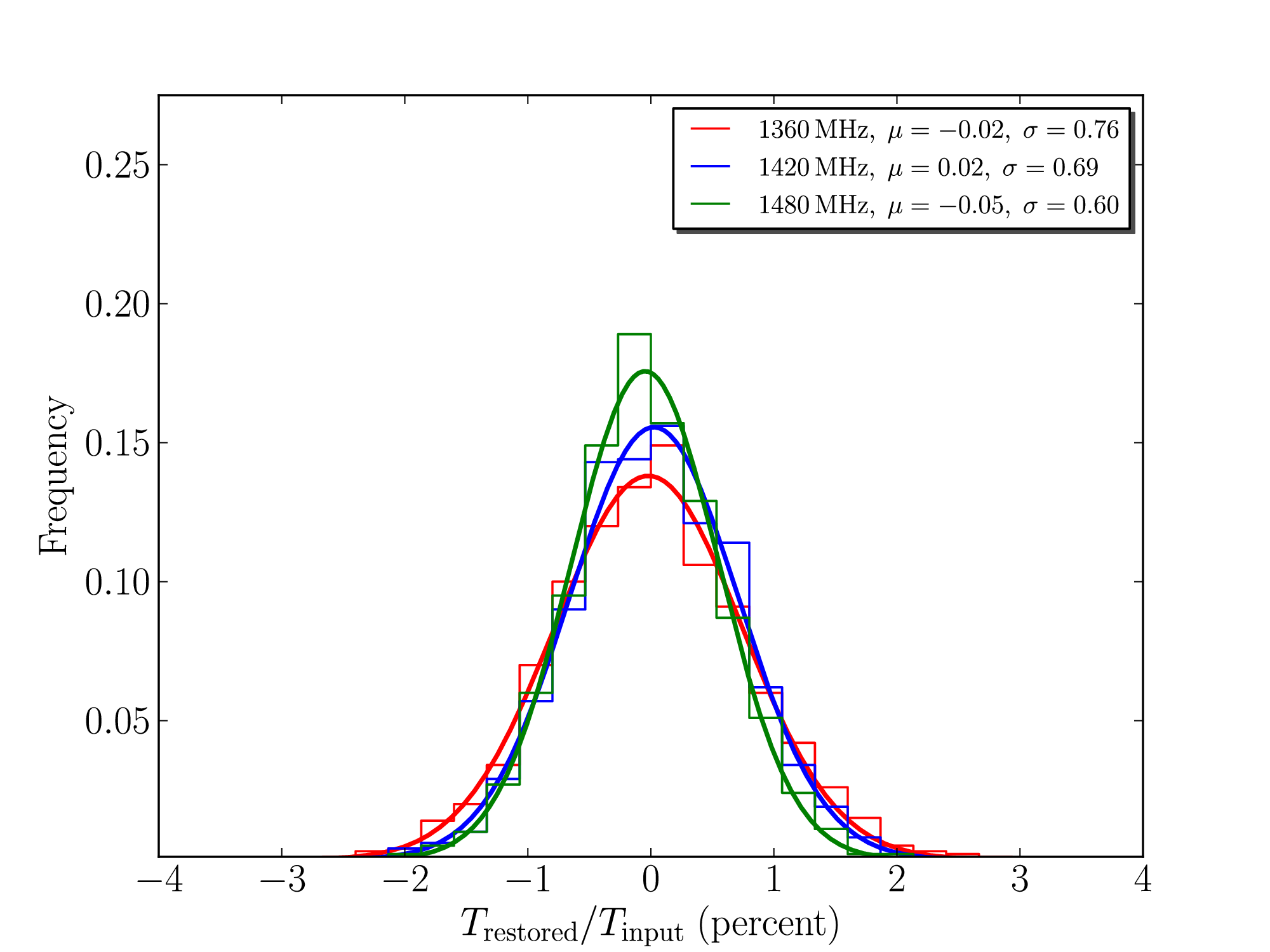}\\[0ex]
   \includegraphics[width=0.4\textwidth,bb=14 1 521 392,clip=]{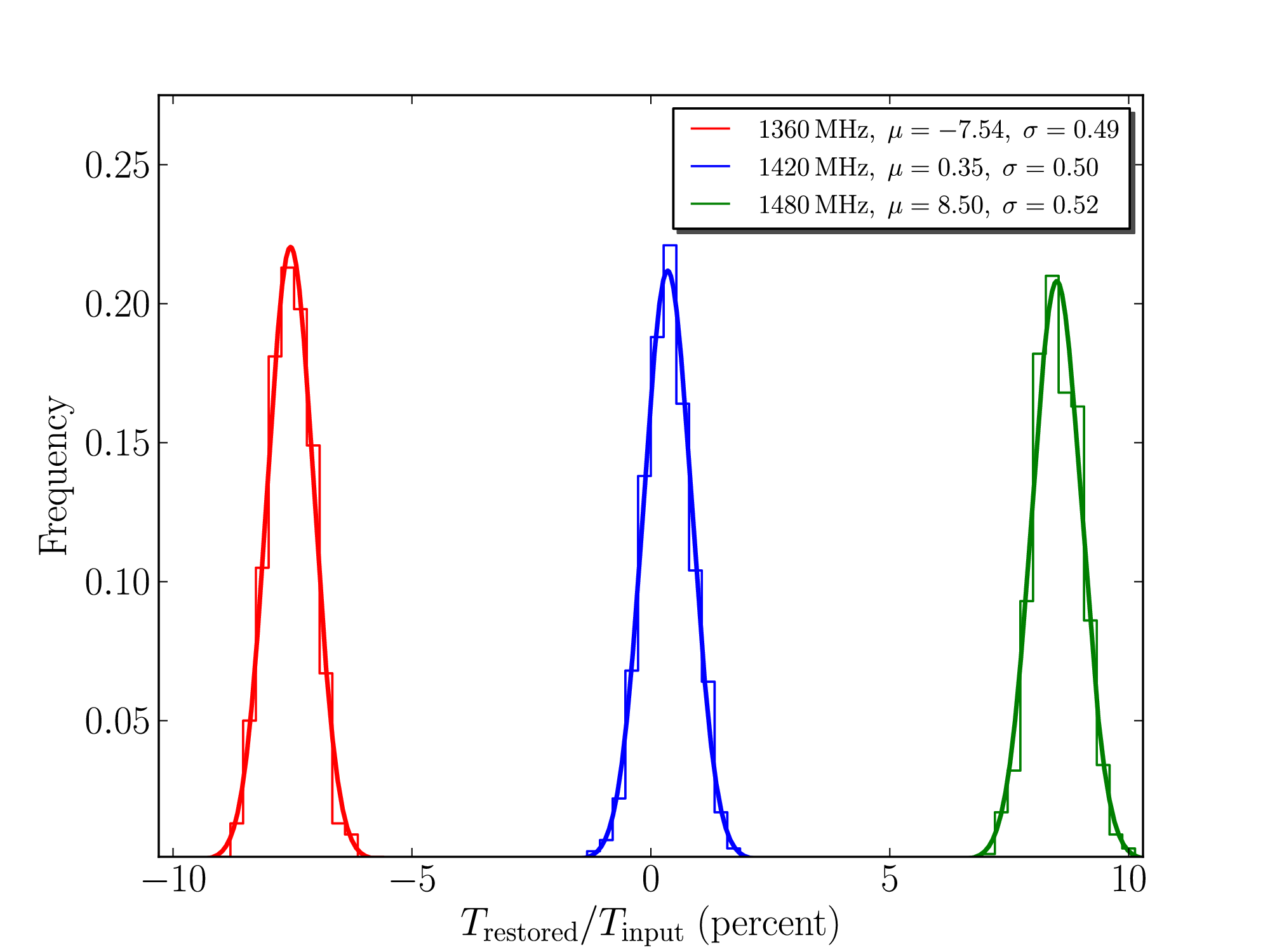}\quad
   \includegraphics[width=0.4\textwidth,bb=14 1 521 392,clip=]{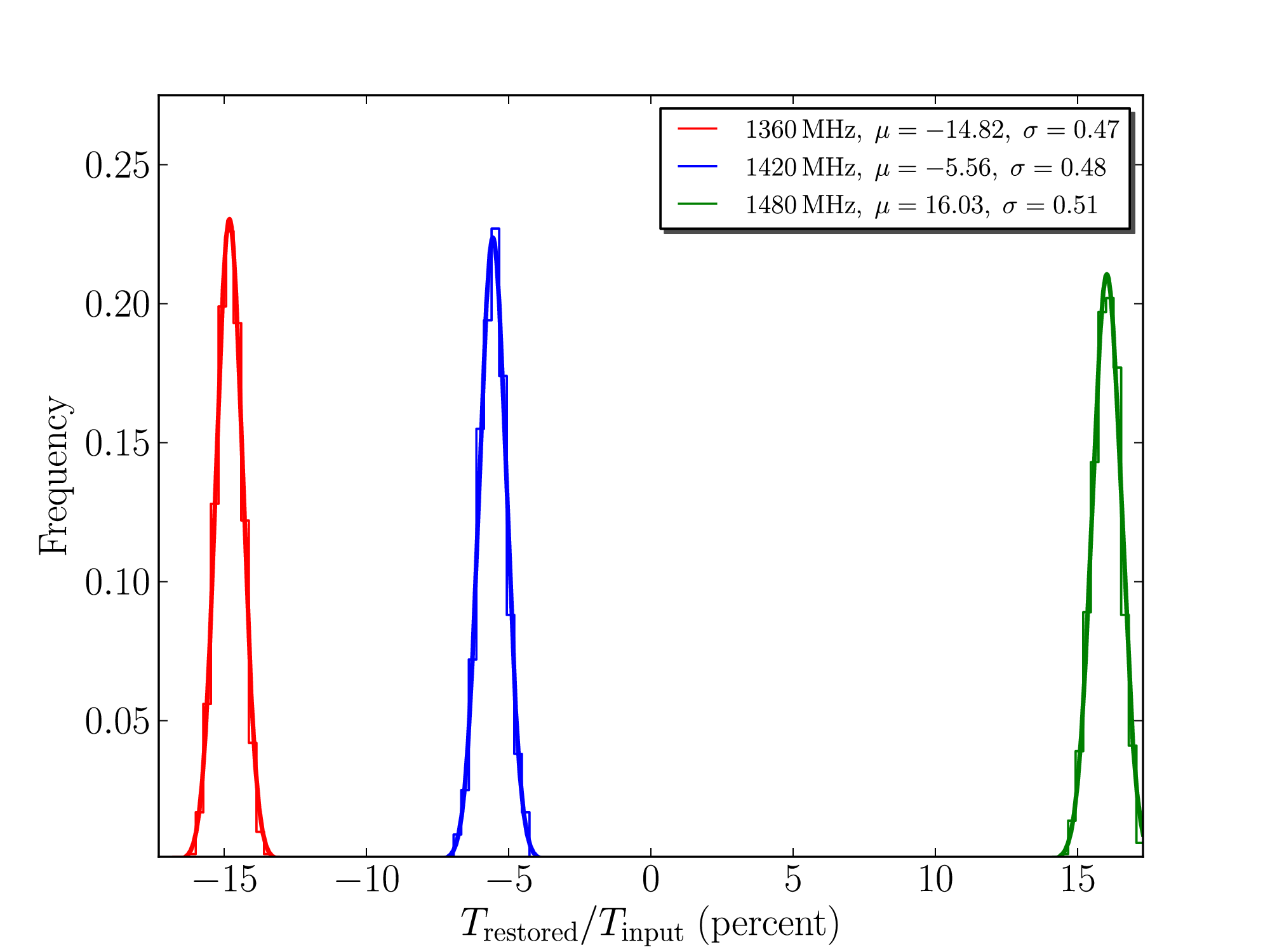}
  \caption{To estimate the robustness of the different methods, we repeated each simulation a thousand times. The resulting flux errors for the three Gaussian components were evaluated. The histograms show the error distribution for the two cases, \textbf{left (right) panels} without (with) SW. The \textbf{upper rows} show the results for the method described in Section\,\ref{subsec:pswitchmethod1} using only \textsc{Off} positions to estimate $T_\mathrm{cal}/T_\mathrm{sys}$. The \textbf{middle panels} were produced using the procedure from Section\,\ref{subsec:pswitchmethod2}. The \textbf{lower row} contains the result for the classical method ($T_\mathrm{sys}=\mathrm{const.}$ and $T_\mathrm{cal}=\mathrm{const.}$). The bin width has been kept constant for each histogram, such that the amplitudes can be directly compared. For each distribution of errors, we fitted a Gaussian distribution, the parameters (mean value $\mu$ and width $\sigma$) of which are stated in the plots.}%
   \label{fig:pswitcherrorhistograms}%
\end{figure*}

The resulting histograms of the relative flux error for the different cases are shown in Fig.\,\ref{fig:pswitcherrorhistograms} for each of the three simulated Gaussian emission lines.  The left (right) panels show the case without (with) standing waves (SW). The upper row contains the results provided by the method described in Section\,\ref{subsec:pswitchmethod1} (using only the \textsc{Off} position), the middle panels those for when both the \textsc{On} and \textsc{Off} positions were used (see Section\,\ref{subsec:pswitchmethod2}), and in the lower panels the classical calibration scheme was applied (Section\,\ref{subsec:pswitchclassicalmethod}). The bin width was the same for each histogram, to enable the amplitudes (normalised number counts) to be directly compared. We fitted a Gaussian to each distribution of errors, the parameters (mean value $\mu$, width $\sigma$) of which are stated in the plots. Using the proposed calibration methods to account for frequency-dependent system temperatures results in an \textit{unbiased} estimate of the true flux values. Interestingly, the classical method has a similar scatter ($\sigma$) in the derived flux values. Nonetheless, intensities are \textit{systematically} wrong in an unpredictable way. In the absence of SW, the two more sophisticated methods provide similar results. This argues in favour of the first method relying only on one fitting model, while the second requires two. In the presence of SW, the second method is still unbiased but has a somewhat larger scatter. We conclude that the calibration scheme described in Section\,\ref{subsec:pswitchmethod1} is optimal for position switching measurements.

\subsection{Frequency switching}\label{subsec:simulationfswitch}

\begin{figure*}[!tp]
   \centering%
   \includegraphics[width=0.4\textwidth,bb=14 24 521 392,clip=]{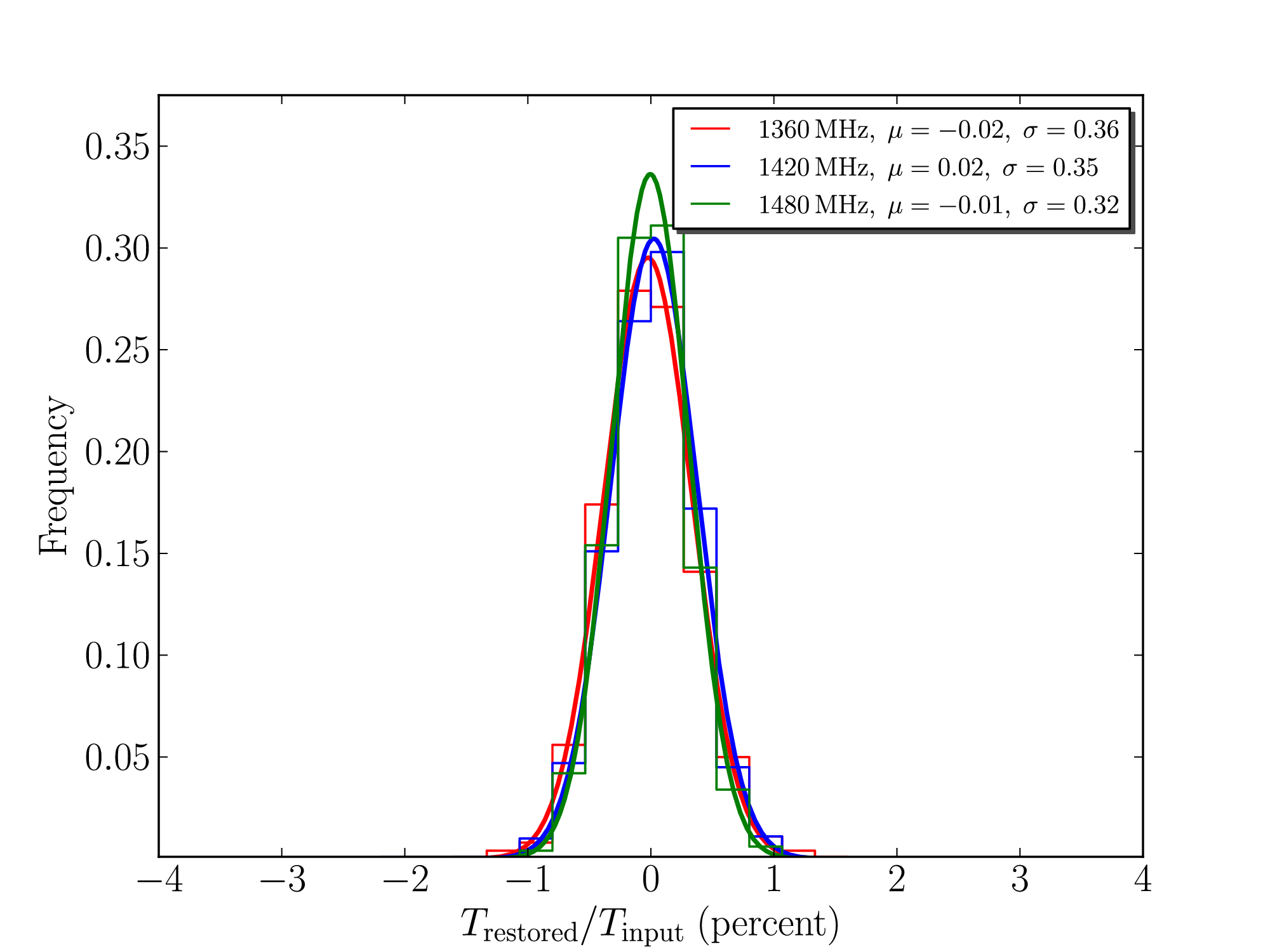}\quad
   \includegraphics[width=0.4\textwidth,bb=14 24 521 392,clip=]{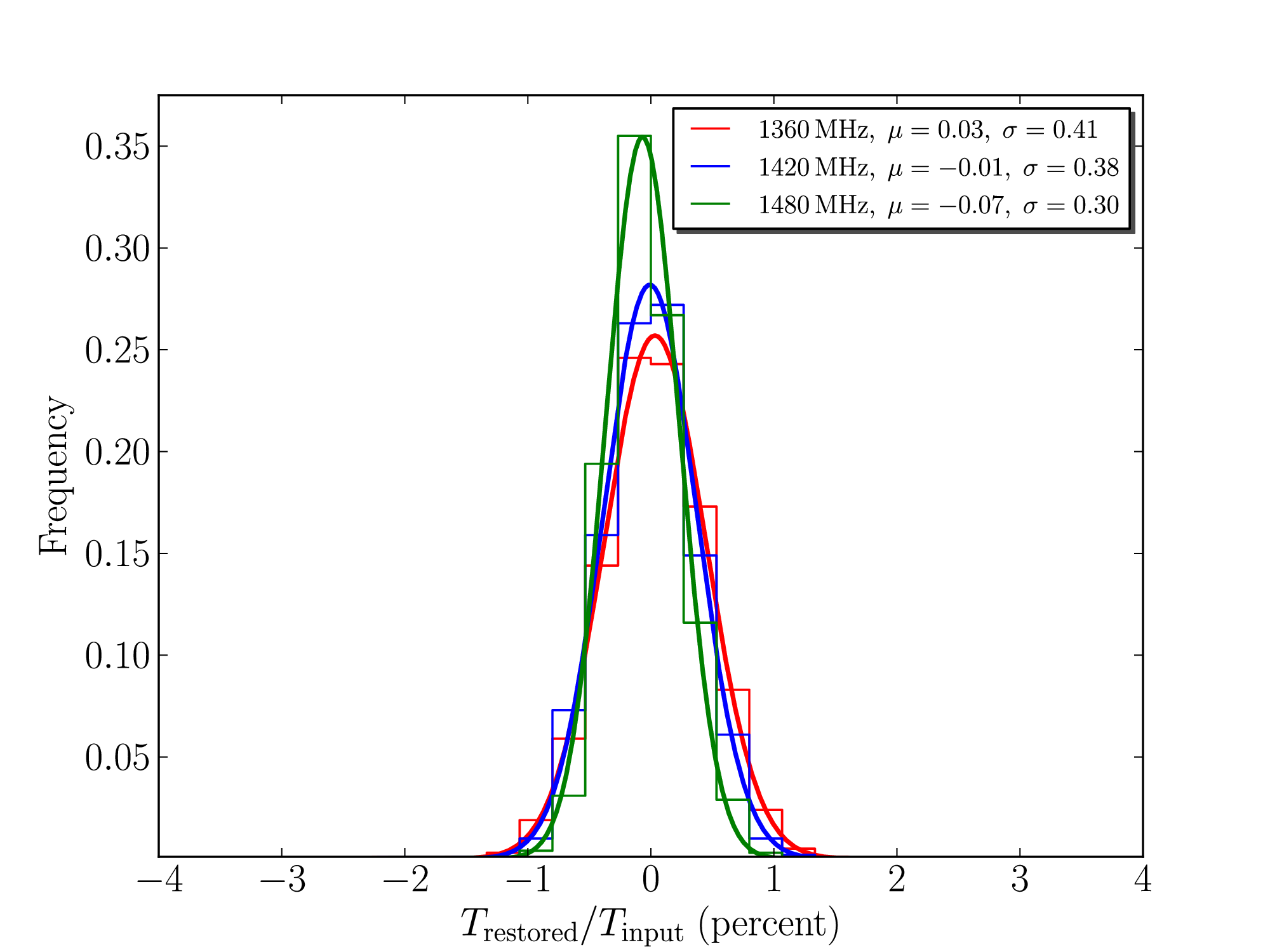}\\[0ex]
   \includegraphics[width=0.4\textwidth,bb=14 24 521 392,clip=]{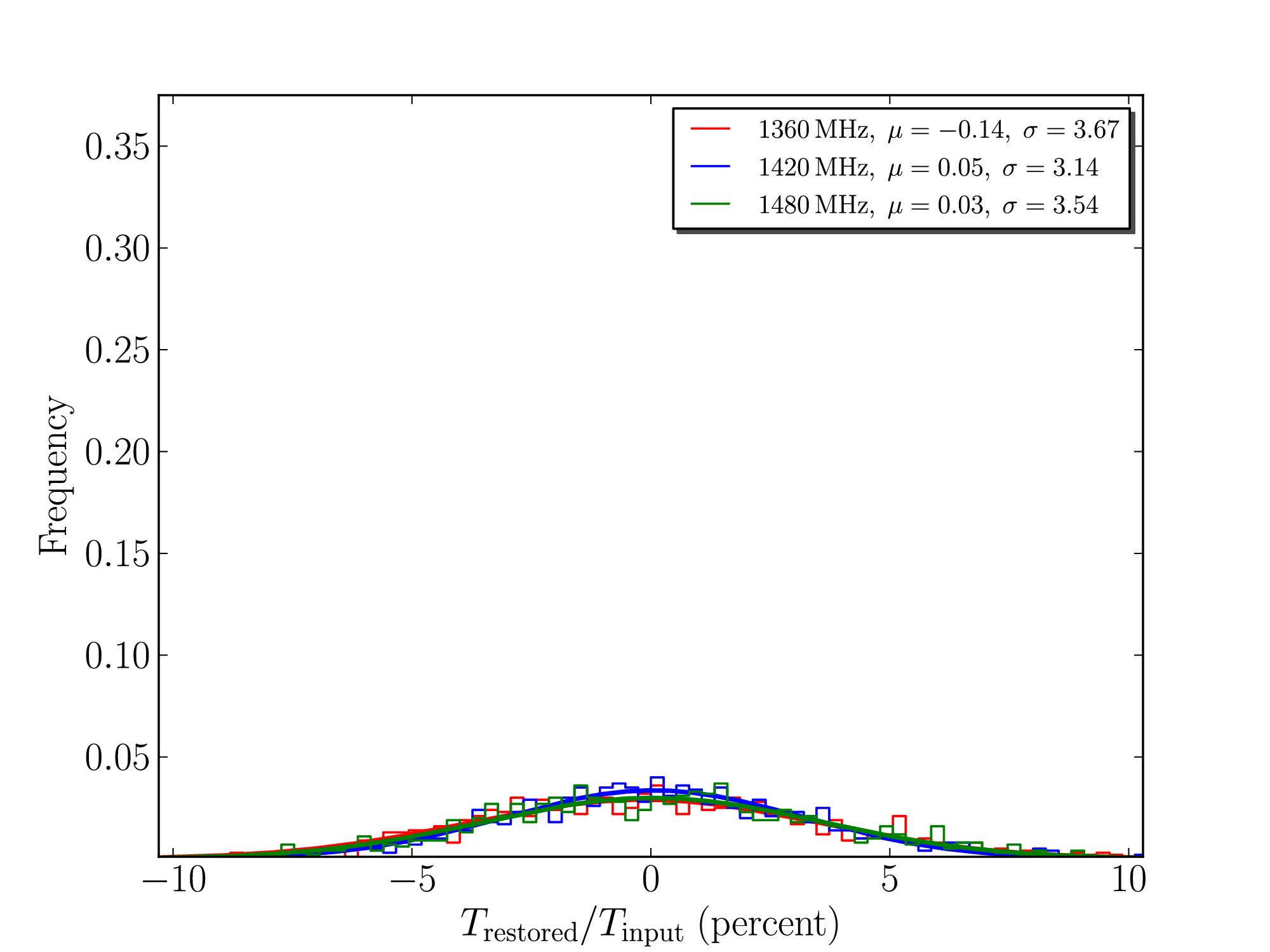}\quad
   \includegraphics[width=0.4\textwidth,bb=14 24 521 392,clip=]{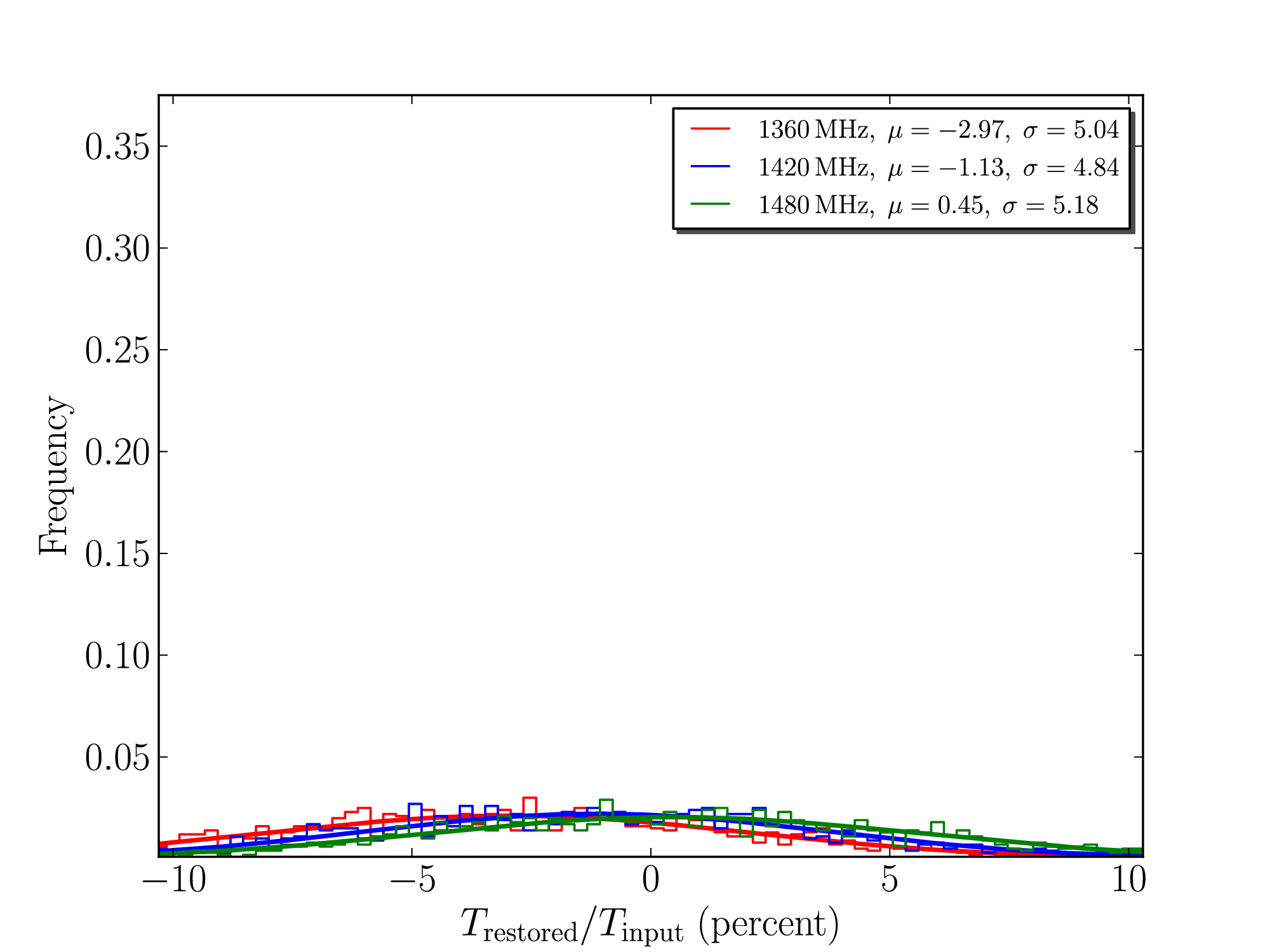}\\[0ex]
   \includegraphics[width=0.4\textwidth,bb=14 1 521 392,clip=]{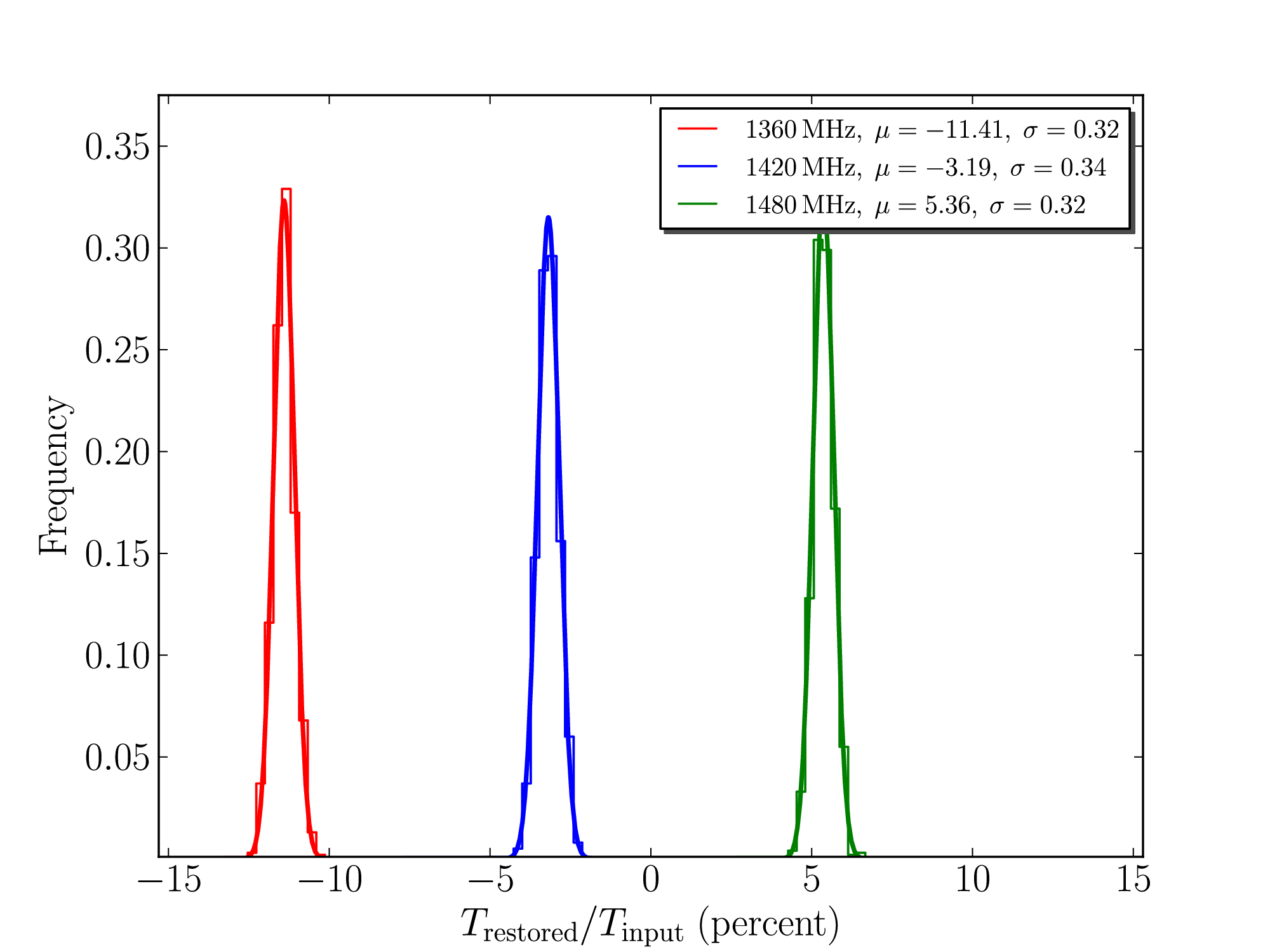}\quad
   \includegraphics[width=0.4\textwidth,bb=14 1 521 392,clip=]{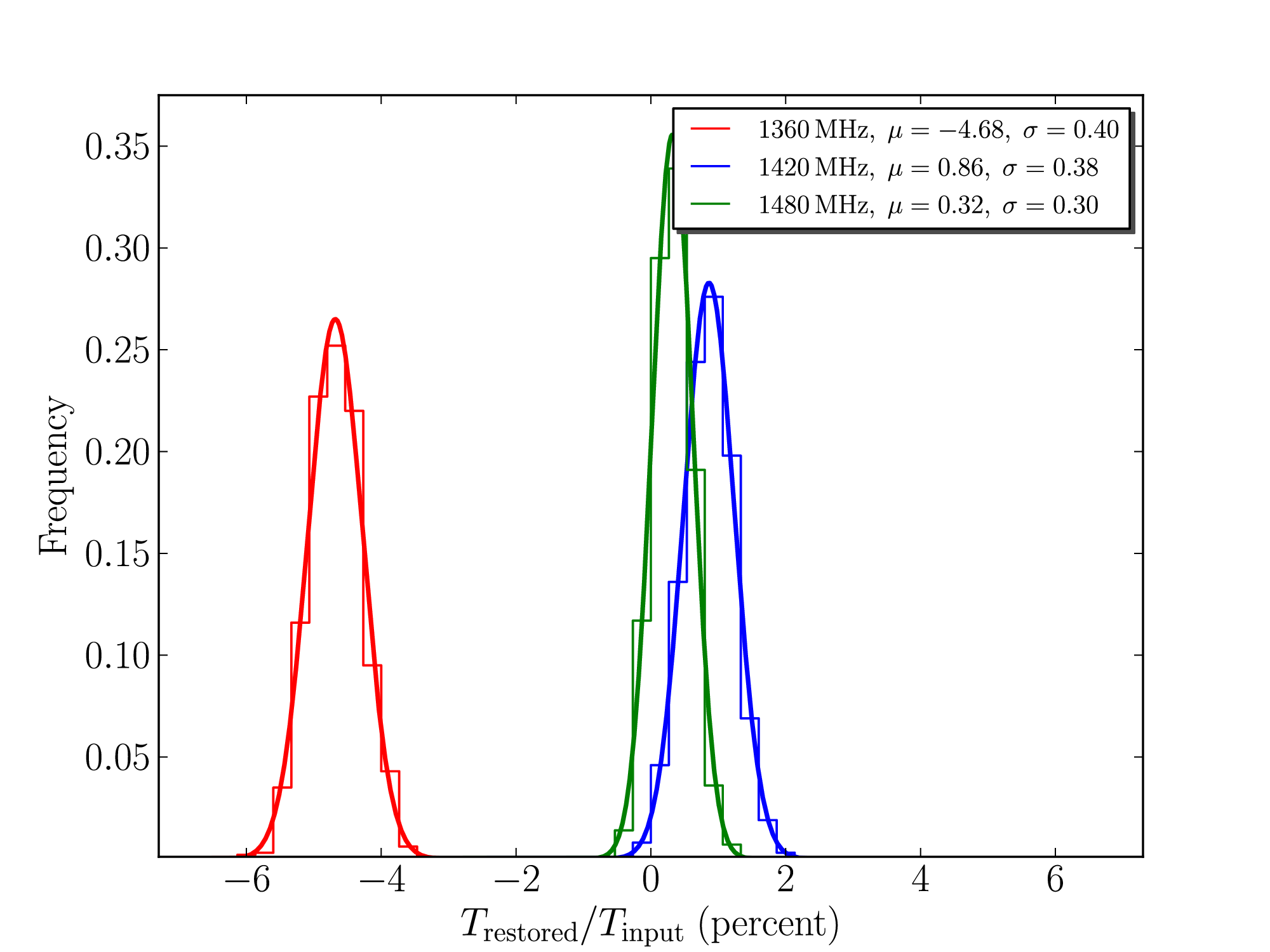}
  \caption{As in Fig.\,\ref{fig:pswitcherrorhistograms}, the plots show the error distribution for a thousand simulations after applying frequency switching. For the first method, Section\,\ref{subsec:fswitchmethod1}, the resulting histograms are somewhat smaller, which can be attributed to the slightly smaller residual baseline RMS ($\sim80\,\mathrm{mK}$ for frequency switching, $\sim120\,\mathrm{mK}$ for position switching) in our simulations. The situation is much worse when using the second calibration scheme, Section\,\ref{subsec:fswitchmethod2}. The distributions are broader, even more so in the SW case, owing to the relatively small difference between $f$ and $f^\mathrm{cal}$.}%
   \label{fig:fswitcherrorhistograms}%
\end{figure*}

As above, Fig.\,\ref{fig:fswitcherrorhistograms} illustrates the results for the three different calibration methods for frequency switching.  The proposed calibration schemes provide \textit{unbiased} intensity estimates, while the classical method is affected by systematic errors. Unfortunately, the two-model approach (middle panels) results in a much higher spread around the true value than for the position switching simulations, as the difference between $f$ and $f^\mathrm{cal}$ is smaller --- a direct consequence of frequency switching. As for position switching, the first method (see Section\,\ref{subsec:fswitchmethod1}) is generally more suitable, as the associated error distribution is narrower. Nevertheless, the second method, Section\,\ref{subsec:fswitchmethod2}, might be considered if baselines become a problem, as the latter approach is less affected by the standing wave shape in the final spectra.

We note that we use the same noise contribution in both the position and frequency switching methods. Owing to the in-band frequency switching, the resulting noise in the final spectrum is smaller by a factor of $\sqrt{2}$. Whether the lower level of noise can compensate for the potentially lower quality of the frequency switching results is a decision that has to be made by the observer, because it depends on the chosen instrumental set-up (e.g. receiver, SW) and observing scheme.

\section{Measuring $T_\mathrm{cal}$}\label{sec:tcalmeasurement}
The reliability of the calibration in all the methods that we have described strongly relies on accurate knowledge of the noise diode's spectral density. Consequently, some effort should be invested in the measurement of $T_\mathrm{cal}$.

\subsection{Using hot--cold calibration}\label{subsec:hotcoldtheory}
The only means of determining $T_\mathrm{cal}$ directly is to apply a hot--cold calibration. This is achieved by pointing the feed horn to an absorber at ambient temperature and subsequently to a tank containing liquid nitrogen at 77\,K. The experiment is time critical as after a while the feed itself  starts to physically react to the low temperatures causing instabilities in the system.

Since the temperatures of the hot and cold phases are known, one can easily compute the transfer function (i.e. bandpass curve and calibration factor) that converts the input spectra (in K) to spectrometer output (in arbitrary units)
\begin{equation}
G(\nu)=\frac{P_\mathrm{hot}^\mathrm{[cal]}-P_\mathrm{cold}^\mathrm{[cal]}}{T_\mathrm{hot}^\mathrm{[cal]}-T_\mathrm{cold}^\mathrm{[cal]}}\label{eqhotcold}, 
\end{equation}
using $T_\mathrm{hot}=300\,\mathrm{K}$ and $T_\mathrm{cold}=77\,\mathrm{K}$. Dividing the input spectra $P$ by $G$, one obtains calibrated data. Finally, after calibrating the spectra we can compute the contribution of the noise diode $T_\mathrm{cal}$ by subtracting the \textit{non-cal} from the \textit{cal} phase
\begin{equation}
T_\mathrm{cal}=T_\mathrm{hot,cold}^\mathrm{cal}-T_\mathrm{hot,cold}=\frac{P_\mathrm{hot,cold}^\mathrm{cal}-P_\mathrm{hot,cold}}{G}\label{eqtcal}\,. 
\end{equation}
In Appendix\,\ref{subsec:tcalhotcold}, we provide an example of a hot--cold calibration measurement of the  6.5-mm receiver at the 100-m telescope Effelsberg.

Since a small error in the determination of $T_\mathrm{cal}$ has a large impact on the later data reduction, hot--cold measurements should be performed with great care. There are some potential issues, which we briefly discuss in the following. 

Firstly, the hot load at room temperature, i.e. $\sim 300\,\mathrm{K}$ emits a much larger amount of power than the sources usually observed. This difference of 20\,dB should not be a problem for any modern receiver and backend\footnote{At least for the FPGA-based backends that are in use at most major facilities today. Older devices, such as auto-correlators with their low input quantisation, might cause problems.}. However, it is important to level the system accordingly, otherwise the amplifiers might be over-driven on the hot load leading to compression of the signal.  

Feed horns in the primary focus of a telescope have a rather large beam width because they need to illuminate the antenna dish. As a consequence, they have large spill-over angles, which require the hot and cold load to be large enough such that all side lobes of the feed also point towards the load. To achieve this, one should place the load, usually an absorber, in a metallic tank and plunge the feed into this as much as possible. For the hot load, side lobes play a lesser role since the environment is usually at the same (ambient) temperature as the absorber, e.g. when the receiver is in the lab or in the focus cabin of the 100-m telescope. However, if the receiver is not in a closed environment, radiation from the sky may be reflected and enter the horn, changing the net radiative temperature of the load. Likewise for the cold load, radiation from the environment must be avoided to as great an extent as possible. In practice, the radiation temperature of the cold load is up to 2\,K or 3\,K higher than the nominal 77\,K \citep[which itself depends on the atmospheric pressure,][]{ulich80} owing to this leakage. 

The cold load is tricky to handle. The absorber must be completely immersed in the liquid nitrogen, the amount of which can be a logistical issue at a remote site or for large receivers, e.g. multi-horn systems. Furthermore, both the vaporisation of nitrogen, and frost that can potentially form on the feed enforces strict time constraints. Frost reflects part of the incident radiation and changes the properties of the feed. If several hot--cold measurements are desired for one receiver, e.g. when the receiver has several sub-bands, one should switch between hot and cold regularly while ensuring a constant height above the absorber.  

The absorption material itself must fit the observed frequency. For low frequencies, below several $100\,\mathrm{MHz}$, it can be a problem to even find an efficient absorber, which should also have a good heat conductivity in order to reach the temperature of liquid nitrogen sufficiently quickly.

Unfortunately, calibration diodes are known to drift with time.  It is therefore essential to repeat hot--cold measurements at least every few months if high precision is desired.

\subsection{Using continuum sources with well-known flux}\label{subsec:contcal}
Another way of obtaining a good estimate of the $T_\mathrm{cal}$ spectrum is to utilise strong continuum sources with a well-known flux density. For example, using position switching one can easily invert Eq.\,(\ref{eqpswitchbaseusingkappa}) 
\begin{equation}
T_\mathrm{cal}=\frac{T_\mathrm{sou}}{\kappa_\mathrm{off}\frac{P_\mathrm{on}-P_\mathrm{off}}{P_\mathrm{off}}}=\frac{T_\mathrm{sou}}{(\kappa_\mathrm{off}+1)\frac{P_\mathrm{on}^\mathrm{cal}-P_\mathrm{off}^\mathrm{cal}}{P_\mathrm{off}^\mathrm{cal}}}\label{eqtcalfromcont}\,.
\end{equation}
Depending on the S/N of the recorded spectra, it may not even be necessary to use a model of $\kappa_\mathrm{off}$. This of course increases the noise in the $T_\mathrm{cal}$ measurement somewhat, but has the advantage that the result is independent of the choice of a specific model. We present an example measurement using the 6-cm secondary-focus receiver at the 100-m in Appendix\,\ref{subsec:tcalcontinuum}.

The true brightness temperature, $T_\mathrm{sou}^\ast=\eta_\mathrm{ap}^{-1}T_\mathrm{sou}\exp\left(\tau_0 \mathrm{AM}\right)$, (or rather the associated flux density $S=2k_\mathrm{B} T_\mathrm{sou}^\ast/A$, where $A$ is the physical area of the dish) is usually tabulated in a continuum flux catalogue, which is of course not the measured antenna temperature $T_\mathrm{sou}$. As a consequence, one not only needs to take into account atmospheric dampening but also the aperture efficiency, $\eta_\mathrm{ap}$. To determine $\eta_\mathrm{ap}$ for a moderately-sized or large single dish telescope, one has no choice but to measure it using astronomical (point-like) sources. This is usually done using a hot--cold measurement (to absolutely calibrate $T_\mathrm{cal}$, which yields $T_\mathrm{sou}$ after application of Eq.\,(\ref{eqtcalfromcont})) and --- in a timely manner --- relate it to the known flux density of a calibration source. Thus, how is the latter determined in the first place? The brightest radio sources in the sky have been observed with well-calibrated horn antennae (or a dipole arrays), the response of which can be theoretically calculated or precisely measured (since they are smaller). This served to establish the (frequency-dependent) flux values for sources such as Cas\,A, Vir\,A, or Cyg\,A. From these primary calibrators, flux scales for additional sources were derived using relative measurements. For a detailed discussion, we refer to the literature\citep[e.g.,][and references there-in]{findlay66,baars77}. It is clear that if the use of continuum flux catalogues is desired, the $\eta_\mathrm{ap}$ values should be regularly updated, using hot--cold calibration. 

Furthermore, to use a continuum source for a measurement of $T_\mathrm{cal}$ the condition $\Delta T_\mathrm{sys}=0$ must be fulfilled (see Eq.\,(\ref{eqdeltatsys})). In practice, this can be an issue, for example the SW contribution can depend on the incident continuum radiation \citep{winkel11techreport}.  The observer should at least ensure that the \textsc{On} and \textsc{Off} measurements track the same elevation angle(s), such that the amount of spill-over (ground radiation) and the atmospheric contribution are similar.

\section{Summary}\label{sec:summary}
When dealing with position- or frequency-switched observations, we have shown that in many cases it is of utmost importance to account for the frequency dependence of the system temperature and noise calibration signal in the flux calibration task. Simplification of the problem, i.e., treating $T_\mathrm{sys}$ and $T_\mathrm{cal}$ as constants, leads to a systematic calibration bias, which can have a strong impact on the scientific results from reduced data. We have proposed several methods to circumvent this bias at the cost of a greater complexity in the data reduction.

The conclusions that we can draw from this paper include:
\begin{enumerate}
\item In general, position switching provides much more accurate results than frequency switching because it is less affected by a variety of potential issues, such as standing waves. If the system temperature and/or calibration signal have strongly fluctuating spectral features on similar or smaller scales than the astronomical signal that is to be observed, frequency switching is not applicable at all, because of the spectral-line ghost issue. \textit{In these cases, least squares frequency switching might well be worth the higher technical and computational effort.}  Unfortunately, 50\% of the observing time is effectively lost when using position switching, such that the final choice between the two techniques is reduced to the question of whether a longer integration time on-source compensates for the potential problems. The answer to this question is of course different for every receiving system and observing condition.  For non-constant $G_\mathrm{RF}(\nu)$, one cannot achieve a correct calibration using frequency switching, but we have shown how a combination of position and frequency switching can be applied in these cases.

\item Another significant advantage of position switching is the ability to reconstruct the continuum flux of the source. This also makes it possible to use known continuum calibrators to determine the $T_\mathrm{cal}$ spectrum, which is essential to track time-dependent drifts.

\item The presence of standing waves make sophisticated calibration more difficult, as the quality of a non-linear least squares fitting involving sine-functions relies heavily on the initial model parameters. This problem is even more pronounced for the second approach, which makes use of two independent functions (see Sections\,\ref{subsec:pswitchmethod2} and \ref{subsec:fswitchmethod2}). The use of signal filters, instead of parametric models, to suppress noise, e.g. in $\kappa^{-1}$, is a solution to this problem but at the cost of introducing an additional (correlated) small noise component.

\item The $T_\mathrm{cal}$ intensity should be sufficiently high. All proposed methods utilise the difference (and its inverse) between the spectra that contain the noise diode's signal and those that do not. A higher S/N helps us to avoid numerical instabilities. If $T_\mathrm{cal}$ is too large, the average system temperature is of course much higher than necessary. Usually $T_\mathrm{cal}/T_\mathrm{sys}\approx20\%$ is a suitable value.

\item Although under certain circumstances, the sophisticated methods presented are not as robust as the classical method, e.g. in the presence of standing waves, they allow for an unbiased calibration that is of utmost importance to all scientific investigations.
\end{enumerate}

\begin{acknowledgements}
Our results are based on observations with the 100-m telescope of the MPIfR (Max-Planck-Institut f\"ur Radioastronomie) at Effelsberg. We sincerely thank the referee, Ron Maddalena, for his numerous and valuable suggestions (especially to combine position and frequency switching, and to apply a weighting scheme in Appendix\,\ref{sec:appendixnoise}) and comments, which helped to greatly improve our work. We are grateful to Axel Jessner, Charlotte Sobey, and Lars Fl\"{o}er for careful reading and editing of the manuscript. We would also like to thank the Effelsberg staff for their continuous support.
\end{acknowledgements}

\bibliographystyle{aa}
\bibliography{references}

\begin{thebibliography}{14}
\expandafter\ifx\csname natexlab\endcsname\relax\def\natexlab#1{#1}\fi

\bibitem[{{Baars} {et~al.}(1977){Baars}, {Genzel}, {Pauliny-Toth}, \&
  {Witzel}}]{baars77}
{Baars}, J.~W.~M., {Genzel}, R., {Pauliny-Toth}, I.~I.~K., \& {Witzel}, A.
  1977, \aap, 61, 99

\bibitem[{{Braatz}(2009)}]{gbtidl}
{Braatz}, J. 2009, Calibration of GBT Spectral Line Data in GBTIDL v2.1, Tech.
  rep., National Radio Astronomy Observatory

\bibitem[{{Findlay}(1966)}]{findlay66}
{Findlay}, J.~W. 1966, \araa, 4, 77

\bibitem[{{Heiles}(2007)}]{heiles07}
{Heiles}, C. 2007, \pasp, 119, 643

\bibitem[{{Kraus}(2009)}]{kraus09techreport}
{Kraus}, A. 2009, Calibration of the Effelsberg 100m telescope, Tech. rep.,
  Max-Planck-Institut f\"{u}r Radioastronomie

\bibitem[{{Maddalena} \& {Johnson}(2005)}]{maddalena05}
{Maddalena}, R.~J. \& {Johnson}, C.~H. 2005, in Bulletin of the American
  Astronomical Society, Vol.~37, American Astronomical Society Meeting
  Abstracts, 173.02

\bibitem[{{Pardo} {et~al.}(2006){Pardo}, {Serabyn}, {Cernicharo}, \&
  {Wiedner}}]{pardo06}
{Pardo}, J.~R., {Serabyn}, E., {Cernicharo}, J., \& {Wiedner}, M.~C. 2006, in
  Seventeenth International Symposium on Space Terahertz Technology, ed.
  {A.~Hedden, M.~Reese, D.~Santavicca, L.~Frunzio, D.~Prober, P.~Piitz,
  C.~Groppi, \& C.~Walker}, 291--293

\bibitem[{{Rohlfs} \& {Wilson}(2004)}]{rohlfs04}
{Rohlfs}, K. \& {Wilson}, T.~L. 2004, {Tools of radio astronomy}, ed. {Rohlfs,
  K.~\& Wilson, T.~L.}

\bibitem[{{Rottmann} \& {Roy}(2007)}]{wvrtechreport}
{Rottmann}, H. \& {Roy}, A. 2007, MPIfR Water Vapor Radiometer. Scientific
  Evaluation, Tech. rep., Max-Planck-Institut f\"{u}r Radioastronomie

\bibitem[{{Ulich} {et~al.}(1980){Ulich}, {Davis}, {Rhodes}, \&
  {Hollis}}]{ulich80}
{Ulich}, B.~L., {Davis}, J.~H., {Rhodes}, P.~J., \& {Hollis}, J.~M. 1980, IEEE
  Transactions on Antennas and Propagation, 28, 367

\bibitem[{{Wiener}(1949)}]{wiener49}
{Wiener}, N. 1949, {Extrapolation, interpolation, and smoothing of stationary
  time series: with engineering applications}, ed. {Wiener, N.}

\bibitem[{{Winkel} {et~al.}(2011){Winkel}, {Fl\"{o}er}, \&
  {Kerp}}]{winkel11techreport}
{Winkel}, B., {Fl\"{o}er}, L., \& {Kerp}, J. 2011, EBHIS Technical Report,
  Tech. rep., Max-Planck-Institut f\"{u}r Radioastronomie

\bibitem[{{Winkel} {et~al.}(2010){Winkel}, {Kalberla}, {Kerp}, \&
  {Fl{\"o}er}}]{winkel10}
{Winkel}, B., {Kalberla}, P.~M.~W., {Kerp}, J., \& {Fl{\"o}er}, L. 2010, \apjs,
  188, 488

\bibitem[{{Winkel} \& {Kerp}(2007)}]{winkel07b}
{Winkel}, B. \& {Kerp}, J. 2007, \apjs, 173, 166

\end{thebibliography}

\appendix

\section{Theoretical noise levels for position and frequency switching}\label{sec:appendixnoise}
\subsection{Position switching}\label{subsec:appendixpswitch}

Using Eq.\,(\ref{eqpswitchbase}), we can calculate the theoretical noise level, $T_\mathrm{rms}$, of the \textit{cal} and \textit{non-cal} phases. Assuming that the system temperature can be modelled (e.g., using one of our methods)  and hence is noise-free, we can rewrite Eq.\,(\ref{eqpswitchbase})
\begin{align}
T_\mathrm{sou}+\Delta T_\mathrm{sys}&=(T_\mathrm{sys,off}^\mathrm{[cal]})_\mathrm{model}\frac{P_\mathrm{on}^\mathrm{[cal]}-P_\mathrm{off}^\mathrm{[cal]}}{P_\mathrm{off}^\mathrm{[cal]}}\nonumber\\
&=(T_\mathrm{sys,off}^\mathrm{[cal]})_\mathrm{model}\left[\frac{T_\mathrm{on}^\mathrm{[cal]}}{T_\mathrm{off}^\mathrm{[cal]}}-1\right],
\end{align}
where we have used the shorter notation $T_\mathrm{on}^\mathrm{[cal]}\equiv T_\mathrm{sou}^\mathrm{[cal]}+T_\mathrm{sys,on}^\mathrm{[cal]}$ and $T_\mathrm{off}^\mathrm{[cal]}\equiv T_\mathrm{sys,off}^\mathrm{[cal]}$. The latter two quantities are affected by noise according to the radiometer equation (see Eq.\,(\ref{eqradiometer})). The resulting RMS is then
\begin{align}
T_\mathrm{rms}^\mathrm{[cal]}&=(T_\mathrm{sys,off}^\mathrm{[cal]})_\mathrm{model}\sqrt{ \frac{ (T^\mathrm{[cal]}_\mathrm{rms,on})^2}{(T^\mathrm{[cal]}_\mathrm{off})^2} + \frac{  (T^\mathrm{[cal]}_\mathrm{rms,off})^2 (T^\mathrm{[cal]}_\mathrm{on})^2}{(T^\mathrm{[cal]}_\mathrm{off})^4}}\nonumber\\
&=\sqrt{  ( T^\mathrm{[cal]}_\mathrm{rms,on})^2 + ( T^\mathrm{[cal]}_\mathrm{rms,off})^2\frac{  (T^\mathrm{[cal]}_\mathrm{on})^2}{(T^\mathrm{[cal]}_\mathrm{off})^2}}\label{eqpswitchsinglerms},
\end{align}
which is at least a factor of $\sqrt{2}$ larger than the noise in each individual phase (if the continuum of the source is negligible). Likewise, the S/N is a factor of $\sqrt{2}$ smaller because in position switching two raw spectra contribute to the noise power but only one of them contains a signal. This is the trade-off one has to pay when applying a switching technique. The final noise level after averaging both phases, $T_\mathrm{aver}=(T_\mathrm{sou}+T_\mathrm{sou}^\mathrm{cal})/2$, is then
\begin{equation}
T_\mathrm{rms,aver}= \frac{1}{2} \sqrt{ (T_\mathrm{rms})^2 + (T_\mathrm{rms}^\mathrm{cal})^2 }\overset{T_\mathrm{sys,off}\gg T^\mathrm{cal} }{\longrightarrow} \frac{T_\mathrm{rms}}{\sqrt{2}}\label{eqpswitchrms},
\end{equation}
which is roughly a decrease by $\sqrt{2}$ relative to each of the \textit{cal}/\textit{non-cal} phases individually. This implies that the final RMS is only about the same level as the (mean) noise in the individual phases! As discussed in Section\,\ref{subsec:pswitchmethod1}, one possible way of improving on the final noise would be to smooth the reference spectrum prior to division, when only the noise of the \textit{on} phase would then propagate into the final result.

There might be cases, where $T_\mathrm{sys,off} \ngg T^\mathrm{cal}$ and Eq.\,(\ref{eqpswitchrms}) is dominated by $T_\mathrm{rms}^\mathrm{cal}$ increasing the noise in the averaged spectrum. The following weighting scheme could then be used to improve the situation
\begin{equation}
T_\mathrm{aver}=\frac{w T_\mathrm{sou}+w^\mathrm{cal}T_\mathrm{sou}^\mathrm{cal}}{w+w^\mathrm{cal}},
\end{equation}
where $(w^\mathrm{[cal]})^{-1}\propto \textrm{var}[T^\mathrm{[cal]}]=(T_\mathrm{rms}^\mathrm{[cal]})^2$. One can easily infer that
\begin{equation}
\left(\frac{1}{T_\mathrm{aver,rms}}\right)^2=\left(\frac{1}{T_\mathrm{rms}}\right)^2+\left(\frac{1}{T^\mathrm{cal}_\mathrm{rms}}\right)^2,
\end{equation}
which leads to smaller noise values in the averaged spectrum than Eq.\,(\ref{eqpswitchrms}). In the limit $T_\mathrm{sys,off}\gg T^\mathrm{cal} $, both equations return the same result.

A practical problem, however, is to estimate the noise spectrum $T_\mathrm{rms}^\mathrm{[cal]}(\nu)$ needed to determine the weighting spectra $w^\mathrm{[cal]}(\nu)$ with sufficient precision, which can be complicated in the presence of either emission lines or radio frequency interference.

\subsection{Frequency switching}
As in the case of position switching, we modify Eq.\,(\ref{eqfswitchbaseeq1}) and \,(\ref{eqfswitchbaseeq2})
\begin{align}
\tilde T_\mathrm{sig}^\mathrm{[cal]} &=\left(T_\mathrm{sou,+}+T_\mathrm{sys,+}^\mathrm{[cal]}\right)_\mathrm{model}\frac{P_\mathrm{sig}^\mathrm{[cal]} - P_\mathrm{ref}^\mathrm{[cal]}}{ P_\mathrm{ref}^\mathrm{[cal]}}\nonumber\\
&=\left(T_\mathrm{sou,+}^\mathrm{cont}+T_\mathrm{sys,+}^\mathrm{[cal]}\right)_\mathrm{model}\left[\frac{T_\mathrm{sig}^\mathrm{[cal]}}{T_\mathrm{ref}^\mathrm{[cal]}}-1\right],\\
\tilde T_\mathrm{ref}^\mathrm{[cal]}&=\left(T_\mathrm{sou,-}+T_\mathrm{sys,-}^\mathrm{[cal]}\right)_\mathrm{model}\frac{P_\mathrm{ref}^\mathrm{[cal]} - P_\mathrm{sig}^\mathrm{[cal]}}{ P_\mathrm{sig}^\mathrm{[cal]}}\nonumber\\
&=\left(T_\mathrm{sou,-}^\mathrm{cont}+T_\mathrm{sys,-}^\mathrm{[cal]}\right)_\mathrm{model}\left[\frac{T_\mathrm{ref}^\mathrm{[cal]}}{T_\mathrm{sig}^\mathrm{[cal]}}-1\right]
\end{align}
where $T_\mathrm{sig}^\mathrm{[cal]}\equiv T_\mathrm{sou,-}+T_\mathrm{sys,-}^\mathrm{[cal]}$ and  $T_\mathrm{ref}^\mathrm{[cal]}\equiv T_\mathrm{sou,+}+T_\mathrm{sys,+}^\mathrm{[cal]}$. We note, that for the modelled (effective) system temperature we assume that only continuum contributions could be inferred, as discussed in Section\,\ref{subsec:fswitchghosts}.

With
\begin{align}
T_\mathrm{sig,rms}^\mathrm{[cal]}&=\left(T_\mathrm{sou,+}^\mathrm{cont}+T_\mathrm{sys,+}^\mathrm{[cal]}\right)_\mathrm{model} \sqrt{ \frac{ ( T^\mathrm{[cal]}_\mathrm{rms,sig})^2}{(T^\mathrm{[cal]}_\mathrm{ref})^2} + \frac{ ( T^\mathrm{[cal]}_\mathrm{rms,ref})^2 (T^\mathrm{[cal]}_\mathrm{sig})^2}{(T^\mathrm{[cal]}_\mathrm{ref})^4}},\\
T_\mathrm{ref,rms}^\mathrm{[cal]}&=\left(T_\mathrm{sou,-}^\mathrm{cont}+T_\mathrm{sys,-}^\mathrm{[cal]}\right)_\mathrm{model} \sqrt{ \frac{ ( T^\mathrm{[cal]}_\mathrm{rms,ref})^2}{(T^\mathrm{[cal]}_\mathrm{sig})^2} + \frac{ ( T^\mathrm{[cal]}_\mathrm{rms,sig})^2 (T^\mathrm{[cal]}_\mathrm{ref})^2}{(T^\mathrm{[cal]}_\mathrm{sig})^4}}
\end{align}
the final result, after shift-and-averaging, is
\begin{equation}
\begin{split}
T_\mathrm{rms,aver}= &\frac{1}{4} \left[ \left(T_\mathrm{sig,rms}(\nu+\Delta \nu)\right)^2 +\left(T_\mathrm{ref,rms}(\nu-\Delta \nu)\right)^2 \right.\\
&\left. + \left(T^\mathrm{cal}_\mathrm{sig,rms}(\nu+\Delta \nu)\right)^2 + \left(T_\mathrm{ref,rms}^\mathrm{cal}(\nu-\Delta \nu)\right)^2 \right]^{-\frac{1}{2}}\label{eqfswitchrms}\,.
\end{split}
\end{equation}
This is roughly a factor of $\sqrt{2}$ lower than for position switching, which is simply because for frequency switching the integration time on-source is twice as long.

As in Section\,\ref{subsec:appendixpswitch}, one could again apply a weighting scheme, which is more suitable for cases where  $T_\mathrm{sys,off} \ngg T^\mathrm{cal} $.

\subsection{Application to the example simulations}

\begin{figure}[!t]
\centering%
\includegraphics[width=0.45\textwidth,bb=15 42 522 392,clip=]{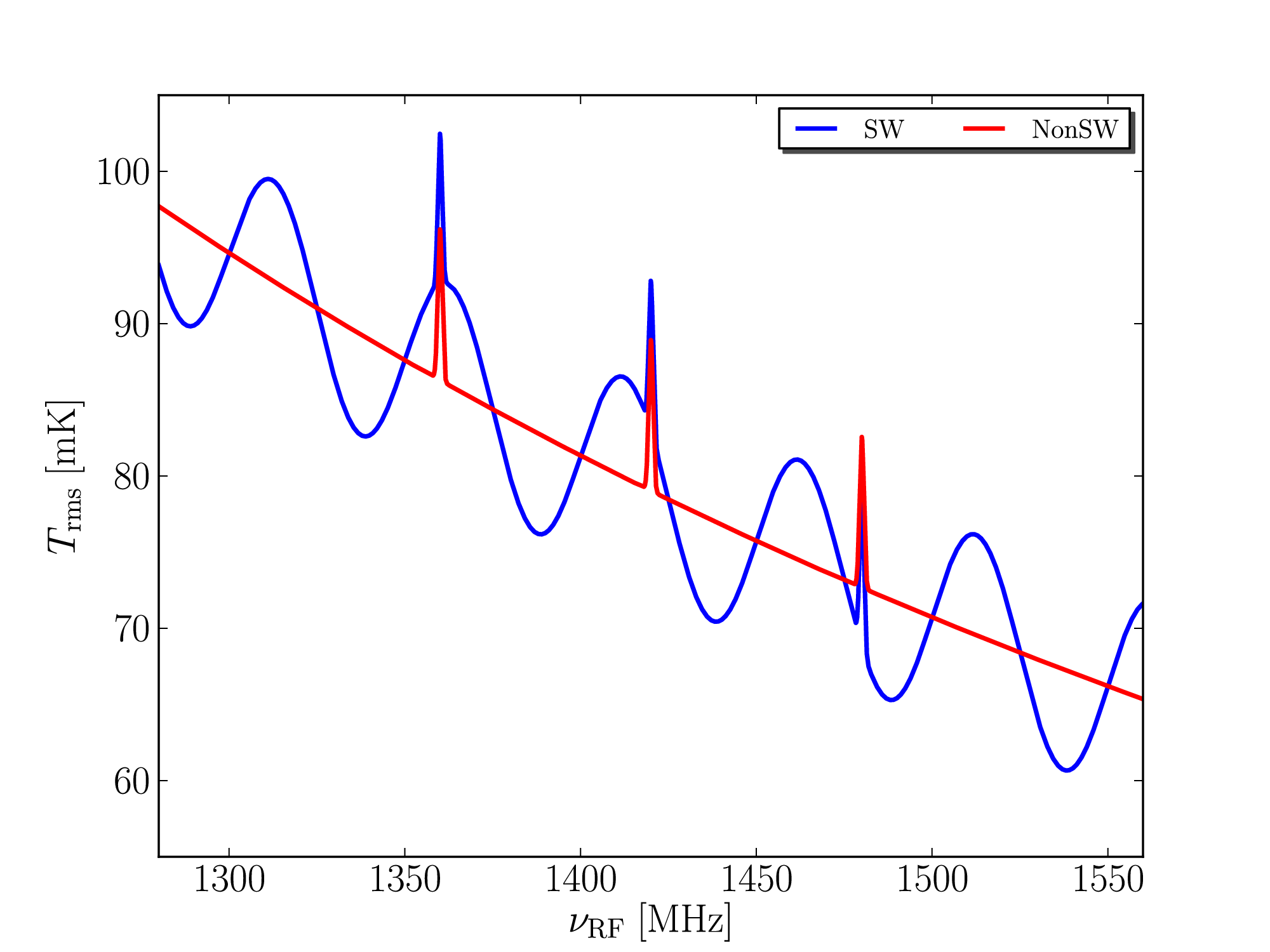}\\[0ex]
\includegraphics[width=0.45\textwidth,bb=15 1 522 392,clip=]{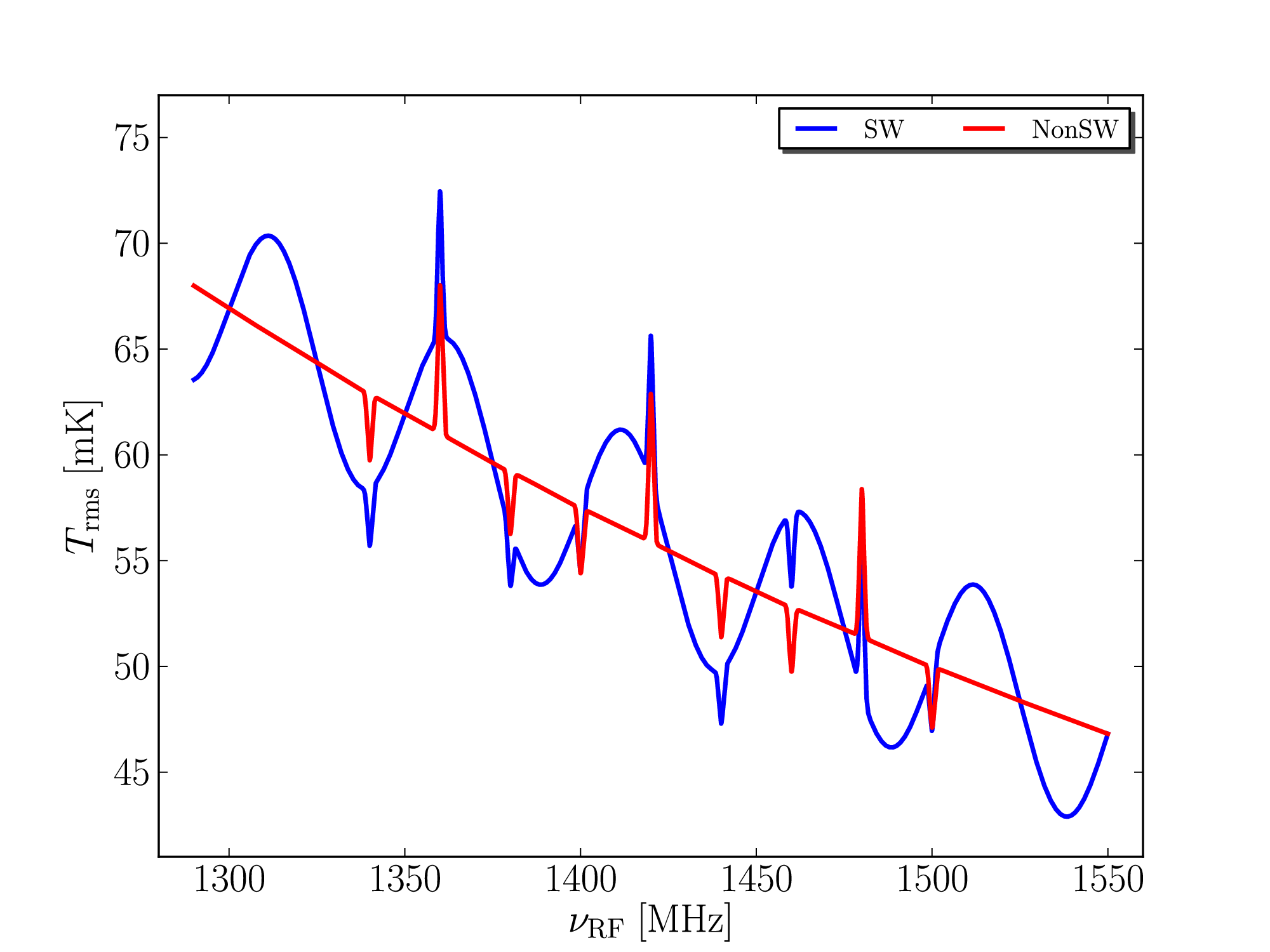}
\caption{Theoretical noise levels for the example simulations. The upper panel shows the resulting RMS for position switching, and the lower panel contains the frequency switching results, which are about a factor of $\sqrt{2}$ lower because of the longer effective integration time on-source.}%
\label{fig:theoreticalrms}%
\end{figure}

To demonstrate that our calibration methods produce the expected noise levels, we plot in Fig.\,\ref{fig:theoreticalrms} $T_\mathrm{rms,aver}$ as calculated with Eq.\,(\ref{eqpswitchrms}) and (\ref{eqfswitchrms}) using the specific numbers for the four temperature phases (which defines the noise in each phase). Comparing the theoretical values with the measured RMS levels (at two spectral positions) in the figures in Section\,\ref{sec:pswitch} and \ref{sec:fswitch}, we find that they are consistent.

\section{Examples of the measurement of $T_\mathrm{cal}$}
\subsection{Using hot--cold calibration}\label{subsec:tcalhotcold}
Here, we present measurements using the 6.5-mm receiver at Effelsberg. This test was undertaken after a hot--cold measurement using a continuum backend, such that the feed had already been exposed to the cold load for some time (about 20\,min). The receiver was tuned to 43.1\,GHz and a bandwidth of 100\,MHz was chosen. Each load was integrated for 10\,min. Furthermore, a measurement of the sky was performed afterwards (toward zenith).

\begin{figure}[!t]
\centering%
\includegraphics[width=0.45\textwidth,bb=15 42 521 392,clip=]{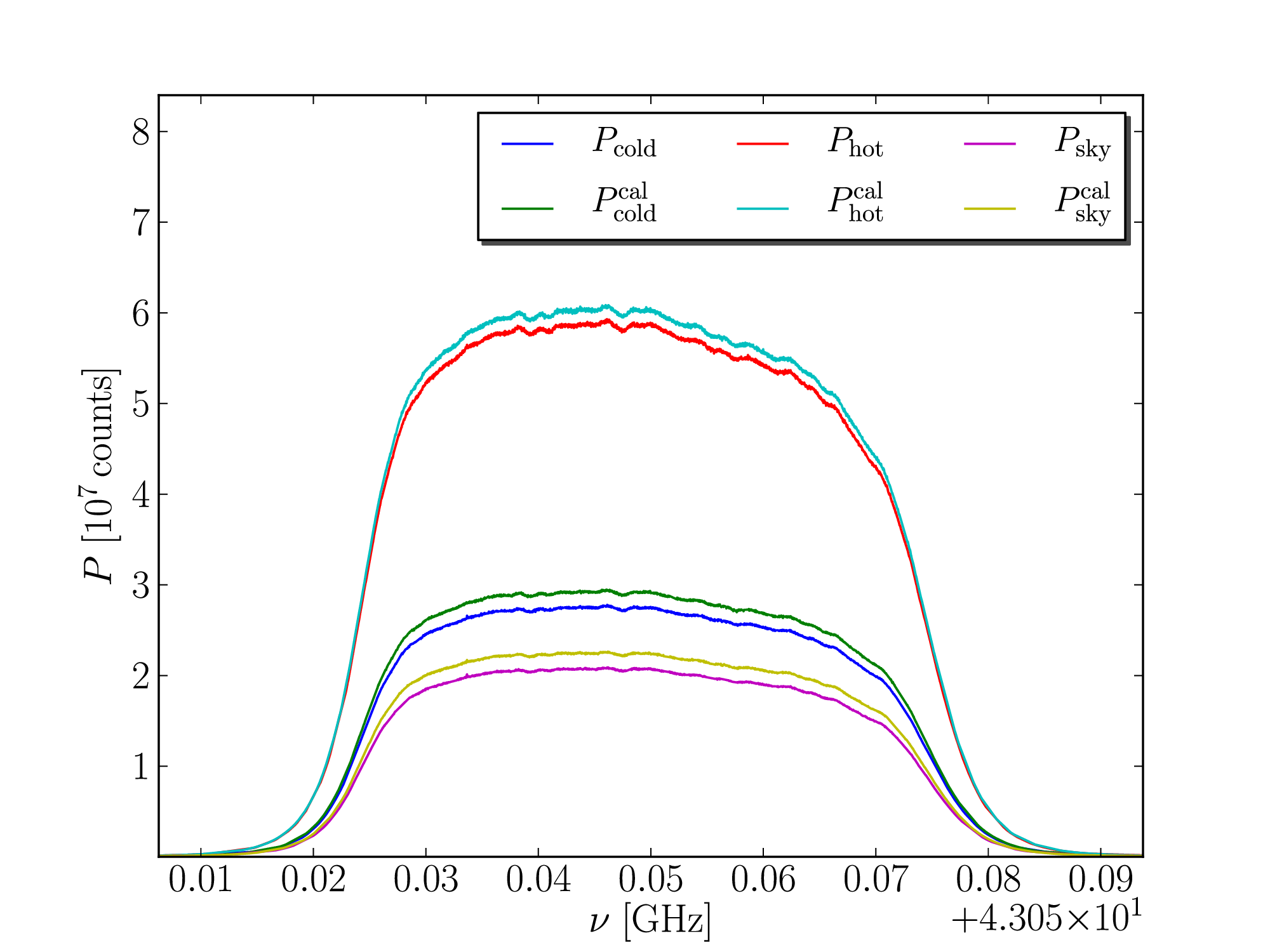}\\[0ex]
\includegraphics[width=0.45\textwidth,bb=15 1 521 392,clip=]{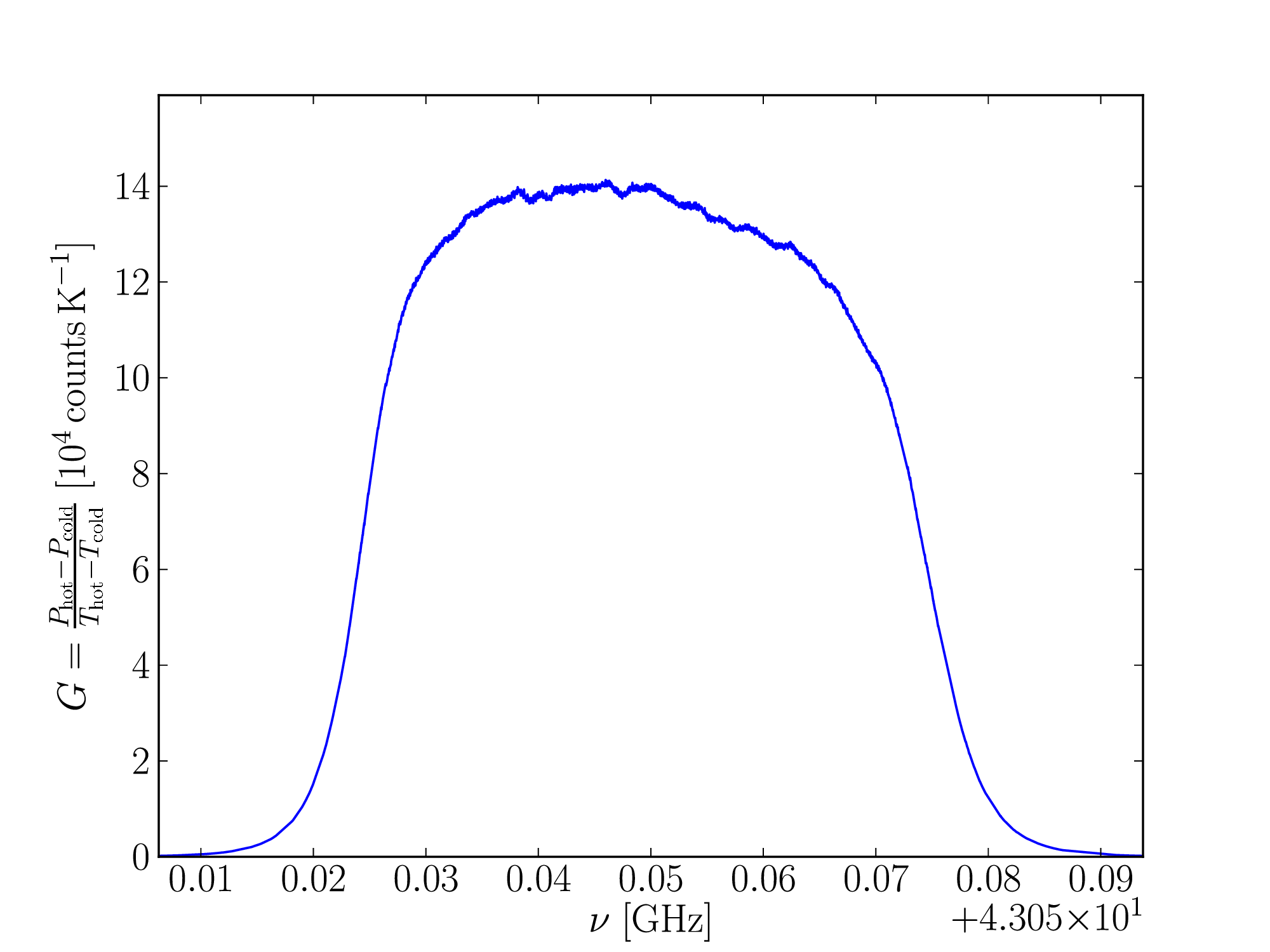}
\caption{\textbf{Upper panel:} Raw spectra measured during a hot--cold experiment with the 6.5-mm receiver in spectroscopy mode. Each spectrum is the average of a 10\,min observation. The sky position was observed afterwards by pointing the telescope to the zenith. \textbf{Lower panel:} Using the raw spectra measured for hot and cold load, one can infer the (calibrated) gain-curve, i.e. the transfer function. Residual noise is present but the overall shape and even several small features are clearly visible.}%
\label{fig:hc65mmraw}%
\end{figure}

Figure~\ref{fig:hc65mmraw} (top panel) shows the integrated raw spectra for the three measurements, separated into the \textit{cal} and \textit{non-cal} phases. The curves reveal the typical footprint of the IF filter of the receiving system. Using Eq.\,(\ref{eqhotcold}) with $T_\mathrm{hot}=300\,\mathrm{K}$ and $T_\mathrm{cold}=77\,\mathrm{K}$, we computed $G(\nu)$, which is averaged over the \textit{cal} and \textit{non-cal} phases (see Fig.\,\ref{fig:hc65mmraw}, lower panel). 

\begin{figure}[!t]
\centering%
\includegraphics[width=0.45\textwidth,bb=15 42 521 392,clip=]{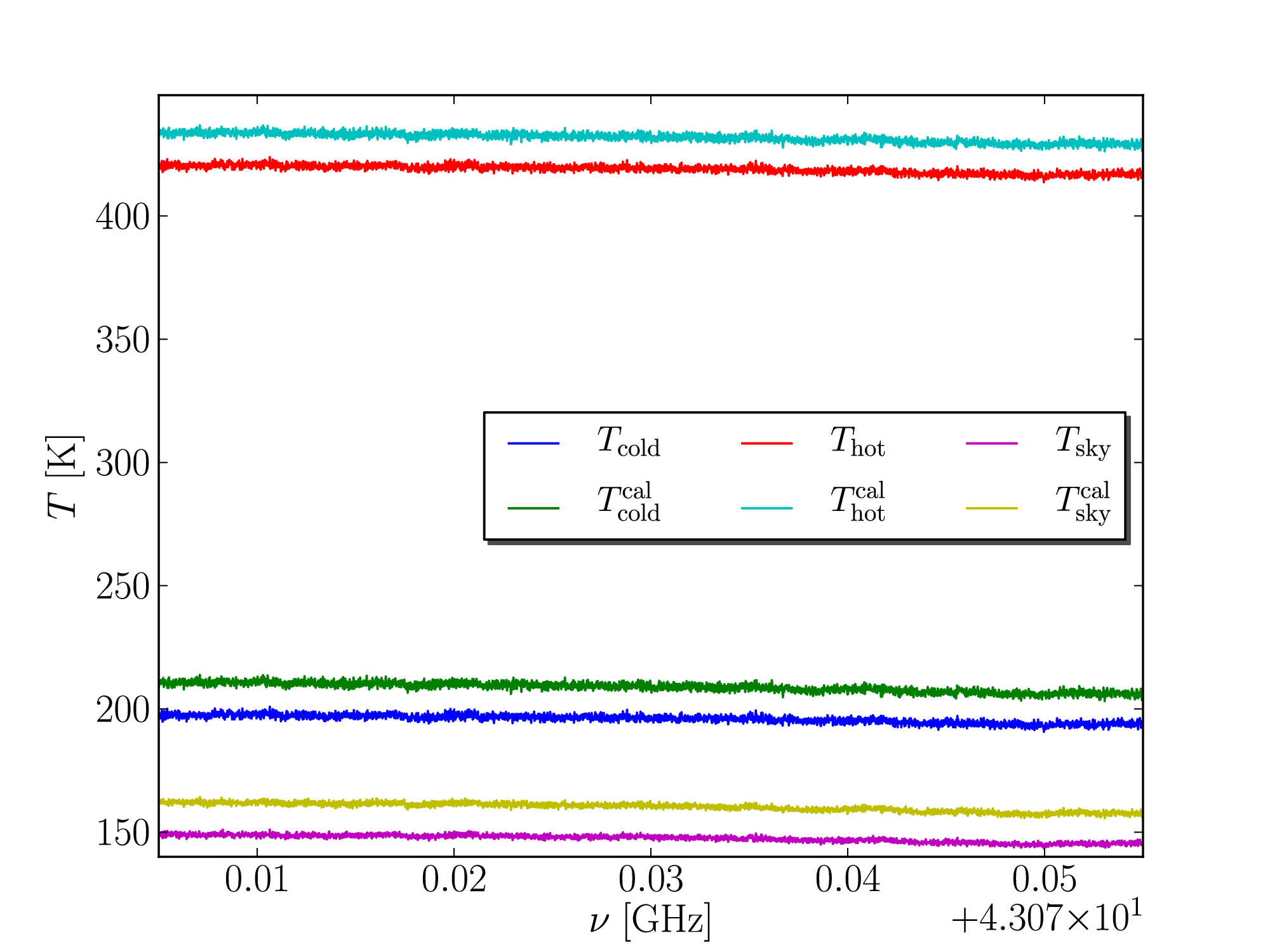}\\[0ex]
\includegraphics[width=0.45\textwidth,bb=15 1 521 392,clip=]{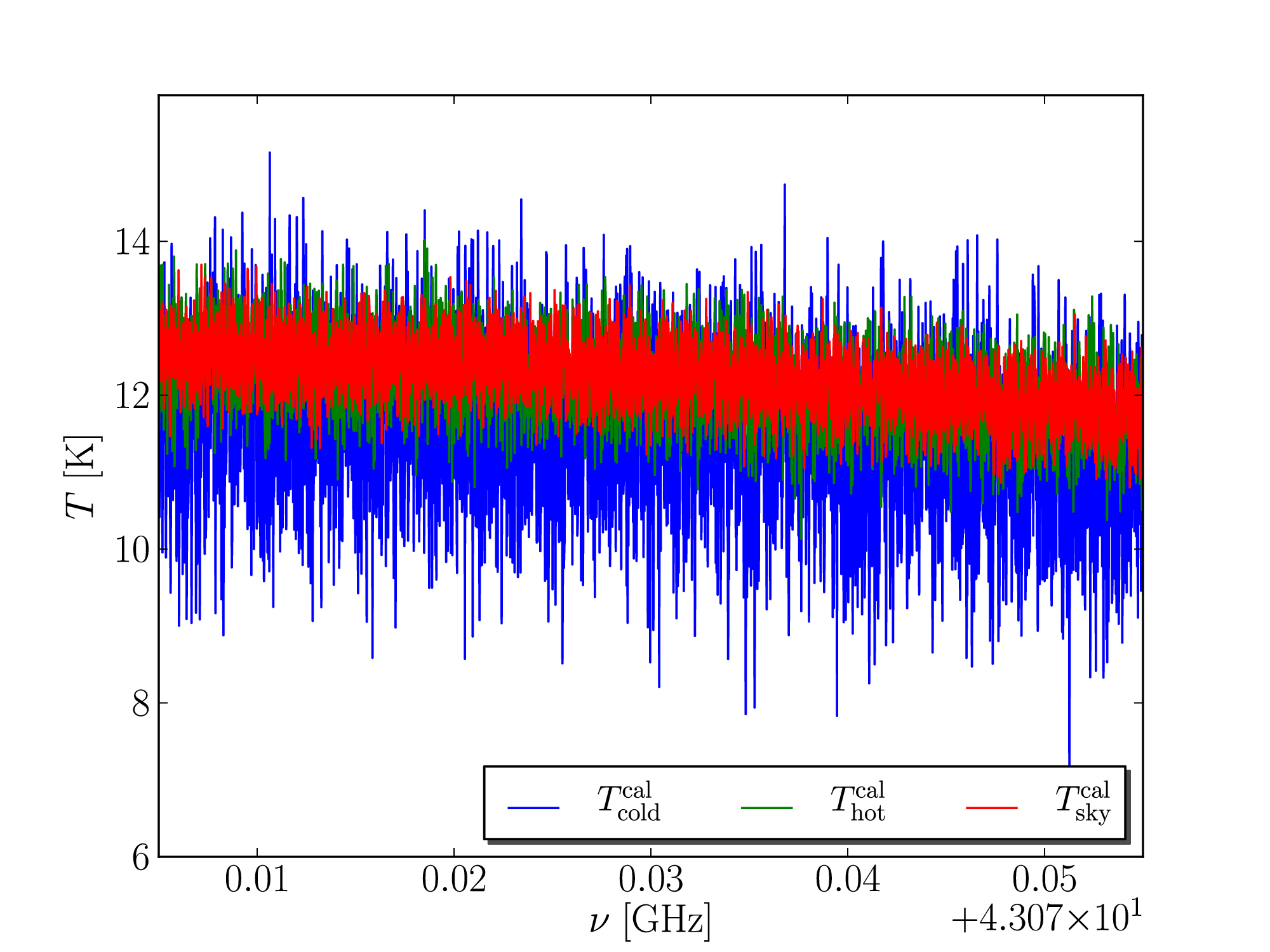}
\caption{\textbf{Upper panel:} With the determined transfer function (Fig.\,\ref{fig:hc65mmraw} lower panel), one can calibrate the raw spectra (Fig.\,\ref{fig:hc65mmraw} upper panel). The individual spectra show a slight slope. $T_\mathrm{sky}$ is equivalent to the overall system temperature $T_\mathrm{sys}$ (which of course is a function of observational parameters as telescope pointing position, temperature, atmospheric conditions, etc). \textbf{Lower panel:} Using the results from the left panel, one can recover the $T_\mathrm{cal}$ spectrum. Note that the blue curve $T_\mathrm{cold}^\mathrm{cal}$ is different from the two other measurements. (For a discussion of this issue, see text.)}%
\label{fig:hc65mmcalibrated}%
\end{figure}
Dividing the input, $P$, by the inferred gain curve, $G$, one obtains the calibrated data as plotted in Fig.\,\ref{fig:hc65mmcalibrated} (top panel).

The value of $T_\mathrm{cal}$ can be computed via Eq.\,(\ref{eqtcal}) (see Fig.\,\ref{fig:hc65mmcalibrated}, lower panel). As can be seen, the $T_\mathrm{cal}$ spectrum obtained from the cold load measurement is much noisier and has a different mean value. This is a very strong indication that the receiver was already drifting, probably because the feed was exposed to the nitrogen for too long time. Nevertheless, the results illustrate the frequency-dependent behaviour of the calibration signal, having no `ripples' or similar defects but a substantial slope considering the rather small bandwidth observed. We note that the effective bandwidth available is smaller than the nominal 100\,MHz owing to edge effects. The plot was computed for the inner 50\,MHz of the spectrum.

Using the calibrated spectra for the hot-, cold-, and sky-measurements, some interesting conclusions can be drawn. As for the hot and cold spectra we can assume that
\begin{equation}
T_\mathrm{sys}^\mathrm{h,c}=T_\mathrm{rx}+T_\mathrm{h,c},
\end{equation}
one can infer the receiver noise contribution, $T_\mathrm{rx}$. From Fig.\,\ref{fig:hc65mmcalibrated} (top panel), we have at $\nu=43.1\,\mathrm{GHz}$ a $T_\mathrm{sys}^\mathrm{h}\approx419\,\mathrm{K}$ and $T_\mathrm{sys}^\mathrm{c}\approx196\,\mathrm{K}$, hence $T_\mathrm{rx}\approx119\,\mathrm{K}$. 
At 6.5-mm wavelength, the system temperature on `blank' sky (in zenith position) is given approximately by
\begin{equation}
T_\mathrm{sys}^\mathrm{sky}=T_\mathrm{rx}+T_\mathrm{atm}+T_\mathrm{CMB}+T_\mathrm{spill,ohmic}.
\end{equation}
The antenna gain efficiency can be neglected, as the contributions can be treated as constant over the beam pattern. With $T_\mathrm{sky}\approx148\,\mathrm{K}$ and $T_\mathrm{CMB}\approx3\,\mathrm{K}$, it follows that $T_\mathrm{atm}+T_\mathrm{spill,ohmic}\approx26\,\mathrm{K}$. Spill-over can be neglected at this frequency (owing to the strong tapering, $T_\mathrm{spill}\lesssim1\,\mathrm{K}$) and ohmic losses are less than 1\,K (the reflectivity of the aluminium panels is $\sim0.999$). The measured value for the $T_\mathrm{atm}\approx25\,\mathrm{K}$ (at zenith) can be converted into the radiation temperature of the atmosphere, $T_\mathrm{atm}^\ast$, using
\begin{equation}
T_\mathrm{atm}=T_\mathrm{atm}^\ast\left[1-\exp(-\tau_0\mathrm{AM})\right].
\end{equation}
For a zenith opacity of $\tau_0=0.1$ \citep[obtained from an atmospheric model;][]{pardo06}, we calculate $T_\mathrm{atm}^\ast\approx263\,\mathrm{K}$ . For the sky measurement, the telescope was pointing toward zenith, hence $\mathrm{AM}=1$. However, since weather conditions were very good, $\tau$ might have been somewhat lower, which would lead to an increase in the inferred $T_\mathrm{atm}^\ast$. We measured an ambient temperature of $T_\mathrm{amb}\approx275\,\mathrm{K}$, air pressure of $P=977\,\mathrm{hPa}$, and humidity of 57\%.

\subsection{Using continuum calibrators}\label{subsec:tcalcontinuum}
\begin{figure}[!t]
\centering%
\includegraphics[width=0.45\textwidth,bb=20 42 531 402,clip=]{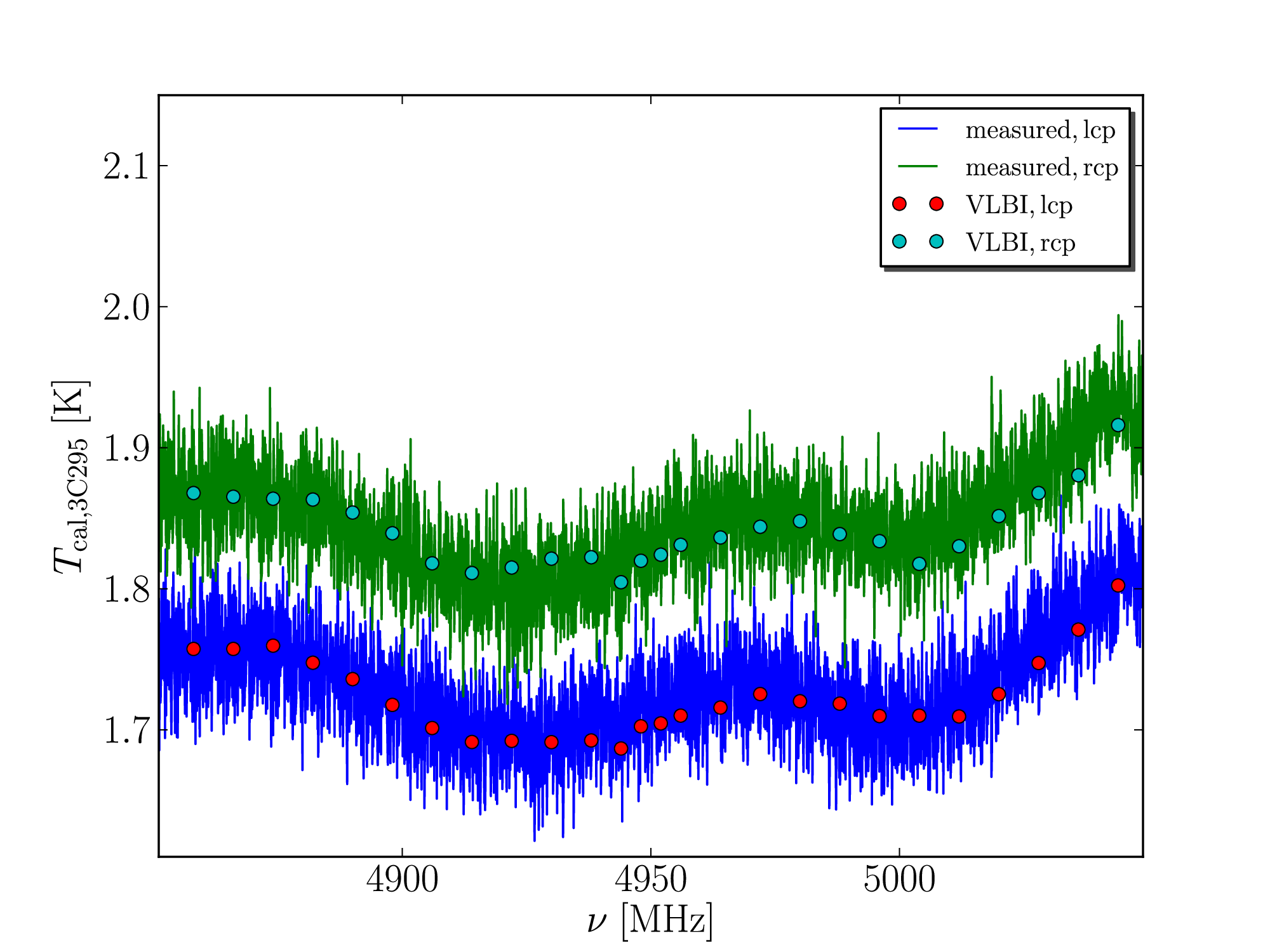}\\[0ex]
\includegraphics[width=0.45\textwidth,bb=20 1 531 402,clip=]{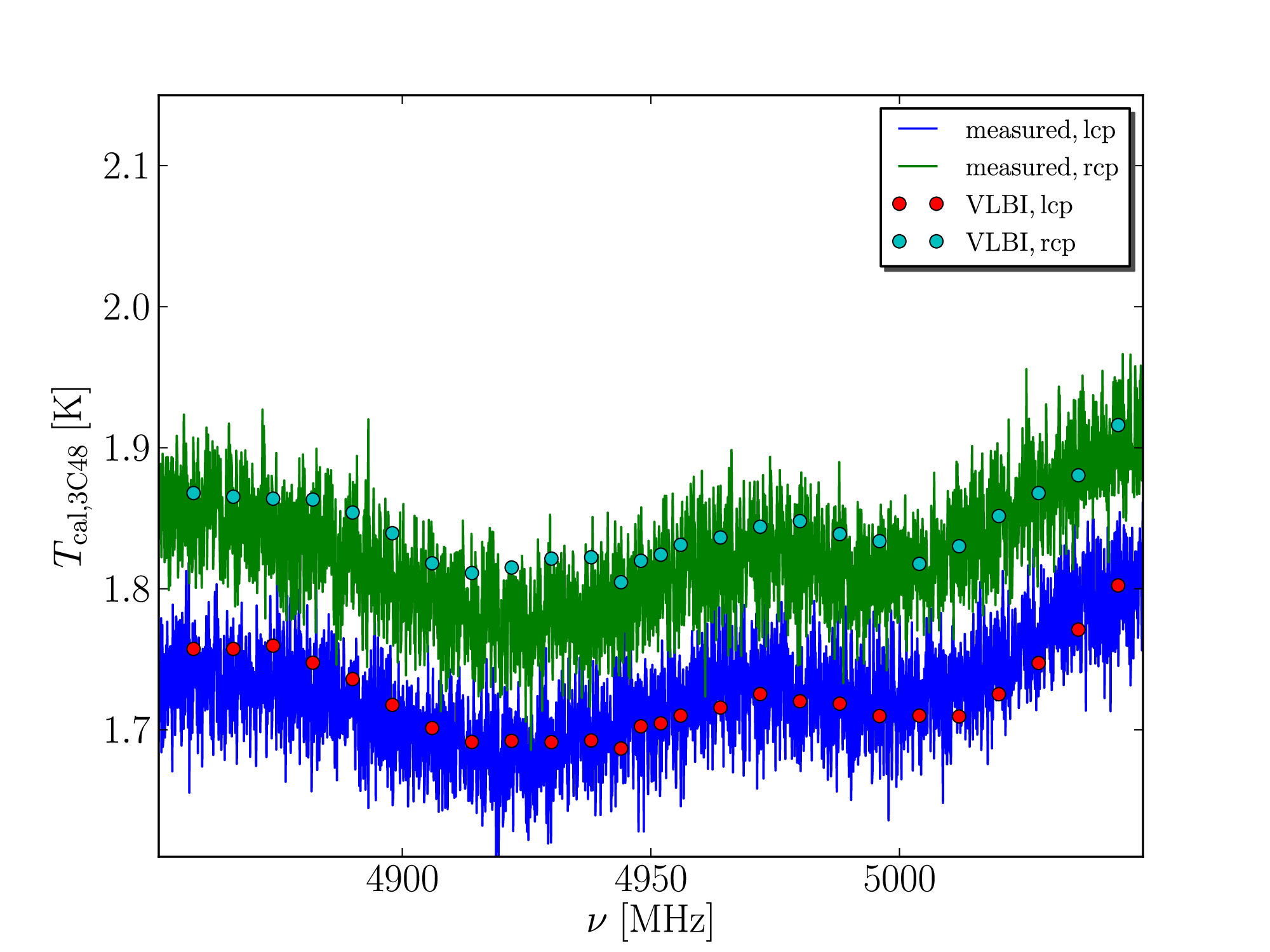}
\caption{Measurements of the $T_\mathrm{cal}$ spectrum for the 6-cm receiver at the 100-m telescope. Two continuum sources (3C\,295, upper panel; 3C\,48, lower panel) were used. Overplotted are values obtained from VLBI calibration experiments. Note that the right-hand side polarisation signals have been shifted upwards by 70\,mK for visualisation purposes.}%
\label{fig:contcal6cm}%
\end{figure}

As for the hot--cold calibration, we present our results of a $T_\mathrm{cal}$ measurement using continuum calibration sources with accurately known flux values. Several continuum sources were observed using the 6-cm receiver in the secondary focus of the 100-m. The receiving system had an effective bandwidth of 250\,MHz. The integration time \textsc{On} and \textsc{Off} source was 5\,min each. For the source fluxes of the continuum calibrators, we used new values derived by Kraus et al. (in prep.). At 6-cm, one also has to correct the spectra for an elevation-dependent antenna gain\footnote{$P_\mathrm{obs}=P\cdot\left(a_0+a_1\mathrm{El}+a_2\mathrm{El}^2\right)^{-1}$.}. Furthermore, we applied an opacity correction for the atmosphere assuming $\tau_\mathrm{zenith}=0.016$ as a typical value for the Effelsberg site at 6-cm. In Fig.\,\ref{fig:contcal6cm}, we show our results for the sources 3C\,295 and 3C\,48, along with measurements of $T_\mathrm{cal}$ from VLBI experiments.  The two methods lead to consistent results with a discrepancy of less than 3\%.

\end{document}